\documentclass[acmsmall, authorversion, screen]{acmart}

\usepackage[utf8]{inputenc} 
\usepackage[nolist]{acronym} 
\usepackage[linesnumbered, ruled, vlined]{algorithm2e} 
\usepackage{multirow} 
\usepackage{makecell} 
\usepackage{subcaption} 
\usepackage{listings} 
\usepackage{graphicx} 

\title[Federated Recommender System]{A Privacy Preserving System for Movie Recommendations Using Federated Learning}

\setcopyright{rightsretained}
\copyrightyear{2023}

\acmJournal{TORS}
\acmYear{2023}
\acmVolume{1}
\acmNumber{1}
\acmArticle{1}
\acmMonth{1}
\acmDOI{10.1145/3634686}

\author{David Neumann}
\orcid{0000-0003-1907-8329}
\affiliation
{
    \position{Scientific Researcher}
    \institution{Fraunhofer Institute for Telecommunications, Heinrich Hertz Institute, HHI}
    \department[0]{Department of Artificial Intelligence}
    \department[1]{Efficient Deep Learning Group}
    \streetaddress{Einsteinufer 37}
    \postcode{10587}
    \city{Berlin}
    \country{Germany}
}
\email{david.neumann@hhi.fraunhofer.de}

\author{Andreas Lutz}
\orcid{0000-0002-2973-0096}
\affiliation
{
    \position{Student Research Assistant}
    \institution{Fraunhofer Institute for Telecommunications, Heinrich Hertz Institute, HHI}
    \department[0]{Department of Artificial Intelligence}
    \department[1]{Efficient Deep Learning Group}
    \streetaddress{Einsteinufer 37}
    \postcode{10587}
    \city{Berlin}
    \country{Germany}
}
\email{andreas.lutz@hhi.fraunhofer.de}

\author{Karsten Müller}
\orcid{0000-0001-8611-7864}
\affiliation
{
    \position{Head of Efficient Deep Learning Group}
    \institution{Fraunhofer Institute for Telecommunications, Heinrich Hertz Institute, HHI}
    \department[0]{Department of Artificial Intelligence}
    \department[1]{Efficient Deep Learning Group}
    \streetaddress{Einsteinufer 37}
    \postcode{10587}
    \city{Berlin}
    \country{Germany}
}
\email{karsten.mueller@hhi.fraunhofer.de}

\author{Wojciech Samek}
\orcid{0000-0002-6283-3265}
\affiliation
{
    \position{Head of Department of Artificial Intelligence and Head of Explainable AI Group}
    \institution{Fraunhofer Institute for Telecommunications, Heinrich Hertz Institute, HHI}
    \department[0]{Department of Artificial Intelligence}
    \department[1]{Explainable AI Group}
    \streetaddress{Einsteinufer 37}
    \postcode{10587}
    \city{Berlin}
    \country{Germany}
}
\affiliation
{
    \position{Professor}
    \institution{Technical University of Berlin}
    \department{Department of Electrical Engineering and Computer Science}
    \streetaddress{Marchstraße 23}
    \postcode{10587}
    \city{Berlin}
    \country{Germany}
}
\email{wojciech.samek@hhi.fraunhofer.de}

\begin{abstract}
    Recommender systems have become ubiquitous in the past years. They solve the \emph{tyranny of choice} problem faced by many users, and are utilized by many online businesses to drive engagement and sales. Besides other criticisms, like creating filter bubbles within social networks, recommender systems are often reproved for collecting considerable amounts of personal data. However, to personalize recommendations, personal information is fundamentally required. A recent distributed learning scheme called federated learning has made it possible to learn from personal user data without its central collection. Consequently, we present a recommender system for movie recommendations, which provides privacy and thus trustworthiness on multiple levels: First and foremost, it is trained using federated learning and thus, by its very nature, privacy-preserving, while still enabling users to benefit from global insights. Furthermore, a novel federated learning scheme, called FedQ, is employed, which not only addresses the problem of non-i.i.d.-ness and small local datasets, but also prevents input data reconstruction attacks by aggregating client updates early. Finally, to reduce the communication overhead, compression is applied, which significantly compresses the exchanged neural network parametrizations to a fraction of their original size. We conjecture that this may also improve data privacy through its lossy quantization stage.
\end{abstract}



\newcommand{\textsb}[1]{{\libertineSB #1}}

\SetCommentSty{CustomCommentFont}
\SetArgSty{textnormal}

\begin{document}

\maketitle

\begin{acks}
    This work was created as part of the COPA EUROPE project (COllaborative Platform for trAnsmedia storytelling and cross channel distribution of EUROPEan sport events), which has received funding from the European Union's \grantsponsor{grant-sponsor:eu-horizon-2020}{Horizon 2020 Research and Innovation Programme}{https://research-and-innovation.ec.europa.eu/funding/funding-opportunities/funding-programmes-and-open-calls/horizon-2020_en} under Grant Agreement No. \grantnum[https://cordis.europa.eu/project/id/957059]{grant-sponsor:eu-horizon-2020}{957059}.
\end{acks}

\section{Introduction}

Due to the ever-increasing sizes of corpora of items such as movies, articles, games, non-digital goods, etc., the task of finding novel and engaging content or products for each individual user or customer becomes increasingly difficult, even with the help of search engines. This problem is known as the \emph{tyranny of choice}~\cite{bibliography:the-tyranny-of-choice}. Therefore, well-engineered \acp{recsys} are one of the most important pieces of technology for the success of many digital enterprises, providing them with the required engagement and sales. Harvard Business Review even calls \acp{recsys} the single most important algorithmic distinction between ``born digital'' enterprises and legacy companies~\cite{bibliography:great-digital-companies-build-great-recommendation-engines}. 80\% of the content people watch on Netflix sources from a \ac{recsys}, and they estimate that recommendations and personalization save them 1 billion USD per year~\cite{bibliography:the-netflix-recommender-system}. 35\% of what customers purchase at Amazon comes from a \ac{recsys}~\cite{bibliography:how-retailers-can-keep-up-with-customers}. At Airbnb, search ranking and similar listing recommendations drive 99\% of all booking conversions~\cite{bibliography:real-time-personalization-using-embeddings-for-search-ranking-at-airbnb}.

Accordingly, a growing number of online businesses are adopting \acp{recsys} to expand customer engagement and sales. This causes a worrying trend of companies gathering and storing continuously increasing amounts of personal customer data. Even with data protection legislation like the \acp{eu} \ac{gdpr}~\cite{bibliography:eu-gdpr} it is opaque to users what data is collected and arduous to take agency over one's personal data. All this gathered and derived personal information is at risk of being misused or leaked.

On one hand, in order to improve the personalization of customer recommendations, personal information is indispensable. On the other hand, the principles of data economy and data avoidance are essential to preserve user's privacy, and provide them with control over their own personal data. Recently, \ac{fl} was introduced as a distributed \ac{ml} method, which avoids the centralized accumulation of user data entirely and thus provides data privacy. Unlike regular \ac{ml} training algorithms with centrally collected data, \ac{fl} is designed to leave the data at its origin and instead train many models or variants of one model on each of these local datasets. The clients only share the training updates, which are then aggregated into an updated global model. As a result, all participating clients benefit from distributively training the model on all data, without ever sharing the data itself. Accordingly, this scheme, first introduced by~\citet{bibliography:federated-optimization}, is aimed towards scenarios in which the local data is privacy sensitive and thus owners do not want to disclose it.

While classical \ac{recsys} approaches usually only require user interaction data as input signals, modern approaches can use more privacy-sensitive input signals, such as age, gender, country of origin, and device information. This has the potential to further improve the predictive power of \acp{recsys}. The privacy-preserving nature of \ac{fl} makes it a perfect fit for training \ac{recsys} models without users having to give up their personal data. Furthermore, \ac{fl} helps to distribute the burden of data storage and the computational overhead of training among many clients. On the contrary, \ac{fl} also has the following disadvantages: (1) \emph{training time} will be increased as compared to traditional central training, because client devices are less capable and not always available, (2) \emph{non-\ac{iid} data} can hinder convergence and result in a model with lower performance than its centrally trained counterpart, (3) \emph{battery usage} of mobile client devices will increase due to the complex computations required to train the model, resulting in shorter battery lives, and (4) the \emph{communication overhead} of continuously exchanging training updates between the clients and the central server, which is especially problematic when clients are on a metered mobile connection.

The exploration of the combination of \ac{fl} and \acp{recsys} towards the subfield of \acp{fedrec} has only recently started and has not yet been fully explored in the literature with only a few publications available on this topic. Therefore, this work introduces an end-to-end, high-performance, scalable \ac{fedrec} solution for movie recommendations, which is entirely driven by \ac{fl} and addresses common issues of \acp{fedrec}. System scalability is verified through experiments conducted on more than 162,000 \ac{fl} clients. To our best knowledge, this is the first work with this client range.

The proposed system inherently provides privacy and thus trustworthiness on several levels: First, through the federated training that only transmits \ac{nn} parameters, while every participating client's personal data remains private. Second, early aggregation of client updates prevents input data reconstruction attacks. And third, we apply lossy \ac{nnc} compression methods that not only provide significant communication reduction, but we also conjecture that its quantization acts as a parameter obfuscation and thus may also strengthen the \ac{fl} setup against input data reconstruction attacks.

A common challenge among \acp{recsys} is that most users only produce extraordinarily little training data, while a tiny fraction of highly engaged users produce a lot of training data. In classical \acp{recsys} this is primarily an issue because these few users dominate the \ac{recsys} and suppress the interests of less frequent users. In \acp{fedrec} this poses the additional problem of small noisy updates, which can hinder global model convergence. To counteract this problem, we introduce a technique to chain client trainings together in a privacy-preserving manner, in order to produce more stable model updates. To summarize, our contributions are as follows:

\begin{enumerate}
    \item A privacy-preserving movie \ac{recsys} trained end-to-end using \ac{fl}
    \item Extreme scalability with experimental evidence for more than 162,000 clients
    \item Compressed communication between central server and clients with state-of-the-art \ac{nnc}
    \item Novel queue-based federated training to address non-\ac{iid} and imbalanced local datasets
\end{enumerate}

\section{Related Work}

\subsection{Recommender Systems}

Initially \acp{recsys} for collaborative filtering tasks were often modeled using matrix factorization techniques. The general idea is to embed input signals, like users and items, in a joint latent space, and quantify the similarity between them using an interaction function, which is the dot product in the simplest case~\cite{bibliography:matrix-factorization-techniques-for-recommender-systems}. Several approaches in the literature were introduced to enhance the predictive power of the model, e.g., incorporating additional features~\cite{bibliography:feature-based-matrix-factorization} or combining it with neighborhood models~\cite{bibliography:factorization-meets-the-neighborhood}. Since matrix factorization relies on linear dependencies between the input signals, substituting any arbitrary function for the inner product led to promising results. \citeauthor{bibliography:neural-collaborative-filtering} utilized a \ac{dnn} for this task, which proved to be better suited for capturing the latent structures in the data, resulting in higher prediction accuracy~\cite{bibliography:neural-collaborative-filtering, bibliography:dnns-for-youtube-recommendations}. Several architectures in the context of \ac{dl} were proposed to further improve the baseline models for \acp{recsys}. \citeauthor{bibliography:recommendation-system-with-hierarchical-recurrent-neural-network-for-long-term-time-series} utilized a \ac{rnn} to include time series data from the previous items the user has interacted with~\cite{bibliography:recommendation-system-with-hierarchical-recurrent-neural-network-for-long-term-time-series}. To address the limits of \acp{rnn} for sequential recommendations~\citeauthor{bibliography:personalized-top-n-sequential-recommendation-via-convolutional-sequence-embedding} proposed a \ac{cnn} to incorporate the fact that dependency relations were not necessarily the consequence of consecutive user-item interactions~\cite{bibliography:personalized-top-n-sequential-recommendation-via-convolutional-sequence-embedding}. \citeauthor{bibliography:autoRec-autoencoders-meet-collaborative-filtering} used an item-based autoencoder to reconstruct ratings received as an input~\cite{bibliography:autoRec-autoencoders-meet-collaborative-filtering}. \citeauthor{bibliography:collaborative-denoising-auto-encoders-for-top-N-recommender-systems} enhanced this approach by employing a denoising autoencoder, which can handle corrupted data~\cite{bibliography:collaborative-denoising-auto-encoders-for-top-N-recommender-systems}. \citeauthor{bibliography:graph-convolutional-neural-networks-for-web-scale-recommender-systems} used a \ac{gcn} to combine graph convolutions and efficient random walks to solve scalability issues faced in web-scale recommendation tasks~\cite{bibliography:graph-convolutional-neural-networks-for-web-scale-recommender-systems}.

\subsection{Federated Learning}
\label{section:related-work-federated-learning}

\Ac{fl} is a recently proposed distributed learning scheme, which was originally proposed by~\citet{bibliography:federated-optimization}, where a set of client devices $C$, jointly train an \ac{ml} model $M_{Global}$ on their private datasets $\mathcal{D}_i $. Usually, \ac{fl} is performed under the supervision of a central coordinating server. In traditional \ac{ml} the local client datasets would be accumulated into a central dataset $\mathcal{D}_{Central} = \bigcup_{i=1}^{\left|C\right|} \mathcal{D}_i$ on which a central model $M_{Central}$ is trained. In \ac{fl}, the local datasets are never disclosed by the clients. Instead, the central server initializes a global model $M_{Global}$ parameterized by a vector $\boldsymbol{\theta} \in \mathbb{R}^d$, that is sent to all clients $c_i$ which train the global model on their local datasets $\mathcal{D}_i$, effectively optimizing their local objective $\mathcal{L}_i(\mathcal{D}_i; \boldsymbol{\theta})$, which results in a local update $U_i$\footnote{Depending on the specific \acs*{fl} algorithm that is being used, these local updates are of different type, e.g., in \acs*{fedsgd} an update is represented by the gradient~\cite{bibliography:federated-optimization}, in \acs*{fedavg} the update is represented by the parametrization of the updated local model~\cite{bibliography:communication-efficient-learning-of-deep-networks-from-decentralized-data}, and in federated distillation the update is represented by the soft labels that were produced by the updated local model on a central training dataset~\cite{bibliography:communication-efficient-on-device-machine-learning}.}. The local update is then sent back to the central server, which uses an aggregation operator to combine the updates into an updated global model $M^{'}_{Global} = \text{Agg} \left\{U_i \mid i \in \left\{1, 2, \dots, |C|\right\} \right\}$. This process is repeated until a suitable convergence metric is met. The objective of \ac{fl} can therefore be stated as the following minimization problem~\cite{bibliography:communication-efficient-learning-of-deep-networks-from-decentralized-data}:

\begin{align}
    \min_{\boldsymbol{\theta} \in \mathbb{R}^d} \sum_{i=1}^{\left|C\right|} \frac{\left|\mathcal{D}_i\right|}{\left |\bigcup_{j=1}^{\left|C\right|} \mathcal{D}_j \right |} \mathcal{L}_i(\mathcal{D}_i; \boldsymbol{\theta}), \text{where} \qquad \mathcal{L}_i(\mathcal{D}_i; \boldsymbol{\theta}) = \frac{1}{\left|\mathcal{D}_i\right|} \sum_{x, y \in \mathcal{D}_i} \ell(x, y; \boldsymbol{\theta}),
\end{align}

with $\ell(x, y; \boldsymbol{\theta})$ denoting the loss of the client model on input $x$ with ground-truth $y$, given the model parametrization $\boldsymbol{\theta}$. \Ac{fl} allows the global model $M_{Global}$ to train on significantly more data than if each client had only trained on its private data. Thus, under ideal conditions, given a performance metric $P$, the performance of the global model $P_{Global}$ should be better than that of each individual client $\forall i \in \left\{1, 2, \dots, |C|\right\}: P_{Global} > P_i$. \Ac{fl} permits a certain degree of deviation from the performance of an equivalent centrally trained model but provides data security and privacy protection in return. Still, the goal is to minimize the deviation $|P_{Central} - P_{Global}|$.

In the original \ac{fl} scheme, \ac{fedsgd}, proposed by~\citet{bibliography:federated-optimization}, the clients perform a training step and send the computed gradient back to the central server, which averages the gradient across all clients and applies it to the global model. Since then, several other methods have been proposed in the literature. \citeauthor{bibliography:communication-efficient-learning-of-deep-networks-from-decentralized-data} proposed \ac{fedavg}, where the clients train for multiple local epochs and send their updated local model to the central server instead of the gradient. The updated parameters are then weighted proportionally by the number of local training samples available to each client and then averaged by the central server~\cite{bibliography:communication-efficient-learning-of-deep-networks-from-decentralized-data}. Furthermore, they employ client sub-sampling, a technique where only a random subset of clients is selected for each communication round~\cite{bibliography:a-general-theory-for-client-sampling-in-federated-learning, bibliography:optimal-client-sampling-for-federated-learning}. \ac{fedavg} can be seen as a generalization of \ac{fedsgd}, which only executes a single iteration of gradient descent in each round of communication~\cite{bibliography:privacy-preserving-deep-learning, bibliography:communication-efficient-learning-of-deep-networks-from-decentralized-data}. Although there were theoretical guarantees for the convergence of \ac{fedavg} in cases of heterogeneous data, impractical assumptions such as strong convexity or smoothness of the objective function needed to hold~\cite{bibliography:on-the-convergence-of-fedavg-on-non-iid-data}. \citeauthor{bibliography:fedeval-a-benchmark} showed experimentally, that \ac{fedavg} could lose up to 9\% accuracy in comparison to \ac{fedsgd}~\cite{bibliography:fedeval-a-benchmark}, when dealing with non-\ac{iid} data. \citeauthor{bibliography:federated-optimization-in-heterogeneous-networks} tackled this problem and presented a generalization of \ac{fedavg}. They introduced a surrogate objective to constrain the locally updated parameters to be close to the current global model. This helped to stabilize convergence behavior resulting in a significant increase in test accuracy by 22\% on average~\cite{bibliography:federated-optimization-in-heterogeneous-networks}. \citeauthor{bilbiography:federated-learning-on-non-iid-features-via-local-batch-normalization} proposed to only share the trainable parameters of \ac{batchnorm} with the central server without communicating their running averages of the batch statistics to the server. Aggregating the trainable parameters from all clients but keeping the running averages local helps to alleviate the problem of feature shift in non-\ac{iid} training scenarios~\cite{bilbiography:federated-learning-on-non-iid-features-via-local-batch-normalization}. \citeauthor{bilbiography:stochastic-controlled-averaging-for-on-device-federated-learning} utilize control variates as a variance reduction technique to approximate the update direction of the server model and each client model. The client drift, which naturally arises from training on different local data distributions, can be estimated by the difference between these update directions and is corrected by adding it in the local training of each client~\cite{bilbiography:stochastic-controlled-averaging-for-on-device-federated-learning}. \citeauthor{bibliography:class-aware-client-selection-for-effective-aggregation-in-federated-learning} rely on clustering the clients according to the classes of data they possess. They only average parameters from the same group while updating the central server model, guaranteeing that parameters are only averaged on a set of clients with a comparable data distribution~\cite{bibliography:class-aware-client-selection-for-effective-aggregation-in-federated-learning}. \citeauthor{bibliography:performance-enhancement-in-federated-learning-by-reducing-class-imbalance-of-non-iid-data} propose a two step approach. Firstly, they use data oversampling to eliminate data class imbalances among clients. In the second step the clients are selected in such a way, that their data distribution is nearly uniform. Furthermore, the central server constantly adjusts the amount of data for local training, the batch size, and the learning rate of the clients to avoid performance degradation~\cite{bibliography:performance-enhancement-in-federated-learning-by-reducing-class-imbalance-of-non-iid-data}. We also address data heterogeneity and introduce our own generalization of \ac{fedavg}, named \ac{fedq}.

Although \ac{fl} operates in a decentralized environment, the participating client's privacy may be compromised by merely transmitting the training update. \citeauthor{bibliography:inverting-gradients} reconstructed high-resolution images by examining the data present in each client's communicated gradients~\cite{bibliography:inverting-gradients}. \citeauthor{bibliography:data-leakage-in-federated-averaging} were also able to extract sensitive information contained in the weights obtained by the \ac{fedavg} procedure. Therefore, the concept of differential privacy~\cite{bibliography:the-algorithmic-foundations-of-differential-privacy} is often applied in the setting of \ac{fl}. When working with aggregated data, differential privacy can be utilized to protect the private information contained in individual data points. Differential privacy achieves this data protection by perturbing the data points with random noise. This exploits the fact that a single data point has relatively little impact on the aggregated data as a whole, but adding random noise alters the individual data points to a degree that no useful information can be extracted from them~\cite{bibliography:differential-privacy-a-survey-of-results}. \citeauthor{bibliography:federated-learning-with-differential-privacy} proposed to add specific noise to the parameters of each client before aggregation by the central server~\cite{bibliography:federated-learning-with-differential-privacy}. This ensures a decent training accuracy while a certain level of privacy is maintained, if there are a sufficiently large number of clients involved~\cite{bibliography:federated-learning-with-differential-privacy}. \citet{bibliography:privacy-preserving-deep-learning-via-additively-homomorphic-encryption} proposed to use homomorphic encryption in the more general setting of distributed training and~\citet{bibliography:privacy-preserving-machine-learning-with-homomorphic-encryption-and-federated-learning} suggested to use it in the setting of \ac{fl}. Homomorphic encryption is a specialized encryption scheme that allows performing certain mathematical operations on the data without decrypting it.

\subsection{Communication-Efficient Federated Learning}
\label{section:related-work-communication-efficient-federated-learning}

When dealing with mobile clients, internet connections may be inconsistent and potentially have high latency. Even when \ac{fl} clients are connected via reliable network connections, mobile connections are usually still bandwidth-constrained and, in many cases, even metered. During the course of \ac{fl}, training updates must be exchanged a multitude of times. Therefore, a central goal in \ac{fl} is communication minimization. When communicating model parametrizations, possible solutions to this include several size reduction techniques: \textbf{Sparsification/Pruning} excludes single neurons (unstructured) or entire layers of neurons (structured) from an \ac{nn}. While sparsification only sets excluded neurons to 0, pruning actually removes them~\cite{bibliography:optimal-brain-damage}. Sparsified models are more amenable to compression, but still have their original size when uncompressed. Pruned models, on the other hand, already have a reduction in size even without compression. The disadvantage of pruned networks is that they may require specialized software and/or hardware to be used, while sparsified models can run on regular software and hardware. \textbf{Distillation} is a technique for transferring the knowledge of a teacher model into a smaller student model. This is done by minimizing the difference between the output of the student model and the output of the teacher model (also known as soft labels) on data points from a separate dataset~\cite{bibliography:distilling-the-knowledge-in-a-neural-network}. In \textbf{quantization} the weights of an \ac{nn} are constrained to a discrete set of values so that they can be represented with fewer bits~\cite{bibliography:a-survey-of-quantization-methods-for-efficient-neural-network-inference}. \textbf{Lossless compression} techniques encode the \ac{nn} data in a way that removes redundancy and thus reduces its size~\cite{bibliography:deep-compression}.

There are many works that have developed communication efficient \ac{fl} solutions using the above-mentioned techniques or combinations of them~\cite{bibliography:sparse-binary-compression, bibliography:federated-learning-strategies-for-improving-communication-efficiency, bibliography:robust-and-communication-efficient-federated-learning-from-non-iid-data}, and even some with specialized techniques, such as federated dropout~\cite{bibliography:expanding-the-reach-of-federated-learning-by-reducing-client-resource-requirements}. \citeauthor{bibliography:federated-learning-strategies-for-improving-communication-efficiency} propose employing quantization, random rotations, and sub-sampling to compress the updated model parameters of the clients before sending them to the central server~\cite{bibliography:federated-learning-strategies-for-improving-communication-efficiency}. \citeauthor{bibliography:communication-efficient-federated-learning-via-knowledge-distillation} adopt an orthogonal strategy: The clients train a teacher model on their local data and distill it into a smaller student model. Instead of communicating the gradients of the teacher models, the clients compress and send the gradients of the smaller student models~\cite{bibliography:communication-efficient-federated-learning-via-knowledge-distillation}. \citeauthor{bibliography:sparse-binary-compression} introduce a compression framework combining communication delay methods, gradient sparsification, binarization, and optimal weight update encoding to reduce the upstream communication cost in distributed learning scenarios~\cite{bibliography:sparse-binary-compression}. To adapt it to the \ac{fl} setting, \citeauthor{bibliography:robust-and-communication-efficient-federated-learning-from-non-iid-data} enhance this approach, taking the compression of the downstream communication and the non-\ac{iid} local data distribution of the clients into account. They construct a framework combining a novel top-$k$ gradient sparsification method with ternarization and optimal Golomb encoding of updated client model parameters~\cite{bibliography:robust-and-communication-efficient-federated-learning-from-non-iid-data}. Another emerging field of research considers combinations of differential privacy and quantization methods in order to reduce communication costs. \citeauthor{bibliography:joint-privacy-enhancement-and-quantization-in-federated-learning} demonstrated that, within their framework, it is possible to quantize data at a given bit rate without sacrificing a specified level of privacy or degrading model performance~\cite{bibliography:joint-privacy-enhancement-and-quantization-in-federated-learning}. They enhanced methods proposed by~\citeauthor{bibliography:fedpaq-a-communication-efficient-federated-learning-method-with-periodic-averaging-and-Quantization} and~\citeauthor{bibliography:federated-learning-strategies-for-improving-communication-efficiency}, which solely use quantization and do not include privacy-related considerations.

\subsection{Federated Recommender Systems}
\label{section:related-work-federated-recommender-systems}

The current public discussion of \acp{recsys} (often just referred to as the algorithm or AI personalization), focuses, among other topics, on their invasive behavior concerning personal data collection~\cite{bibliography:artificial-intelligence-and-mass-personalization-of-communication-content, bilbiography:public-attitudes-towards-algorithmic-personalization-and-use-of-personal-data-online, bibliography:user-privacy-concerns-with-common-data-used-in-recommender-systems, bibliography:do-you-trust-your-recommendations}. This might create a negative relationship between user and \ac{recsys} potentially resulting in anything from user discontent to ``algorithmic hate''~\cite{bibliography:recommender-systems-and-algorithmic-hate}. \Acp{recsys} are arguably a vital part of the user experience on the internet since, without them, the flood of content would be barely manageable. Therefore, \ac{fl} may be part of the solution to the privacy problem of \acp{recsys} by training the recommender models directly on user devices and thereby entirely circumventing the need for gathering private information.

\Ac{fl} has already been proven to work well in many other domains, e.g., cancer research~\cite{bibliography:larynx-cancer-survival-model-developed-through-open-source-federated-learning}, natural language processing~\cite{bibliography:fednlp}, graph \acp{nn}~\cite{bibliography:fedgraphnn}, image classification~\cite{bibliography:real-world-image-datasets-for-federated-learning}, transfer learning~\cite{bibliography:a-secure-federated-transfer-learning-framework}, language models~\cite{bibliography:learning-differentially-private-recurrent-language-models}, mobile keyboard prediction~\cite{bibliography:federated-learning-for-mobile-keyboard-prediction}, and keyword spotting~\cite{bibliography:federated-learning-for-keyword-spotting}, so it is reasonable to anticipate that it is likewise effective in the domain of \acp{recsys}. In fact, there are numerous methods in the literature to incorporate current \ac{recsys} frameworks into \ac{fl}. They can be classified as either focusing on learning algorithms~\cite{bibliography:federated-collaborative-filtering-for-privacy-preserving-personalized-recommendation-system}, security~\cite{bibliography:federating-recommendations-using-differentially-private-prototypes}, or optimization models~\cite{bibliography:fedfast-going-beyond-average-for-faster-training-of-federated-recommender-systems}, depending on the task's objective~\cite{bibliography:federated-recommenders-methods-challenges-and-future}. Matrix factorization is a commonly utilized approach in the first scenario. \citeauthor{bibliography:federated-collaborative-filtering-for-privacy-preserving-personalized-recommendation-system} were among the pioneers in this emerging field by introducing this model to address collaborative filtering tasks in the context of \ac{fl}. They constructed a \ac{recsys} that gives personalized recommendations based on users' implicit feedback~\cite{bibliography:federated-collaborative-filtering-for-privacy-preserving-personalized-recommendation-system}. \citeauthor{bibliography:fedrec-federated-recommendation-with-explicit-feedback} designed a new federated rating prediction mechanism for explicit responses. They employed user averaging and hybrid filling in order to keep the system computationally efficient and the communication costs moderately low~\cite{bibliography:fedrec-federated-recommendation-with-explicit-feedback}.

To increase the model capabilities for each client, \citeauthor{bibliography:personalized-recommendation-algorithm-for-mobile-based-on-federated-matrix-factorization} incorporated a bias term for the input signals. Additionally, weights on the local devices were adjusted, so that any unreasonable user rating is removed~\cite{bibliography:personalized-recommendation-algorithm-for-mobile-based-on-federated-matrix-factorization}. On the other hand, \citeauthor{bibliography:federated-multi-view-matrix-factorization-for-personalized-recommendations} employed a similar strategy, enhancing the model's capacity by incorporating input from other data sources~\cite{bibliography:federated-multi-view-matrix-factorization-for-personalized-recommendations}. \citeauthor{bibliography:demystifying-model-averaging-for-communication-efficient-federated-matrix-factorization} introduced a new algorithmic approach by combining matrix factorization with \ac{fedavg}. They demonstrated, that the cost of communication with the central server for non-\ac{iid} data was decreased by limiting the number of local training iterations~\cite{bibliography:demystifying-model-averaging-for-communication-efficient-federated-matrix-factorization}.

As previously shown, private information can be reconstructed from the clients' transmitted parameters. In order to remedy this, a variety of privacy preserving techniques based on encryption, obfuscation, or masking can be utilized~\cite{bibliography:a-comprehensive-survey-on-privacy-preserving-techniques-in-federated-recommendation-systems}. Communication of encrypted data between the central server and its clients is made possible through the use of homomorphic encryption, allowing for intermediate calculations without the need to first decrypt the data. As a result, the central server is unable to infer the data it is working with~\cite{bibliography:efficient-privacy-preserving-matrix-factorization-for-recommendation-via-fully-homomorphic-encryption}. For this reason, \citeauthor{bibliography:secure-federated-matrix-factorization} propose a secure matrix factorization framework to handle data leakage. They showed how privacy could be compromised by intercepting the clients' gradient updates sent in two consecutive communication rounds to the central server. To address this problem, they encrypted the clients' gradients before sending them to the central server~\cite{bibliography:secure-federated-matrix-factorization}. \citeauthor{bibliography:a-vertical-federation-recommendation-method-based-on-clustering-and-latent-factor-model} enhanced the approach by clustering the encrypted user embeddings to reduce the dimension of the user-item matrix, improving the recommendation's accuracy~\cite{bibliography:a-vertical-federation-recommendation-method-based-on-clustering-and-latent-factor-model}. \citeauthor{bibliography:federated-recommendation-via-fake-marks-and-secret-sharing} utilized a different cryptographic technique: they applied secret sharing, wherein a group of clients can only reconstruct sensitive information if they collaborate by combining their shares~\cite{bibliography:how-to-share-a-secret}. By applying this concept to the clients' locally computed gradients, the authors managed to construct a \ac{fedrec} framework that provides strong privacy guarantees on the clients' individual data~\cite{bibliography:federated-recommendation-via-fake-marks-and-secret-sharing}. Another technique concerns secure multi-party computation, that refers to a protocol for computing a function based on the data of a group of clients without disclosing private information to one another~\cite{bibliography:secure-multiparty-computation-and-secret-sharing}. \citeauthor{bibliography:federated-neural-collaborative-filtering} utilized this approach in the setting of federated \ac{ncf}. They demonstrated that employing a secure multi-party computation protocol for \ac{fedavg} protects privacy when dealing with an honest but curious entity without compromising the quality of the \ac{recsys}~\cite{bibliography:federated-neural-collaborative-filtering}.

Differential privacy falls in the category of privacy preservation techniques that use obfuscation. \citeauthor{bibliography:federating-recommendations-using-differentially-private-prototypes} added differential privacy to \ac{fl} utilizing a matrix factorization technique. They succeeded in balancing the privacy loss posed by the repetitive nature of the \ac{fl} process by only requiring a few rounds of communication~\cite{bibliography:federating-recommendations-using-differentially-private-prototypes}. \citeauthor{bibliography:federated-collective-matrix-factorization-for-heterogeneous-collaborative-filtering} designed a matrix factorization-based \ac{recsys} that adds Laplacian random noise to the users' encrypted item embeddings, ensuring a high level of security~\cite{bibliography:federated-collective-matrix-factorization-for-heterogeneous-collaborative-filtering}. \citeauthor{bibliography:stronger-privacy-for-federated-collaborative-filtering-with-implicit-feedback} proposed a system combining differential privacy and implicit user feedback. They constrained the number of local gradient updates sent by the users by the level of privacy each user tries to maintain~\cite{bibliography:stronger-privacy-for-federated-collaborative-filtering-with-implicit-feedback}. We also address the problem of privacy preservation by obfuscation: Instead of applying random noise to the weight updates that are sent to the central server, the weights are quantized, which is both conducive to privacy preservation and reducing the communication overhead. We later provide a detailed attack analysis of the exchanged model parameters, that are potentially susceptible to leak information about the underlying datasets of the participating clients. We present specific attacks applicable to our scenario and examine how their requirements and assumptions do not apply to our approach to privacy preservation, thus rendering them ineffective.

Another method of achieving data security is by introducing pseudo interactions in order to mask user behavior in \acp{fedrec}. This protection mechanism is implemented by adding artificial interactions with randomly selected items to users. This causes the central server to be unable to determine the real set of items a user has interacted with, as the uploaded gradient was computed with respect to both real and artificial interactions.~\cite{bibliography:fedrec-federated-recommendation-with-explicit-feedback}. Since this method produces noisy gradients, degrading the model performance, \citeauthor{bibliography:lossless-federated-recommendation-with-explicit-feedback} introduced denoising clients in the training process~\cite{bibliography:lossless-federated-recommendation-with-explicit-feedback}. Another approach that hits the same mark, but entirely foregoes \ac{fl} was presented by \citet{bibliography:efficient-privacy-preserving-recommendations-based-on-social-graphs}. They employ a random walk-based approach to decentralized optimization, where a randomly chosen client trains its local model for one or multiple epochs before sending its updated parameters to a randomly selected neighboring client according to the underlying graph structure~\cite{bibliography:adaptive-random-walk-gradient-descent-for-decentralized-optimization, bibliography:decentralized-learning-with-random-walks-and-communication-efficient-adaptive-optimization}. \citeauthor{bibliography:efficient-privacy-preserving-recommendations-based-on-social-graphs} adapt this approach to account for privacy by introducing the anonymous random walk technique where clients, instead of training a model, can choose to add their own data to an existing dataset that was sent by a neighboring client in a prior round. The accumulated data can then be uploaded to the central server for centralized training. Due to the nature of the random walk, neither the clients nor the central server know where the individual samples of the accumulated dataset originate from, thus effectively masking the users' identities.

Dealing with the statistical heterogeneity of the clients' local data in the context of \acp{fedrec} is a different area of research. There are various proposed strategies for addressing this issue, which primarily include clustering and meta learning~\cite{bibliography:a-survey-on-federated-recommendation-systems}. \citeauthor{bibliography:personalized-federated-recommendation-system-with-historical-parameter-clustering} designed a \ac{fedrec} utilizing a clustering approach based on historical parameters to form homogeneous groups of clients, in which a personalized model can be trained. These parameters are retrieved by averaging the model parameters from the clients' last communication rounds with the central server~\cite{bibliography:personalized-federated-recommendation-system-with-historical-parameter-clustering}. \citeauthor{bibliography:federated-meta-learning-with-fast-convergence-and-efficient-communication} proposed a different method based on model-agnostic meta-learning, which is a training paradigm where a meta-learner is employed to rapidly train models on new tasks. The meta-learner itself is a trainable algorithm that trains a model on a task, which consists of a support set and a query set. The model is trained using the support set and then evaluated on the query set. Based on this evaluation, a loss is computed, which reflects the ability of the meta-learner to train the model. The meta-learner is then updated to minimize this loss. For example, the meta-learner in the \ac{maml}~\cite{bibliography:model-agnostic-meta-learning-for-fast-adaptation-of-deep-networks} algorithm is used to provide an initial set of parameters for the model that is trained on the task. Meta-learning algorithms are known to generalize effectively to new tasks, which makes them well-suited for tackling the non-\ac{iid} problem in \ac{fl}. For this reason, \citeauthor{bibliography:federated-meta-learning-with-fast-convergence-and-efficient-communication} adapted \ac{maml}, as well as another meta-learning algorithm called Meta-SGD, to the \ac{fl} setting, which enabled them to reach higher model performance than the \ac{fedavg} baseline~\cite{bibliography:federated-meta-learning-with-fast-convergence-and-efficient-communication}. Our \ac{fedrec} was not only affected by heterogeneous client data but also by exceedingly small local datasets. Our approach to non-\ac{iid}-ness, \ac{fedq}, therefore differs greatly from the two above-mentioned approaches, as neither clustering nor meta-learning are capable of handling truly small local datasets.

The clients' potentially constrained resources are the subject of another line of research. Therefore, \citeauthor{bibliography:fedfast-going-beyond-average-for-faster-training-of-federated-recommender-systems} utilized a simple \ac{dnn} with small embedding sizes to balance the number of learnable parameters and the accuracy of the resulting recommendations. Additionally, they presented a new sampling technique coupled with an active aggregation method, which reduced communication costs and produced more accurate models even at an early stage of training~\cite{bibliography:fedfast-going-beyond-average-for-faster-training-of-federated-recommender-systems}. \citeauthor{bibliography:lightweight-federated-recommendation-with-privacy-preserving-matrix-factorization} addressed related problems and developed a new framework that effectively integrates a novel matrix factorization technique with privacy via a federated discrete optimization algorithm. Although the model's RAM, storage, and communication bandwidth requirements were modest, performance was not affected and was even superior to related state of the art techniques~\cite{bibliography:lightweight-federated-recommendation-with-privacy-preserving-matrix-factorization}. Our suggested approach combines all three of the aforementioned sorts of objectives: We balance the model complexity and capacity by opting for a simple, yet scalable \ac{dnn} architecture. This results in remaining resource-efficient on the client side, while still maintaining the possibility of scaling up. Additionally, we anticipate that applying quantization will provide a specific amount of privacy while also lowering the burden associated with exchanging parameters with the central server via potentially bandwidth-constrained network connections.

\section{Method}

In this work, we propose a framework for a \ac{recsys} that is trained end-to-end using \ac{fl}. Before examining the design of the \ac{fedrec} and its components, we want to motivate our decisions with a problem statement. Then, we will explore the general architecture of many complex information retrieval systems on which the architecture of our \ac{recsys} is based and show how each of these components is constructed. Finally, we will demonstrate how all of this translates into an \ac{fl} setting and how we alleviate the problems that arise from such a setup.

\subsection{Problem Statement}
\label{section:problem-statement}

The research documented in this work was conducted as part of the COPA EUROPE project, which is a beneficiary of the \acp{eu} Horizon 2020 Research and Innovation Programme. The project aims to create a live-streaming and \ac{vod} platform that provides users with sports and esports content. To keep users engaged, discoverability of the content is key, therefore, one part of the project aims at developing a \ac{recsys}. Specifically, the objective was to develop a \ac{recsys} in an \ac{fl} setting to provide high-quality recommendations while preserving the user's privacy. From the project's goals and objectives, the following requirements for the \ac{fedrec} can be derived:

\begin{itemize}
    \item \textbf{Large Client Population} -- A live-streaming and \ac{vod} platform for sports and esports may build a large user base, which results in an \ac{fl} client population that comprises hundreds of thousands or even millions of clients.
    \item \textbf{Large Video Catalog} -- With dozens of types of sports and esports games covered, and hundreds of leagues, tournaments and events, the catalog of live streams and \ac{vod} content may grow substantially over time.
    \item \textbf{Increased Personalization} -- The \ac{fl} setup is meant to enable the \ac{recsys} to leverage more personal user data in addition to user-item interactions for higher personalization without requiring the data to ever leave the user's device. The requirement to take advantage of more personal user data implies that the employed \ac{ml} model must be able to handle multiple data modalities and learn complex, non-linear dependencies between features contained in this data.
    \item \textbf{Substantial Communication Overhead} -- The potentially large client population leads to a very significant communication overhead for the central server. Furthermore, the clients are expected to use mobile devices that may lack a reliable, high-bandwidth internet connection. Therefore, it is of paramount importance to reduce the communication overhead incurred by the constant communication between the central server and its clients.
\end{itemize}

The following sections will detail how these requirements were translated into the architecture of the \ac{fedrec} and the design of its components. All decisions concerning architecture and design, as well as the research into privacy-preservation, scalability, \ac{nnc}, and the handling of non-\ac{iid} and imbalanced local datasets were motivated and informed by these requirements.

\subsection{Recommender System Architecture}

\begin{figure}[ht]
    \centering
    \includegraphics[width=0.5\textwidth]{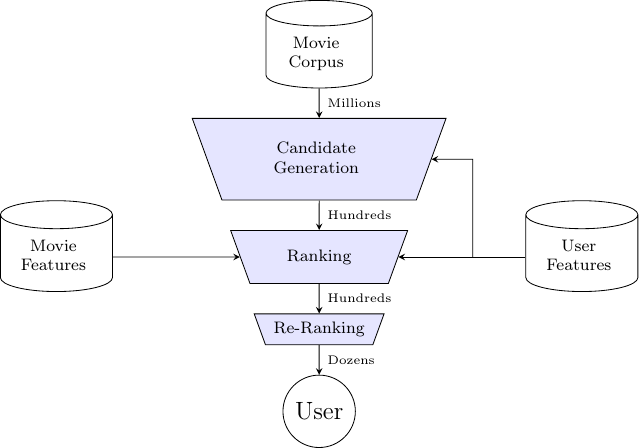}

    \caption{Flow diagram of the ``funnel-like'' three-stage \ac{recsys} architecture of the proposed \ac{recsys}, consisting of candidate generation, ranking, and re-ranking stages (inspired by Figure 2 in~\cite{bibliography:dnns-for-youtube-recommendations}).}
    \Description{Flow diagram of the ``funnel-like'' three-stage \ac{recsys} architecture of the proposed \ac{recsys}, consisting of candidate generation, ranking, and re-ranking stages (inspired by Figure 2 in~\cite{bibliography:dnns-for-youtube-recommendations}).}
    \label{figure:recommender-system-architecture}
\end{figure}

As the \ac{recsys} is required to handle a large user base and movie catalog, we decided to follow the well-known three-stage funnel-like architecture, which is also employed by other forms of information retrieval systems. These three phases comprise: candidate generation, ranking, and re-ranking (cf. Figure~\ref{figure:recommender-system-architecture}). The candidate generation phase takes the entire corpus of movies and narrows it down to usually a couple hundred movies that are somewhat relevant to the user. This phase must be fast because it must sift through possibly millions of movies, which in turn means that not all of the resulting elements are 100\% relevant to the user. The ranking phase has a more complex model of the user's interest. It scores each of the candidate movies and ranks them by their scores. This two-step approach to the generation of recommendations greatly expedites the retrieval process. If each item in the corpus had to be ranked individually, this process would not scale well to the large item corpora. Finally, the re-ranking phase is an optional phase, which can implement hand-crafted rules to improve recommendations. This can include rules such as removing click-bait content, enforcing age restrictions, ensuring freshness, and promoting predefined content. These systems will be further explored in the following sections.

\subsection{Candidate Generation}
\label{section:candidate-generation}

Candidate generation is comprised of an algorithm that is trained to select a small number (usually in the order of hundreds) of items from a vast corpus of items (usually in the order of millions) that are generally relevant to the user. One classical approach to candidate generation is matrix factorization. Non-linear models, such as \acp{nn}, however, are capable of forming a much deeper ``understanding'' of the latent structures in the data and \acp{nn} have been used in \acp{recsys} since at least 2016~\cite{bibliography:dnns-for-youtube-recommendations}. Although there have been attempts to adapt classical \ac{ml} algorithms for the use in \ac{fl}, e.g., matrix factorization~\cite{bibliography:federated-collaborative-filtering-for-privacy-preserving-personalized-recommendation-system}, gradient-based learning algorithms are much better suited and well-researched within the framework of \ac{fl}. Furthermore, \acp{nn} allow for much more fine-grained control over model architecture decisions and are capable of handling a diverse set of input data modalities, which is one of the project's requirements. For this reason, we decided to use a \ac{dnn} architecture for our candidate generation model.

Prior to choosing a specific design, the training objective must be formulated. For \acp{recsys} there are many different objectives that are commonly used, e.g., rating prediction, watch time prediction, click-through-rate prediction, and watch prediction. Since the algorithm has to be able to sift through millions of items, the underlying model must be simple and, most importantly, fast. Therefore, we decided to train the candidate generation model on next watch prediction. This means that it receives a list of past movie watches of a user as input and predicts a probability distribution over all movies in the corpus. The top-$k$ movies can then be interpreted as the movies that the user will most likely watch next. So instead of performing inference on all movies in the corpus, the model only has to be invoked once to retrieve a list of candidate recommendations.

\begin{figure}[ht]
    \centering
    \includegraphics[height=0.4\textheight]{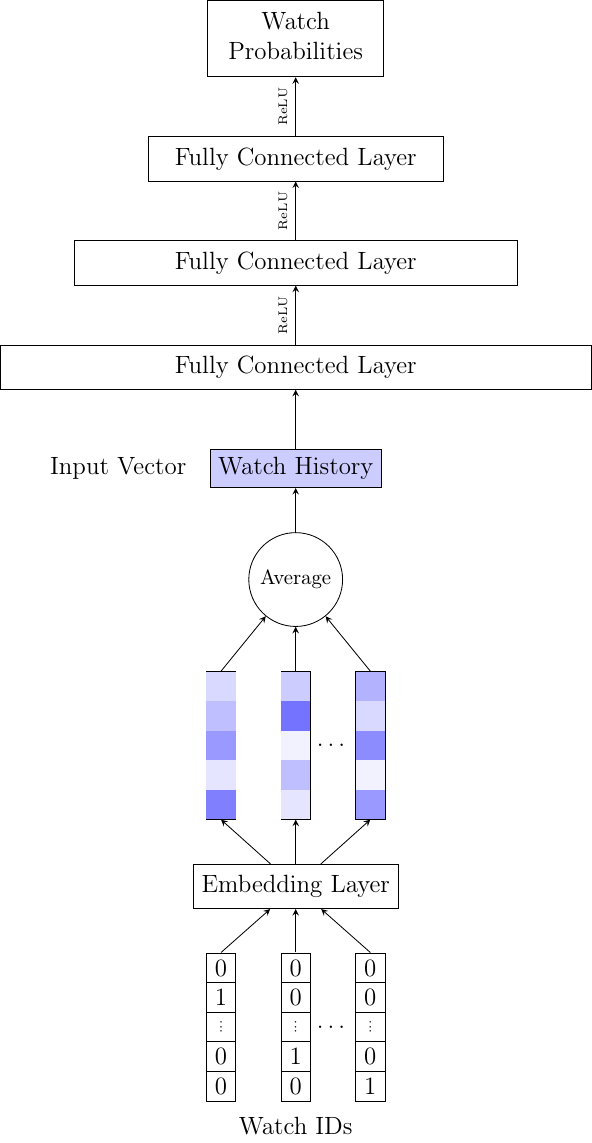}

    \caption{\Ac{dnn} candidate generator model architecture of the \ac{recsys}~\cite{bibliography:dnns-for-youtube-recommendations}.}
    \Description{\Ac{dnn} candidate generator model architecture of the \ac{recsys}~\cite{bibliography:dnns-for-youtube-recommendations}.}
    \label{figure:candidate-generator-model}
\end{figure}

The chosen architecture for the candidate generator model is shown in Figure~\ref{figure:candidate-generator-model} and inspired by the architecture used in~\cite{bibliography:dnns-for-youtube-recommendations}. An experiment using various recurrent architectures was conducted, but the chosen \ac{dnn} architecture is the best tradeoff between model performance and size. The results of this experiment can be found in Appendix~\ref{appendix:candidate-generator-model-type-experiment}. The first layer of the model is an embedding layer, which takes the sparse one-hot encoded movie watches and embeds them into a 64-dimensional dense vector space. The size of the embedding vectors was experimentally determined. The experiment results can be found in Appendix~\ref{appendix:candidate-generator-number-of-movie-embedding-dimensions}. In contrast to recurrent \acp{nn}, non-recurrent \acp{nn} require inputs of a fixed size. However, the watch histories have variable length and can consist of any number between 1 and \emph{window size} movie watches. To provide the required fixed-length input for the model, the embedded movie watches are then averaged. In practice, other input features could be added here and concatenated to the watch history vector. For example, user-level information could be utilized to improve predictions, if past movie watches are not available or a user only has a few of them, thereby solving the cold-start problem for new users. Unfortunately, we are restrained by the lack of a suitable dataset, which includes user-level information.

The inputs are then fed into a funnel, or tower-like architecture of multiple fully-connected layers with \ac{relu} activations. The final fully-connected layer prior to the output layer is of size 256 and each preceding layer doubles this number, i.e., for a three-layer architecture, the first fully-connected layer is of size 1024, the second of size 512, and the final layer of size 256. As already mentioned, the size of the model has a substantial impact in an \ac{fl} setting. Consequently, an experiment was conducted to determine the optimal number of hidden layers. The results of this experiment are presented in Appendix~\ref{appendix:candidate-generator-number-of-hidden-layers}.

Finally, the next-watch prediction is realized in terms of a classification task, therefore, the output layer of the candidate generator model has as many outputs as there are movies in the corpus. The model is then trained using the softmax cross-entropy loss. A detailed breakdown of the layers that comprise the \ac{nn} architecture of the candidate generator model is presented in Table~\ref{table:candidate-generator-model-architecture} in Appendix~\ref{appendix:candidate-generator-model-architecture}.

\subsection{Ranking}
\label{section:ranking}

The ranking phase of the \ac{recsys} receives the candidate recommendations from the candidate generator phase and ranks them by user relevance. Since it only has to be invoked for a small subset of all movies in the corpus, processing speed is less crucial in contrast to the previous candidate generator model. Therefore, a more precise and complex representation of the user's interests can be learned. Note that the model must be trained within the \ac{fl} environment and thus should not be selected too large.

Learning to rank is a well-studied~\cite{bibliography:learning-to-rank} problem within \ac{ml} and there are numerous approaches, ranging from simple point-wise models, that directly predict a rank, and pair-wise models, which learn to rank two items relative to each other, to more elaborate list-wise models, which learn to rank items in a list~\cite{bibliography:learning-to-rank-for-information-retrieval}. In the case of a movie \ac{recsys}, the ranker model can be implemented as a rating prediction, where the predicted rating is used to sort the items. We decided on this simple approach. It turned out that a simple regression model tended to learn to predict the mean rating if trained without any constraints. Therefore, we decided to re-formulate the problem as a classification task, as the dataset being used contains a discrete set of possible ratings between 0.5 and 5.0 in steps of 0.5, resulting in 10 distinct classes. This approach performs considerably better.

\begin{figure}[ht]
    \centering
    \includegraphics[height=0.4\textheight]{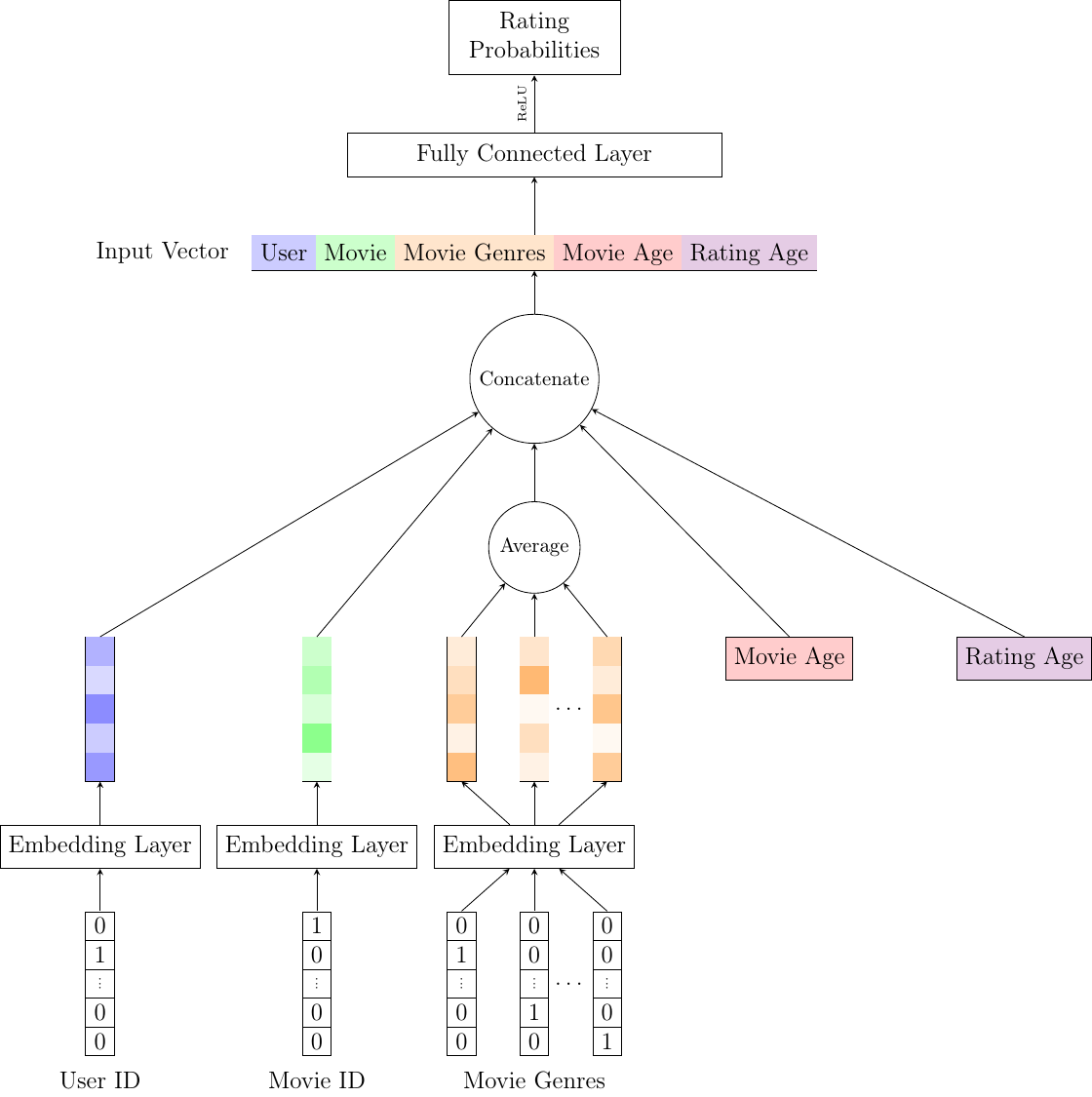}

    \caption{\Ac{dnn} ranker model architecture of the \ac{recsys}.}
    \Description{\Ac{dnn} ranker model architecture of the \ac{recsys}.}
    \label{figure:ranker-model}
\end{figure}

The base architecture of the ranker model is almost equivalent to the design of the candidate generator. The input features, user ID, movie ID, and movie genres are embedded using embedding layers. The optimal embedding sizes were experimentally determined to be 32 for users, 128 for movies, and 16 for genres. A detailed description of these experiments can be found in Appendix~\ref{appendix:ranker-embedding-layer-sizes}. The genre embeddings are then averaged and the resulting vectors are concatenated to form the input of a tower-like classifier, which consists of a single fully-connected layer that outputs a probability distribution over the set of possible ratings. Just like in the case of the candidate generator, we considered adding multiple hidden layers, but experiments with varying numbers of hidden layers determined that a single layer is sufficient. The hidden layer experiments are described in Appendix~\ref{appendix:ranker-number-of-hidden-layers}. Again, more movie-level or user-level information could be added as input features here. Only rating timestamps are provided in the dataset which can be utilized as additional user-level information. By correlating the movies in the dataset with an online movie database, further movie-level information can be retrieved. Therefore, we decided to add the age of the rating and the age of the movie as further input features to determine the efficacy of adding more input signals to the model. A detailed discussion of this can be found in Section~\ref{section:dataset-preprocessing}. The architecture of the ranker model is shown in Figure~\ref{figure:ranker-model}.

Since the classes, distinguished by the ranker model, have a hierarchical relation to each other, we considered using other loss functions than softmax cross-entropy. We have experimentally tested other loss functions, but in practice, softmax cross-entropy provides the best results. The results of the experiment can be found in Appendix~\ref{appendix:ranker-loss-function}. A detailed breakdown of the layers that comprise the \ac{nn} architecture of the ranker model is presented in Table~\ref{table:ranker-model-architecture} in Appendix~\ref{appendix:ranker-model-architecture}.

\subsection{Re-ranking}

The re-ranking phase is an optional step that is often overlooked in \ac{recsys} research, but plays a crucial role in real-world applications. It implements hand-crafted rules to improve recommendations. Examples are the removal of click-bait content, enforcing age restrictions, ensuring freshness, and promoting predefined content. Here, ensuring freshness is probably one of the most important aspects. The candidate generation and ranking phases do not take freshness of the recommended content into consideration, as the ratio between novel and more established content is often hand-tuned (also described as \emph{exploration} vs. \emph{exploitation} trade-off). Age restrictions are also important, as the candidate generator model has no filter in place to prevent recommending age-restricted movies to underage users. Both the candidate generator and the ranker models are static, i.e., given the same input, they will always produce the same output (unless further trained in the meantime). Therefore, the re-ranking phase should also randomly select a subset of the final recommendations, e.g., weighted by the rank predicted by the ranker model, in order to ensure that the user will see something different every time they are presented with recommendations. Mixing in some predefined content, for example movies that have just been released, is an effective way of overcoming the cold-start problem for new content. This would increase the chances of new movies being watched and thus generating training data that can be used to recommend the movies later. Finally, the topic of click-bait detection is an interesting one, but it is considered out-of-scope in this work. As the re-ranking phase only consists of hand-crafted rules and thus does not affect the proposed method, we will abstain from delving deeper into its implementation.

\subsection{Federated Recommender Systems at Scale Using Queue-Based Federated Learning}
\label{section:federated-recommender-systems-at-scale-using-queue-based-federated-learning}

\begin{figure}[ht]
    \centering
    \includegraphics[width=0.65\textwidth]{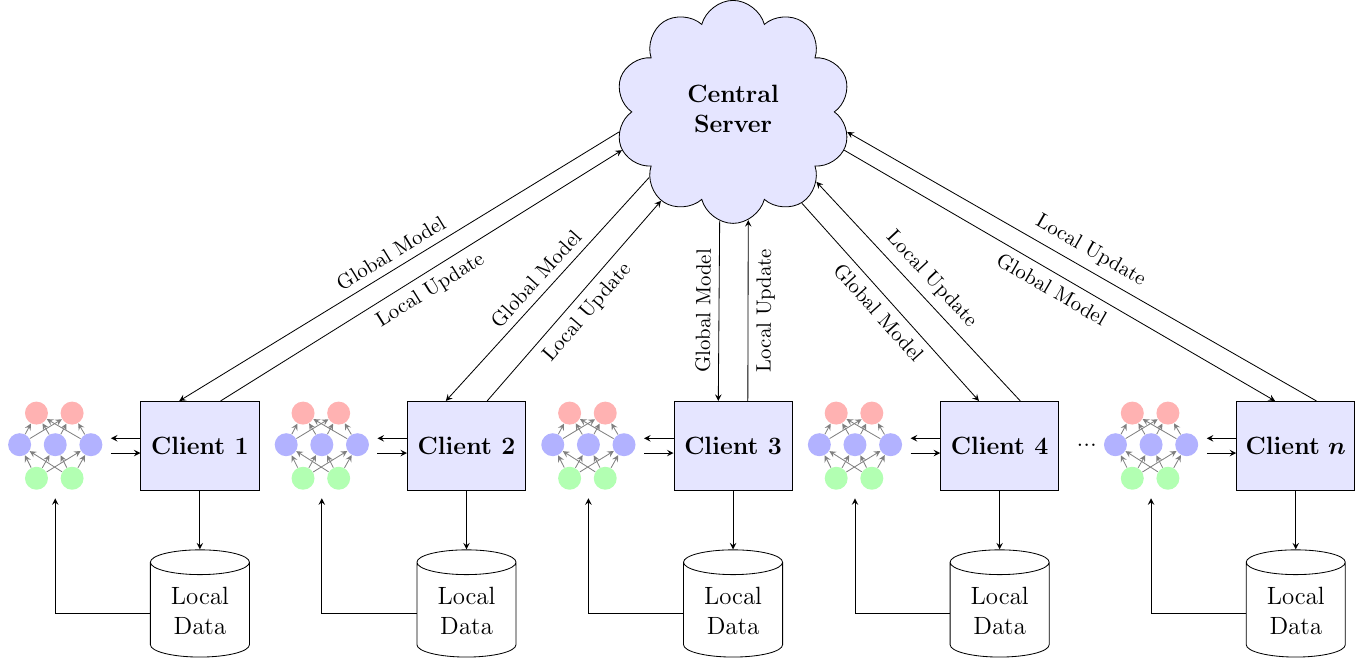}

    \caption{The typical \ac{fedavg} scenario with a central coordinating server and several clients with their local data. The central server sends a global model to the clients, which then perform training on local data. The resulting updated local models are sent back to the central server, which aggregates them into a new global model by averaging the model weights.}
    \Description{The typical \ac{fedavg} scenario with a central coordinating server and several clients with their local data. The central server sends a global model to the clients, which then perform training on local data. The resulting updated local models are sent back to the central server, which aggregates them into a new global model by averaging the model weights.}
    \label{figure:federated-averaging}
\end{figure}

Many variants and adaptations were introduced to \ac{fl}, among which \ac{fedavg}~\cite{bibliography:communication-efficient-learning-of-deep-networks-from-decentralized-data} is one of the most prevalent. In \ac{fedavg}, the server initializes a global model, which is sent to all clients. The clients then proceed to train the model on their local data and send the updated model back to the central server. The central server then aggregates the client models into a new global model by averaging them (usually the mean weighted by the number of samples that the clients trained on is used). The process can be seen in Figure~\ref{figure:federated-averaging} and the algorithm is detailed in Algorithm~\ref{algorithm:fedavg}. \Ac{fedavg} has been proven successful in many \ac{fl} tasks despite theoretical predictions suggesting otherwise~\cite{bibliography:on-the-unreasonable-effectiveness-of-federated-averaging-with-heterogeneous-data}.

{\small
    \begin{algorithm}[ht]
        \caption{Federated Averaging~\cite{bibliography:communication-efficient-learning-of-deep-networks-from-decentralized-data}}
        \label{algorithm:fedavg}

        \DontPrintSemicolon
        \SetNoFillComment

        \SetKwInOut{Input}{Input}
        \SetKwInOut{Output}{Output}
        \SetKwFunction{FMain}{UpdateClient}
        \SetKwProg{Fn}{}{:}{}
        \SetKw{KwEach}{each}
        \SetKw{KwInParallel}{in parallel}

        \Input{
            $C$ is the set of all clients, $\mathcal{D}_i$ is the local dataset of client $c_i \in C$, $T$ is the number of communication rounds, $N$ is the number of clients per communication round, $B$ is the batch size, $E$ is the number of local epochs, and $\eta$ is the learning rate
        }
        \Output{Global Model Parametrization $\boldsymbol{\theta}_T$}

        \vspace{0.25cm}

        \tcc{Runs on the central server}
        Initialize $\boldsymbol{\theta}_0$\;
        \For{\KwEach communication round $t=1, \dots, T$}
        {
            $C_t \gets N$ random clients sub-sampled from $C$ \;
            \For{\KwEach client $c_i \in C_t$ \KwInParallel}
            {
                $\boldsymbol{\theta}^i_{t} \gets$ \texttt{UpdateClient}$(\boldsymbol{\theta}_{t-1}, i)$ \;
            }
            $\boldsymbol{\theta}_{t} \gets \sum_{c_i \in C_t} \frac{\left|\mathcal{D}_i\right|}{\left| \bigcup_{c_j \in C_t} \mathcal{D}_j \right|}\boldsymbol{\theta}^i_{t}$\;
        }
        \KwRet $\boldsymbol{\theta}_T$\;

        \vspace{0.25cm}

        \tcc{Runs on client $i$}
        \Fn{\FMain{$\boldsymbol{\theta}$, $i$}}{
            \For{\KwEach local epoch $e=1, \dots, E$}
            {
                split $\mathcal{D}_i$ into $\left\lceil \frac{\left|\mathcal{D}_i\right|}{B} \right\rceil$ batches of size $B$ \;
                \For{\KwEach batch $b = 1, \dots, \left \lceil \frac{\left|\mathcal{D}_i\right|}{B} \right \rceil$}
                {
                    $\boldsymbol{\theta} \gets \boldsymbol{\theta} - \eta \nabla \mathcal{L}_i(\mathcal{D}_{i, b}; \boldsymbol{\theta})$\;
                }
                \KwRet $\boldsymbol{\theta}$\;
            }
        }
    \end{algorithm}
}

A significant challenge of \ac{fedavg} lies, however, in dealing with non-\ac{iid} client data. The data generating distribution may be different for each client, i.e., the data is not independent and identically distributed between the clients. This means that the local objective of each client may differ, sometimes even significantly, from the global training objective, which may lead to conflicting model updates being sent to the central server that hinders the convergence of the global model. There are different types of non-\ac{iid}-ness, which include:

\begin{itemize}
    \item \textbf{Covariate Shift} -- Local samples may have a different statistical distribution compared to the samples of other clients
    \item \textbf{Prior Probability Shift} -- The labels of the local samples may have a different statistical distribution compared to the samples of other clients
    \item \textbf{Concept Shift} -- Local samples have the same labels as other clients, but they correspond to different features, or local samples have the same features as other clients, but they correspond to different labels
    \item \textbf{Imbalanced Data} -- The data available at the clients may vary significantly in size
\end{itemize}

Many different techniques have been proposed to alleviate the problems associated with non-\ac{iid} data, cf.~\citet{bibliography:federated-learning-on-non-iid-data-a-survey} for a timely overview of different techniques.

Clients with limited local data are another issue that has a comparable effect to non-\ac{iid}-ness. In the case of movie \acp{recsys}, it is common that most users have only watched a few dozen or maybe a few hundred movies. This can lead to exceedingly small, noisy updates of the local model, which result in the global model not converging. Both the problem of imbalanced data and small local datasets can be attenuated by weighting the local model updates during aggregation by the local dataset size of the client. But this also has the unwanted effect of suppressing the interests of many users with little training data and amplifying the interests of a few users with a lot of training data.

We address both problems of non-\ac{iid}-ness and small local datasets by chaining client trainings together. The central server selects a random subset of the client population for each communication round before further subdividing them into small queues of a specified size. The clients constituting a specific queue are assigned uniformly at random from the client subset. The first client in each queue receives the global model for local training, while each consecutive client receives the local model of the client prior to it. The local models of the last client in each of these queues are then aggregated by the central server, similar to \ac{fedavg}. The goal of chaining multiple client trainings is that the resulting model updates are less noisy because they were not only exposed to more data but also to data from multiple different distributions, in contrast to what would normally be possible. Since no client in a queue has information about the origin of its local model nor about its position in the queue, this method is still at least as privacy-preserving as regular \ac{fl}. We call this technique \ac{fedq}. Algorithm~\ref{algorithm:fedq} shows the exact training protocol that we follow.

{\small
    \begin{algorithm}[ht]
        \caption{\ac{fedq}}
        \label{algorithm:fedq}

        \DontPrintSemicolon
        \SetNoFillComment

        \SetKwInOut{Input}{Input}
        \SetKwInOut{Output}{Output}
        \SetKwFunction{FMain}{UpdateClient}
        \SetKwProg{Fn}{}{:}{}
        \SetKw{KwEach}{each}
        \SetKw{KwInParallel}{in parallel}

        \Input{
            $C$ is the set of all clients, $\mathcal{D}_i$ is the local dataset of client $c_i \in C$, $T$ is the number of communication rounds, $N$ is the number of clients per communication round, $L$ is the client queue length, where $L$ divides $N$, $B$ is the batch size, $E$ is the number of local epochs, and $\eta$ is the learning rate
        }
        \Output{Global Model Parametrization $\boldsymbol{\theta}_T$}

        \vspace{0.25cm}

        \tcc{Runs on the central server}
        Initialize $\boldsymbol{\theta}_0$ \;
        \For{\KwEach communication round $t=1, \dots, T$}
        {
            $C_t \gets N$ random clients sub-sampled from $C$ \;
            \For{$k=1, \dots, \frac{N}{L}$ \KwInParallel}
            {
                \tcc{First client in the $k^{\text{th}}$ queue receives the global model}
                $\boldsymbol{\theta}^k_{t} \gets \boldsymbol{\theta}_{t-1}$\;

                \tcc{Dataset sizes of the clients in the queue are aggregated for the weighted mean}
                $s_k \gets 0$ \;
                \For{\KwEach client $c_i$ in $C_t[(k - 1)L + 1 : kL]$}
                {
                    $\boldsymbol{\theta}^k_{t} \gets$ \texttt{UpdateClient}$(\boldsymbol{\theta}^k_{t}, i)$ \;
                    $s_k\gets s_k + \left|\mathcal{D}_i\right|$ \;
                }
            }

            $\boldsymbol{\theta}_{t} \gets \sum_{k = 1}^{\frac{N}{L}} \frac{s_k}{\left| \bigcup_{c_i \in C_t}^{} \mathcal{D}_i \right|}\boldsymbol{\theta}^k_{t}$ \;
        }
        \KwRet{$\boldsymbol{\theta}_T$} \;

        \vspace{0.25cm}

        \tcc{Runs on client $i$}
        \Fn{\FMain{$\boldsymbol{\theta}, i$}}
        {
            \For{\KwEach local epoch $e=1, \dots, E$}
            {
                Split $\mathcal{D}_i$ into $\left \lceil \frac{\left|\mathcal{D}_i\right|}{B} \right \rceil$ batches of size $B$ \;
                \For{\KwEach batch $b = 1, \dots, \left \lceil \frac{\left|\mathcal{D}_i\right|}{B} \right \rceil$}
                {
                    $\boldsymbol{\theta} \gets \boldsymbol{\theta} - \eta \nabla \mathcal{L}_i(\mathcal{D}_{i, b}; \boldsymbol{\theta})$ \;
                }
                \KwRet $\boldsymbol{\theta}$ \;
            }
        }
    \end{algorithm}
}

For the complexity analysis, we compare \ac{fedq} to its baseline, \ac{fedavg}, with respect to the expected time the central server needs to wait before it can aggregate the updated model parameters of the clients in each communication round. The number of local update steps on the $i$th client are given by $E \cdot\frac{\left|\mathcal{D}_i\right|}{B}$. This implies, that each client performs $E \cdot \left [ \frac{\mathbb{E}[\left|\mathcal{D}_i\right|]}{B} \right ] = E \cdot \left [ \frac{\sum_{i=1}^{|C|} \left|\mathcal{D}_i\right|}{|C| \cdot B} \right ]$ steps on average, where the expectation is over the random selection of a client, which follows the uniform distribution~\cite{bibliography:communication-efficient-learning-of-deep-networks-from-decentralized-data}. Therefore, the expected time complexity for a single communication round, depending on the utilized algorithm, can be expressed as:

\begin{displaymath}
    \begin{aligned}
        &\text{\ac{fedavg}:} && \mathcal{O} \left ( P \cdot E \cdot \left[\frac{\sum_{i=1}^{|C|} \left|\mathcal{D}_i\right|}{|C| \cdot B}\right] \right ) \\
        &\text{\ac{fedq}: } && \mathcal{O} \left ( \mathbf{L} \cdot P \cdot E \cdot \left [ \frac{\sum_{i=1}^{|C|} \left|\mathcal{D}_i\right|}{|C| \cdot B} \right ] \right ),
    \end{aligned}
\end{displaymath}

where $P$ denotes the time of a forward and backward pass on the client's local model on a batch of data~\cite{bibliography:data-leakage-in-federated-averaging}. Furthermore, it was assumed that the communication time with the central server is dominated by the average local training time for each client. In summary, \ac{fedq} requires $L$ times as much time as \ac{fedavg} for each communication round.

During the development of \ac{fedq}, further techniques for addressing non-\ac{iid}-ness and small local datasets in \ac{fl} that are partially comparable to \ac{fedq} have emerged, for which the similarities with and differences to \ac{fedq} are discussed in Appendix~\ref{appendix:fedq-and-other-client-chaining-techniques}.

\subsection{Achieving Communication Efficiency}
\label{section:achieving-communication-efficiency}

Besides the problems of data heterogeneity and clients having very little local data, constantly communicating model parametrizations can also lead to a significant overhead. The candidate generator and the ranker models are, depending on the sizes of the embeddings and the number of hidden layers, between 60MB and 120MB in size. Given the massive scale of the user base of a typical movie \ac{recsys}, using \ac{fl} can result in multiple gigabytes of data that must be communicated in each communication round, even at low client sub-sampling rates. Furthermore, the clients are relatively resource constrained, so communication reduction techniques that require complex processing, such as pruning or learned quantization, are not an option.

A recent standard for \ac{nnc}, ISO/IEC 15938-17:2022 (MPEG-7 part 17)~\cite{bibliography:mpeg-7-compression-of-neural-networks-for-multimedia-content-description-and-analysis, bibliography:iso-standard-compression-of-neural-networks-for-multimedia-content-description-and-analysis, bibliography:encoder-optimization-for-the-nnr-standard,bibliography:overview-of-the-neural-network-compression-and-representation-standard}\footnote{A standards-compliant implementation of the \ac{nnc} standard, under a permissive license, is available on GitHub: \url{https://github.com/fraunhoferhhi/nncodec}.}, which is based on the \ac{deepcabac} \ac{nn} compression algorithm~\cite{bibliography:deepcabac-a-universal-compression-algorithm-for-deep-neural-networks}, has shown excellent compression results and requires only little or no preprocessing. Furthermore, it has already been shown to exhibit remarkably high performance in an \ac{fl} setup~\cite{bibliography:deepcabac-plug-and-play-compression-of-neural-network-weights-and-weight-updates}. In its coding core, \ac{nnc} combines specific quantization methods that are adapted to the \ac{nn} layers, followed by a context-adaptive binary arithmetic coding method, which reduces data redundancy.

Employing the \ac{nnc} standard to compress the upstream and downstream communication in our proposed \ac{fedrec} is motivated by the fact that the coding engine at its core, \ac{deepcabac}, permits higher compression performance on a variety of \ac{nn} architectures than comparable techniques in the literature~\cite{bibliography:a-universal-compression-algorithm-for-deep-neural-networks}. \citeauthor{bibliography:a-universal-compression-algorithm-for-deep-neural-networks} showed that the \acp{nn} can be compressed by a factor of \textbf{50.6} on average with negligible loss in performance. Comparable coders based on the weighted Lloyd algorithm~\cite{bibliography:least-squares-quantization-in-pcm, bibliography:towards-the-limit-of-network-quantization} or uniform quantization~\cite{bibliography:fixed-point-quantization-of-deep-convolutional-networks, bibliography:towards-the-limit-of-network-quantization} only managed to compress the models by factors of 13.6 and 5.7 respectively. For example, the authors obtained a compression ratio of 1.58\% with an accuracy of 69.43\% for the VGG16 architecture, whereas comparable literature reports only a compression ratio of 2.05\% with an accuracy of 68.83\%. Similar results are obtained for the MobileNet-v1 and MixNet architectures, resulting in a compression ratio gain of 3.6 and 92.1 percentage points, respectively, without affecting the model performance. These results are obtained by simply applying \ac{deepcabac}, they do not require the use of any optimization techniques, such as bias correction, distillation, or fine-tuning, rendering the \ac{nnc} standard a straightforward plug-and-play procedure~\cite{bibliography:a-universal-compression-algorithm-for-deep-neural-networks, bibliography:deepcabac-plug-and-play-compression-of-neural-network-weights-and-weight-updates}.

There are more specialized techniques for reducing the communication overhead in \ac{fl} that are, however, less comparable to \ac{nnc} as they are not based on entropy coding. For example, FedFast~\cite{bibliography:fedfast-going-beyond-average-for-faster-training-of-federated-recommender-systems} is an alternative to \ac{fedavg}, which increases convergence speeds of the models and thus reduces the number of times updates have to be communicated between server and client. \citeauthor{bibliography:fedfast-going-beyond-average-for-faster-training-of-federated-recommender-systems} provide an experimental evaluation of their method on MovieLens 1M~\cite{bibliography:the-movielens-datasets-history-and-context}, MovieLens 100K~\cite{bibliography:the-movielens-datasets-history-and-context}, TripAdvisor hotel reviews~\cite{bibliography:joint-multi-grain-topic-sentiment-modelling-semantic-aspects-for-online-reviews}, and the Yelp dataset~\cite{bibliography:yelp-dataset}. On MovieLens 100K, FedFast required \textasciitilde24.2\%\footnote{\citeauthor{bibliography:fedfast-going-beyond-average-for-faster-training-of-federated-recommender-systems} claim that for MovieLens 100K FedFast already reached the same performance as \ac{fedavg} at communication round 30, which would correspond to approximately 33 times less communicated data, but their own training graphs suggest that this only happened at approximately communication round 196, which corresponds to the factor of approximately 4 that we reported here.}\textsuperscript{,}\footnote{\citeauthor{bibliography:fedfast-going-beyond-average-for-faster-training-of-federated-recommender-systems} do not publish communication cost savings, so the values presented here were read from the training curves in Figure 3~\cite{bibliography:fedfast-going-beyond-average-for-faster-training-of-federated-recommender-systems} and are therefore only approximations.\label{footnote:fedfast-results-disclaimer}} of the communication rounds to achieve the highest performance of \ac{fedavg}, which corresponds to around 4 times less data communicated. On MovieLens 1M, FedFast reached the best performance of \ac{fedavg} even faster, i.e., after only \textasciitilde1.13\%\footref{footnote:fedfast-results-disclaimer} of the communication rounds that \ac{fedavg} required, which means that approximately 88 times less data was communicated. For TripAdvisor, FedFast only required \textasciitilde7.5\%\footref{footnote:fedfast-results-disclaimer}\textsuperscript{,}\footnote{Again, \citeauthor{bibliography:fedfast-going-beyond-average-for-faster-training-of-federated-recommender-systems} claim that FedFast was 20 times faster than \ac{fedavg}, although their own training curves suggest it was closer to the factor of 13 reported here.} of time to reach the highest performance of \ac{fedavg} as compared to the time that \ac{fedavg} required, which resulted in around 13 times less communication cost. Finally, FedFast required only \textasciitilde17.8\%\footref{footnote:fedfast-results-disclaimer} of the communication rounds to reach the highest performance of \ac{fedavg}, in contrast to how many communication rounds \ac{fedavg} required to reach the performance. This reduces the communication cost of FedFast by almost a factor of 6. These results are, however, not comparable to the compression performance of other methods, as they measure the communication cost required to reach the highest accuracy of \ac{fedavg}, which, however, performs very poorly as compared to FedFast and does not even converge in the case of the TripAdvisor and Yelp datasets. Under realistic conditions, one would not stop the training there, but train the model until convergence, which in some cases happened much later. For example, the training curves presented in Figure 3~\cite{bibliography:fedfast-going-beyond-average-for-faster-training-of-federated-recommender-systems} seem to suggest that for MovieLens 100K and the Yelp dataset FedFast only reached its own highest accuracy at the very end of the training, after 1,000 communication rounds.

Another interesting approach is that of FedKD~\cite{bibliography:communication-efficient-federated-learning-via-knowledge-distillation}, where the clients train a teacher model, which is then distilled into a smaller student model. \Ac{fl} clients communicate the compressed gradients of the student models, which substantially reduces the communication overhead. \citeauthor{bibliography:communication-efficient-federated-learning-via-knowledge-distillation} report that they accrued 18.6 times less communication cost per client on the MIND~\cite{bibliography:mind-a-large-scale-dataset-for-news-recommendation} dataset and 19.9 times less communication cost per client on the ADR~\cite{bibliography:overview-of-the-third-social-media-mining-for-health-workshop} dataset as compared to directly using the larger teacher model, with no loss in performance. Both FedFast and FedKD, however, require substantial changes to the \ac{fl} pipeline, while \ac{deepcabac} consistently offers high, in many cases even the highest reduction in size, while being a plug-and-play solution, that only needs to be applied to the \ac{nn} model. This justifies our choice of utilizing the \ac{nnc} framework for our \ac{fedrec}, since we can expect to have higher compression performances than the previously proposed coding techniques in the literature, without having to integrate any complex optimization techniques.

\subsection{Data Security \& Privacy Protection}
\label{section:data-security-and-privacy-protection}

To achieve the goal of data security and privacy protection \ac{fl}, incorporates the principles of \emph{data minimization}, i.e., processing the data as early as possible (data processing is carried out on the client's device), only collecting data that is absolutely necessary (e.g., in \ac{fedavg} only model parametrizations are transmitted), and discarding any obtained data as soon as possible (after the client models were aggregated into an updated global model, the local models are discarded). Furthermore, \ac{fl} employs the principle of \emph{anonymization}, i.e., no conclusions about the originator shall be drawn from the respective data. In terms of \ac{fl}, this implies that, ideally, only sending training updates should prevent the central server from deriving any further information about its clients. In practice, however, it has been shown that local samples can be reverse-engineered from the gradients~\cite{bibliography:inverting-gradients} in \ac{fedsgd}. To alleviate this problem, anonymization techniques, such as differential privacy, where random noise is added to client data communication~\cite{bibliography:federated-learning-with-differential-privacy}, or homomorphic encryption, where encrypted client updates can be aggregated without decrypting them~\cite{bibliography:privacy-preserving-deep-learning-via-additively-homomorphic-encryption, bibliography:privacy-preserving-machine-learning-with-homomorphic-encryption-and-federated-learning}, can be utilized.

Generally, these kinds of attacks are performed by the central server, who has access to the gradient updates sent by the clients. The attacks reconstruct the client's input data by starting with some arbitrary, e.g., randomly initialized input data, and adapting this dummy data in such a way that the distance between its gradient and the actual gradient received from the client is minimized, for example, by solving the following optimization problem~\cite{bibliography:inverting-gradients, bibliography:data-leakage-in-federated-averaging}:

\begin{align}
    \underset{\widetilde{x}}{\text{argmin}} \; \text{dist}(\nabla_{\boldsymbol{\theta}} \, \ell(\widetilde{x}, y; \boldsymbol{\theta}), \nabla_{\boldsymbol{\theta}} \, \ell(x, y; \boldsymbol{\theta}))\text{,}
\end{align}

where $\widetilde{x}$ denotes the dummy input data, $x$ the unseen training sample of client $c$, $\nabla_{\boldsymbol{\theta}} \, \ell(x, y; \boldsymbol{\theta})$ the intercepted gradient of client $c$, $\nabla_{\boldsymbol{\theta}} \, \ell(\widetilde{x}, y; \boldsymbol{\theta})$ the gradient computed on the dummy input data $\widetilde{x}$ with the ground-truth $y$, which can, for example, be extracted from the gradient of the output layer~\cite{bibliography:improved-deep-leakage-from-gradients}, $\boldsymbol{\theta}$ the parametrization of the updated local model of client $c$, and $\text{dist}(\cdot)$ a distance function. For example,~\citet{bibliography:inverting-gradients} show that in many cases it is possible to use such a technique to reconstruct training images almost perfectly from the gradient, thus demonstrating that \ac{fedsgd} is not as privacy-preserving as thought. A related method, proposed by~\citeauthor{bibliography:secure-federated-matrix-factorization}, is able to reverse-engineer a user's rating information from two consecutive gradient updates in a \ac{fedrec} based on matrix factorization, where the factorization is learned by the users using \ac{sgd}~\cite{bibliography:secure-federated-matrix-factorization}. The attack proposed by~\citeauthor{bibliography:secure-federated-matrix-factorization} is specifically tailored towards federated matrix factorization and is therefore not applicable to our scenario. Furthermore, both methods need to intercept the client's gradient updates and are therefore only pertinent to \ac{fedsgd}. \Ac{fedq} on the other hand, which is employed by us, does not share the gradient but the updated local model and is thus not vulnerable to these kinds of attacks.

And still, \citet{bibliography:data-leakage-in-federated-averaging} showed that it is possible to reconstruct training images in realistic \ac{fedavg} settings. Despite the method's success with a single client relying on many local training rounds, attacking aggregated parameter updates from multiple clients, even if only a few of them are used, significantly degrades the reconstruction performance. Using the \ac{femnist} dataset for demonstration, they specifically showed that attacking the averaged updates of just four clients instead of one significantly lowers the average reconstruction performance of images with \acp{psnr} of 20 or above by 35.8 percentage points, which is evaluated on 100 randomly selected clients from the training set. When conducting this experiment, they chose an optimal configuration of local epochs and batch sizes for the clients. In addition, they rely on the unrealistic assumption that the label counts are known. Having to estimate them, degraded the attack performance by 17 percentage points using the updated parameters of only a single client. In \ac{fedavg}, an attacker can easily retrieve the parameter updates of individual clients, thus making this kind of attack highly effective. But, by the very nature of \ac{fedq}, a potential attacker usually only receives aggregated parameter updates from multiple clients. Thus, using a reasonably large queue of clients should guarantee a high level of data security.

Some recent works have tried to employ gradient/parameter obfuscation to counteract these kinds of attacks. For example, differential privacy is an obfuscation method, where random noise is added to the client updates. While differential privacy is one of the most prevalent obfuscation schemes, others, like gradient quantization and magnitude pruning, have been proposed. For example, \citet{bibliography:a-framework-for-evaluating-privacy-leakages-in-federated-learning} and~\citet{bibliography:deep-leakage-from-gradients} showed that gradient sparsification is a well-functioning approach to mitigate data reconstruction attacks. \citet{bibliography:mixed-quantization-enabled-federated-learning-to-tackle-gradient-inversion-attacks} demonstrated the efficacy of using mixed precision quantization to counteract gradient leakage attacks. They quantized the model gradients of the clients after local training to 16-bit and 8-bit integers before sending them to the central server and showed experimentally that no information was leaked. They ran the attack for 450 iterations both on \ac{fl} setups where the communicated gradients were quantized, as well as baseline \ac{fl} setups without gradient quantization using the \ac{mnist}, Fashion-\ac{mnist}, and CIFAR-10 datasets. In the baseline experiments, training images could be extracted from the gradients after 20 iterations for \ac{mnist}, 20 iterations for Fashion-\ac{mnist}, and 40 iterations for CIFAR-10. In the experiments that applied gradient quantization they were not able to extract any training samples, even after 450 iterations of their attack. Our \ac{nnc} module is capable of using an arbitrary number of quantization points for quantizing model updates, where the number of quantization points is fine-tuned for each layer. This results in a quantization that goes well below 16-bit and in many cases even below 8-bit quantization, which should result in better obfuscation.

These works are good indicators that gradient obfuscation techniques can be successfully employed to counteract attacks such as those proposed by~\citet{bibliography:inverting-gradients} and~\citet{bibliography:data-leakage-in-federated-averaging}. \citet{bibliography:gradient-obfuscation-gives-a-false-sense-of-security-in-federated-learning}, however, call the effectiveness of gradient obfuscation into question, by proposing a novel data reconstruction attack scheme. However, they have only shown their attack to be effective in the domain of image classification, which is a special case, as even reconstructed images that diverge a lot from the actual input image, may contain enough visual information for human observers, while the same amount of reconstruction error on tabular data, as used in \acp{fedrec}, would not be usable with the same amount of error. Also, they have only tried small batch sizes with a small number of local epochs and have only shown uniform quantization. Therefore, we still conjecture that the error induced by the mixed-precision quantization of our \ac{nnc} module, may successfully obfuscate the information contained in the parameter updates sent to the central server or at least make it much harder for attackers to recover any useful information. This, however, remains to be tested in future work.

Finally, most of these attacks assume the central server to be the culprit who wants to reconstruct the input data of its clients. We want to note that outside attackers are usually incapable of intercepting any data from the \ac{fl} process as simple techniques, such as employing SSL/TLS, can effectively mitigate these kinds of attacks.

This concludes our new proposed \ac{fedrec}, which consists of a three-staged recommendation architecture, including a candidate generation, ranking, and re-ranking stage. Furthermore, the \ac{recsys} was extended to use \ac{fl}, applying the developed \ac{fedq} method to effectively operate with extremely high numbers of heterogeneous clients. The communication overhead introduced by constantly communicating parameter updates between the central server and the clients is alleviated by compressing the model parametrizations using a state-of-the-art \ac{nn} compression scheme. Finally, we have discussed the data security and privacy protection capabilities of the proposed architecture. In the following Section~\ref{section:experiments}, we will evaluate the performance of the \ac{fedrec} system experimentally.

\section{Experiments}
\label{section:experiments}

In this section, we describe the experiments performed using our \ac{fedrec} and evaluate its performance. We will first start by describing how the dataset was acquired and processed. Then, we will lay out a non-\ac{fl} baseline to which we will compare the performance of the \ac{fl} system. Then we will demonstrate that standard \ac{fedavg} only yields a moderate performance before applying the \ac{fedq} algorithm to improve performance to equal or even exceed the performance of the non-\ac{fl} baseline. Finally, we will show how the new \ac{nnc} standard can be utilized to significantly decrease the communication overhead. All experiments were performed using PyTorch~\cite{bibliography:pytorch}.

\subsection{Dataset}
\label{section:dataset}

Among datasets suitable for movie \acp{recsys}, the MovieLens dataset by~\citet{bibliography:the-movielens-datasets-history-and-context} is one of the most widely known and used datasets. MovieLens comes in multiple different flavors, among which the \emph{25M} variant is the latest stable benchmark dataset. It contains more than 25 million ratings across almost 60,000 movies made by more than 162,000 users. The MovieLens datasets consist of users, movies, ratings, and tags. As the 25M flavor of the MovieLens dataset is a stable benchmark dataset, it was chosen for our experiments.

\subsubsection{Dataset Analysis}

\begin{figure}[ht]
    \centering

    \includegraphics[height=0.4\textheight]{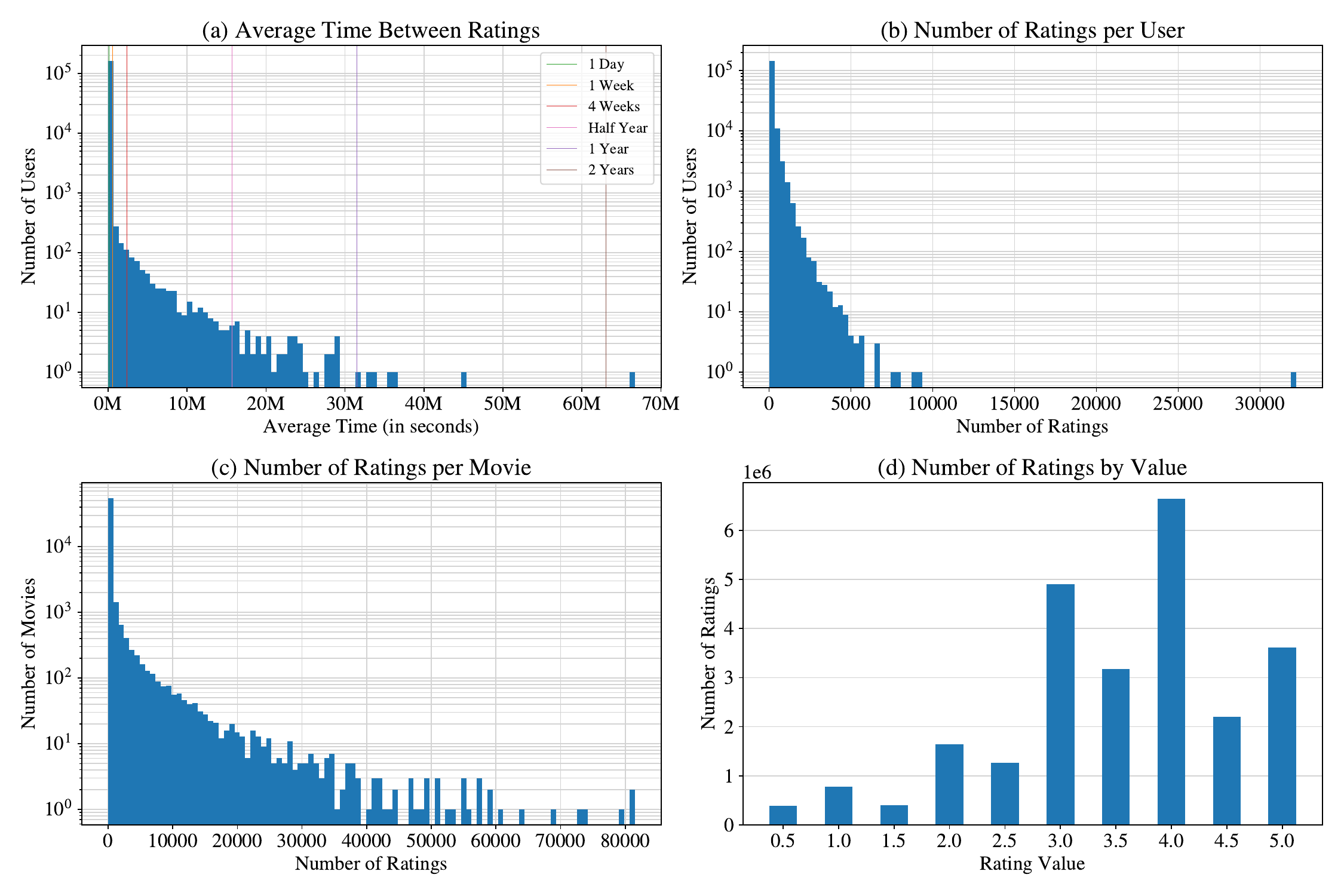}

    \caption{In-depth analysis histograms of the MovieLens 25M dataset: \textbf{(a)} average times between ratings of all users in the dataset, \textbf{(b)} number of ratings per user, \textbf{(c)} number of ratings per movie, and \textbf{(d)} number of ratings of a specific value that were cast by the users.}
    \Description{In-depth analysis histograms of the MovieLens 25M dataset: \textbf{(a)} average times between ratings of all users in the dataset, \textbf{(b)} number of ratings per user, \textbf{(c)} number of ratings per movie, and \textbf{(d)} number of ratings of a specific value that were cast by the users.}
    \label{figure:movielens-dataset-inspection}
\end{figure}

For the candidate generation model, we treat the ratings of the MovieLens dataset as movie watches to predict future watches from past viewing behavior. Therefore, the temporal cohesion of the ratings is particularly important. During an initial screening of the dataset, we observed that the data was inconsistent with ``normal'' viewing behavior, at least for a small number of random samples. For example, some users rated an infeasible number of movies in a single day, while other users had an impossibly high number of total ratings. Therefore, the MovieLens 25M dataset was inspected more closely in terms of four different metrics: (1) average times between ratings of all users in the dataset, i.e., the speed at which users have rated movies, (2) number of ratings per user, (3) number of ratings per movie, and (4) number of ratings cast by rating value. The results are shown in Figure~\ref{figure:movielens-dataset-inspection}.

The MovieLens 25M dataset contains \textbf{59,047} movies that have been rated \textbf{25,000,095} times by \textbf{162,541} users. \textbf{87.1\%} of users have an average time of less than \textbf{1 minute} and  \textbf{97.3\%} have an average time of less than \textbf{1 hour} between two ratings. On average, there are \textbf{32.7} minutes between two ratings. The smallest number of ratings per user is \textbf{20} and the highest number of ratings of any user is \textbf{32,202}. On average, each user has \textbf{153.8} ratings. \textbf{58.8\%} of movies have less than \textbf{10} ratings and \textbf{82.5\%} have less than \textbf{100} ratings. On average, each movie has \textbf{423.4} ratings. The smallest number of ratings per movie is \textbf{1} and the highest number of ratings of any movie is \textbf{81,491}. The \textbf{top-10} most-rated movies have amassed \textbf{2.8\%} of all ratings.

These findings suggest that most of the ratings were performed in a way that indicates that the users of the MovieLens website have mass-rated movies, rather than individually casting the ratings after watching each movie. The ratings per movie are also highly imbalanced, as most movies have few ratings and a few movies have a large number of ratings. This is actually somewhat expected, as there are only a few ``blockbuster'' movies that many people watch, while most movies are only watched by very few people. Finally, the ratings are heavily skewed towards more positive evaluations: ratings of 3.0 and higher are significantly more prevalent than those of 2.5 and below.

\begin{figure}[ht]
    \centering

    \includegraphics[height=0.225\textheight]{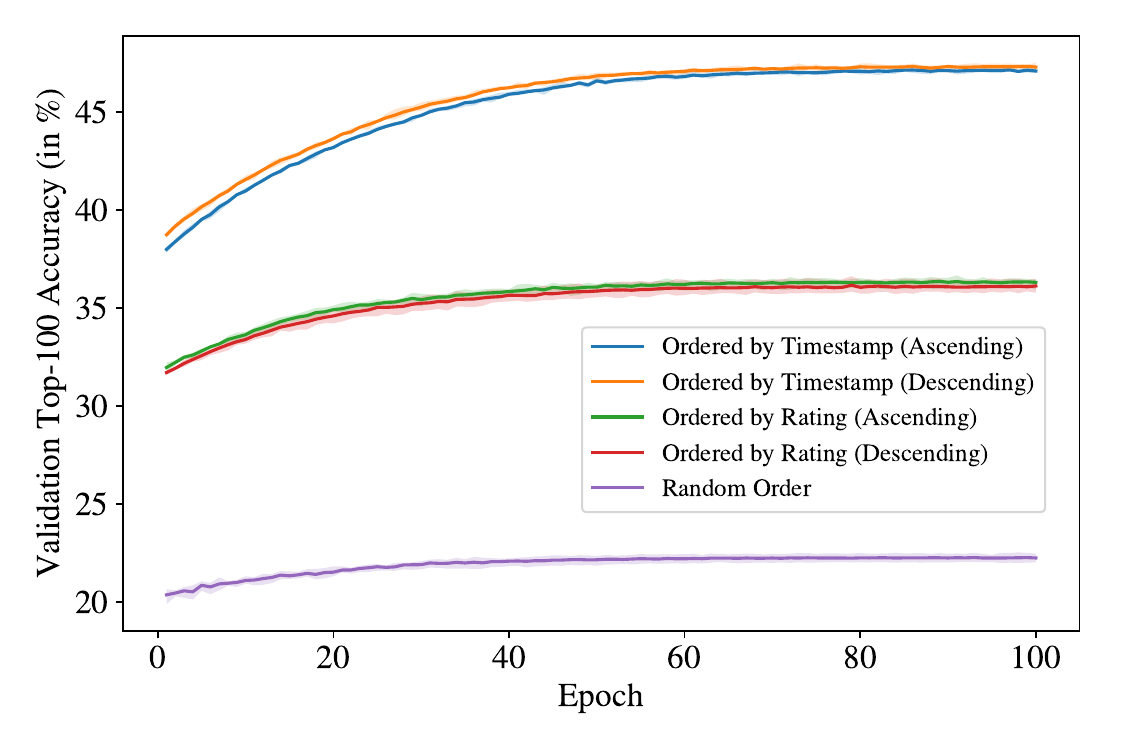}

    \caption{Validation top-100 accuracy results vs. number of epochs for different temporal orderings in the dataset.}
    \Description{Validation top-100 accuracy results vs. number of epochs for different temporal orderings in the dataset.}
    \label{figure:watch-history-dataset-temporal-order-experiment-results}
\end{figure}

The in-depth analysis suggests that the MovieLens dataset may not be suitable for next watch predictions, as the mass-ratings imply that the temporal order does not necessarily coincide with the order in which the movies were watched. To avoid an ill-posed task from the start, a set of experiments were performed, where the user ratings were sorted in multiple ways: by timestamp in ascending order, by timestamp in descending order, by rating in ascending order, by rating in descending order, and in random order. The results of this experiment are shown in Figure~\ref{figure:watch-history-dataset-temporal-order-experiment-results}. The findings reveal that ordering the movie watches by timestamp yields a higher prediction performance, which is measured in terms of top-100 accuracy, than any other ordering\footnote{Surprisingly, ordering the ratings from future to present, i.e., predicting past movie watches from future watch behavior, yields a slightly higher performance than the regular temporal ordering. A statistical fluke can be ruled out, as the experiment was repeated five times and the plot shows the minimum, maximum, and mean top-100 accuracy. We have no explanation for this interesting result.}. Ordering by rating already gives a lower prediction performance, but it is still higher than the performance for random order. This means that, despite the ratings not conforming to ``normal'' viewing behavior, the dataset is actually suitable for the purposes of training the candidate generator, because the assumption of temporal cohesion holds.

\subsubsection{Dataset Preprocessing}
\label{section:dataset-preprocessing}

The candidate generator and the ranker models each have different inputs and outputs, and therefore require a custom dataset that has to be derived from MovieLens. We refer to the dataset for the candidate generation model as the watch history dataset, and the dataset for the ranker model as the rating dataset.

\begin{figure}[ht]
    \centering

    \includegraphics[width=\textwidth]{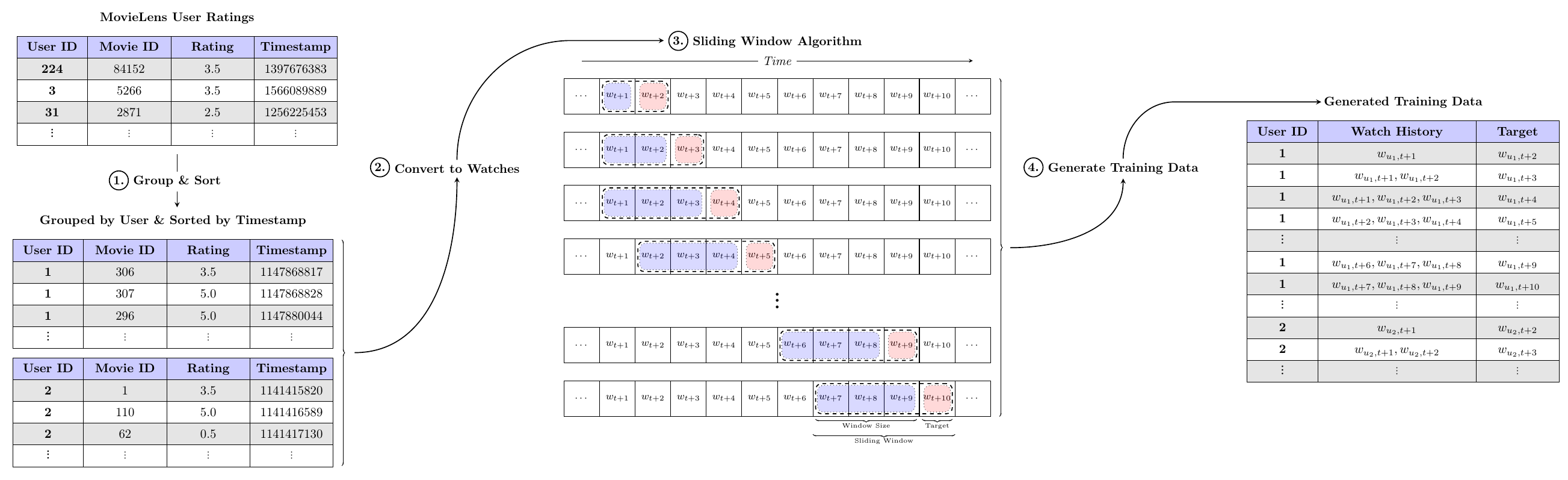}

    \caption{Preprocessing the MovieLens 25M dataset for the training of the candidate generator model.}
    \Description{Preprocessing the MovieLens 25M dataset for the training of the candidate generator model.}
    \label{figure:watch-history-dataset-preprocessing}
\end{figure}

The samples of the former consist of a list of previous movies that a user has watched and a single future movie as prediction target. Since the movie prediction is performed on a per-user basis, the dataset is first grouped by users. Watch histories are made from consecutive movie watches; therefore, the ratings are then ordered by their timestamp. A sliding window is used to extract watch history samples from the movie watches of the users. The preprocessing of the dataset is visualized in Figure~\ref{figure:watch-history-dataset-preprocessing}. The created samples are then stored in a suitable data format for the training, validation, and testing of the candidate generator model.

\begin{figure}[ht]
    \centering

    \includegraphics[height=0.225\textheight]{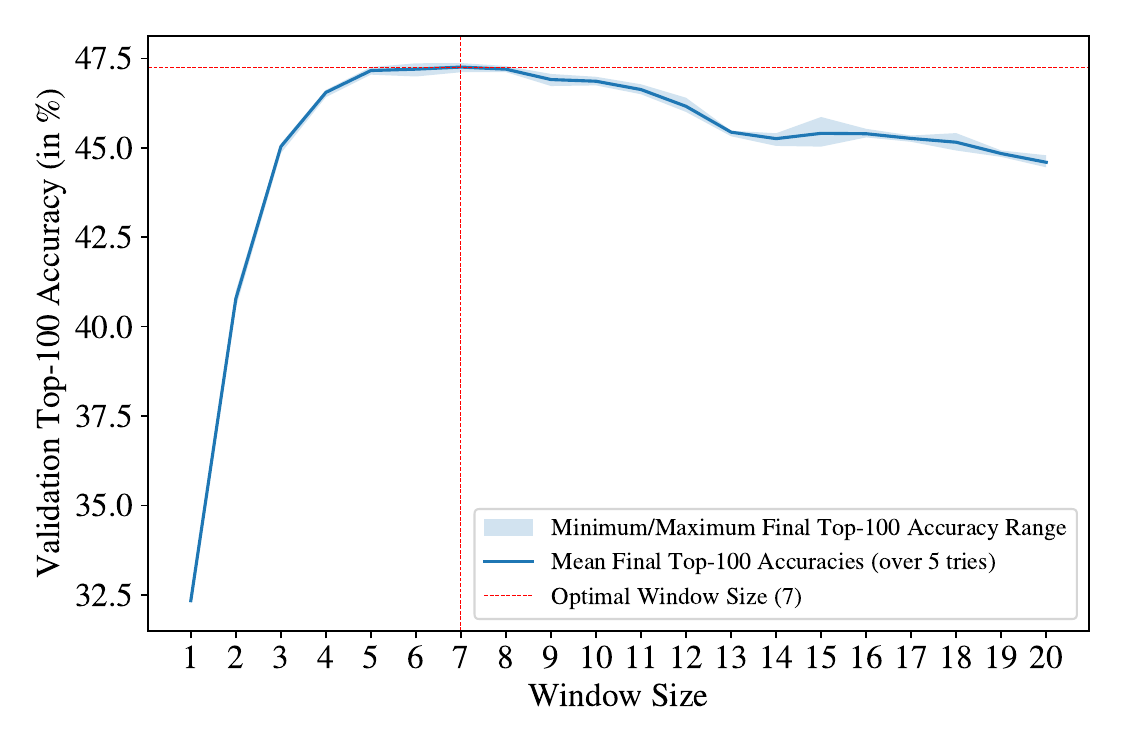}

    \caption{Determining the optimal window size for the MovieLens 25M dataset.}
    \Description{Determining the optimal window size for the MovieLens 25M dataset.}
    \label{figure:watch-history-dataset-window-size-experiment-results}
\end{figure}

The use of a sliding window with a defined upper limit for the number of movies in a watch history is based on the premise that the users' tastes change over time. This implies that a watch becomes less predictive of subsequent watches the longer it lies in the past. Furthermore, depending on the dataset size and the number of trainable parameters, the candidate generator model has an upper capacity limit for learning structure. For too high values of \emph{window size} the candidate generator model performs worse as it is unable to learn the complex correlations in the input data. To determine the optimal window size, multiple datasets with different window sizes were created and used for training the models. The results suggest an optimal window size of 7 (cf. Figure~\ref{figure:watch-history-dataset-window-size-experiment-results}).

\begin{figure}[ht]
    \centering

    \includegraphics[height=0.225\textheight]{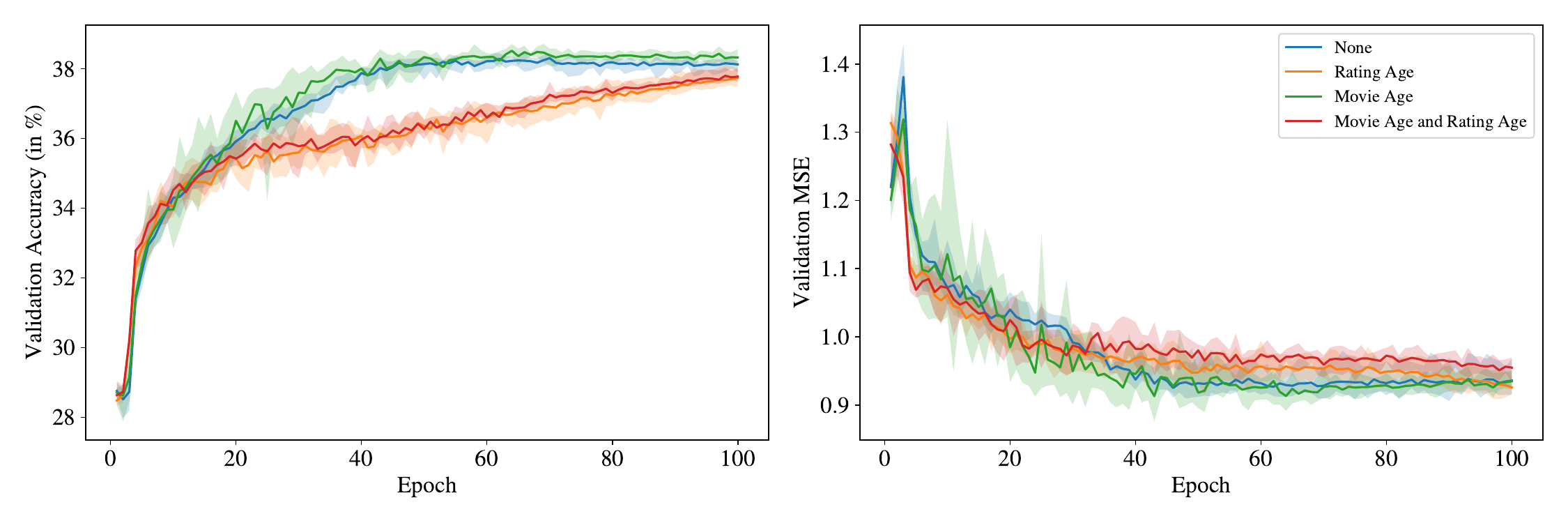}

    \caption{Validation accuracy and \ac{mse} results vs. number of epochs for movie age, rating age, both, and none.}
    \Description{Validation accuracy and \ac{mse} results vs. number of epochs for movie age, rating age, both, and none.}
    \label{figure:ranker-input-features-experiment-results}
\end{figure}

The rating dataset is much simpler, as the MovieLens samples do not have to be reinterpreted. Instead, the rating samples can be directly inferred from each MovieLens sample. Each of it consists of user ID, movie ID, genres of the movie, and user rating. Optionally, the age of the movie and the rating age can be added. The rating age is computed from the rating timestamps, while the movie age is derived from the movie release date, which was retrieved by cross-referencing the MovieLens movies with their corresponding entries in \ac{tmdb}. The movie age and the rating age are both normalized between -1 and 1. Adding the movie age should encourage the model to learn that certain users prefer older or newer movies. The rating age is used to provide the model with an understanding of the temporal component of ratings. During inference, the rating age can be set to 1 to ensure that the model does not take old information about the user into consideration, and thus makes predictions right at the end of the training window. A similar technique has been proposed by~\citet{bibliography:dnns-for-youtube-recommendations}. In order to determine the efficacy of adding these two features, experiments were performed, whose results are presented in Figure~\ref{figure:ranker-input-features-experiment-results}.

Using the movie age yields the best overall accuracy, closely followed by using neither movie nor rating age. Utilizing either the rating age alone or the rating age and the movie age together, results in slower convergence of the model, as well as lower overall accuracy. In terms of \ac{mse}, using movie age, rating age, and using neither yield almost the same overall performance, while using both performs slightly worse. For this reason, we decided to only use the movie age and discard the rating age.

In order to perform \ac{fl} experiments, both the watch history, as well as the rating datasets, were split into much smaller subsets for each \ac{fl} client. Since the movie IDs, user IDs, and genres are fed into embedding layers, the datasets were not simply split randomly, but in a way, that the training data still contained all possible IDs. Otherwise, the validation and test subsets may end up containing IDs that the model was not trained on. For testing the \ac{fl} pipeline, the datasets were randomly split into equal-sized subsets for all \ac{fl} clients, which ensures that the client datasets are balanced and somewhat \ac{iid}. As the MovieLens dataset also contains user IDs, the samples could be split such that each \ac{fl} client receives samples of a single MovieLens user. This allows for properly simulating real-world conditions with non-\ac{iid} data.

\subsection{Baseline Experiments}

\begin{figure}[ht]
    \centering

    \begin{subfigure}{0.33\textwidth}
        \centering
        \includegraphics[height=0.152\textheight]{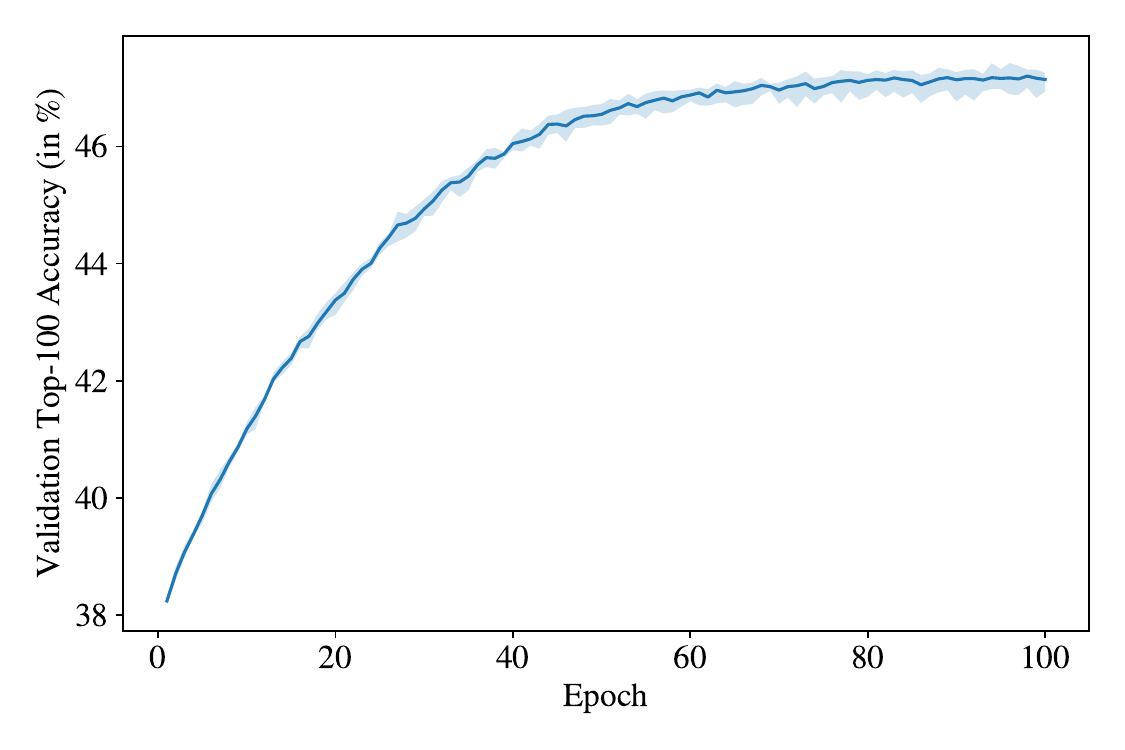}
        \caption{Candidate generator}
        \label{subfigure:candidate-generator-baseline-experiment-results}
    \end{subfigure}%
    \begin{subfigure}{0.66\textwidth}
        \centering
        \includegraphics[height=0.152\textheight]{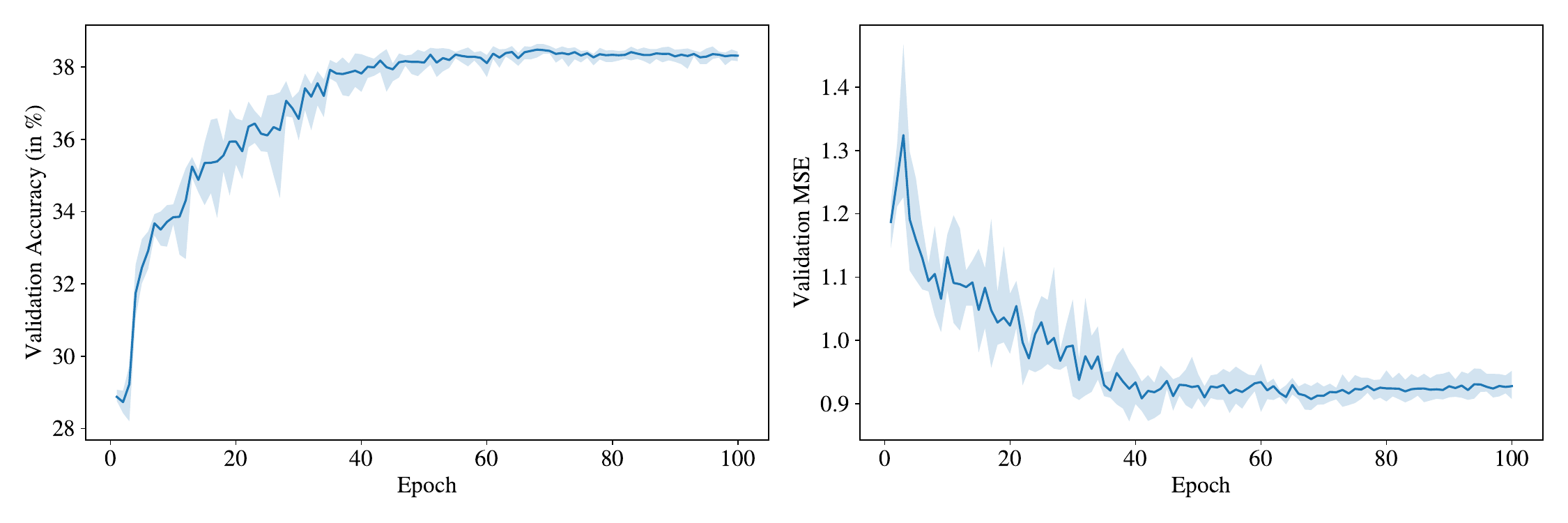}
        \caption{Ranker}
        \label{subfigure:ranker-baseline-experiment-results}
    \end{subfigure}

    \caption{Non-\ac{fl} baseline experiment results for \textbf{(\subref{subfigure:candidate-generator-baseline-experiment-results})} the candidate generator and \textbf{(\subref{subfigure:ranker-baseline-experiment-results})} the ranker.}
    \Description{Non-\ac{fl} baseline experiment results for \textbf{(\subref{subfigure:candidate-generator-baseline-experiment-results})} the candidate generator and \textbf{(\subref{subfigure:ranker-baseline-experiment-results})} the ranker.}
    \label{figure:baseline-experiment-results}
\end{figure}

We first conducted a baseline experiment using the hyperparameters that were selected based on the experiments described in Appendices~\ref{appendix:candidate-generator} and \ref{appendix:ranker}, as well as in Section~\ref{section:dataset}. These experiments are used as a baseline for the \ac{fl} experiments. We trained the candidate generator and the ranker models five times each and present their minimum, maximum, and mean performance in Figure~\ref{figure:baseline-experiment-results}.

The candidate generator outputs a probability distribution over the entire corpus of movies in the MovieLens dataset, which means that it has to distinguish between almost 60,000 classes. Therefore, we report top-100 accuracy (also sometimes referred to as hit-ratio@$k$, where $k = 100$), which rates a classification result as ``correct'' if the ground-truth next watched movie is among the 100 movies with the highest classification scores. For the ranker, we report accuracy, as well as \ac{mse}, which measures how much the predicted rating differs from its ground-truth. The performance was measured on a validation subset of the dataset, which is distinct from the training subset. The highest final top-100 accuracy that was achieved by any of the five trained candidate generators was \textbf{47.26\%}, with an average top-100 accuracy of \textbf{47.15\%}. The best performing ranker model achieved a final accuracy of \textbf{38.43\%} and a final \ac{mse} of \textbf{0.91}, with a mean final accuracy of \textbf{38.31\%} and a mean final \ac{mse} of \textbf{0.93} across all five tries.

\subsection{Federated Learning Experiments}
\label{section:federated-learning-experiments}

\begin{figure}[ht]
    \centering

    \begin{subfigure}{0.33\textwidth}
        \centering
        \includegraphics[height=0.152\textheight]{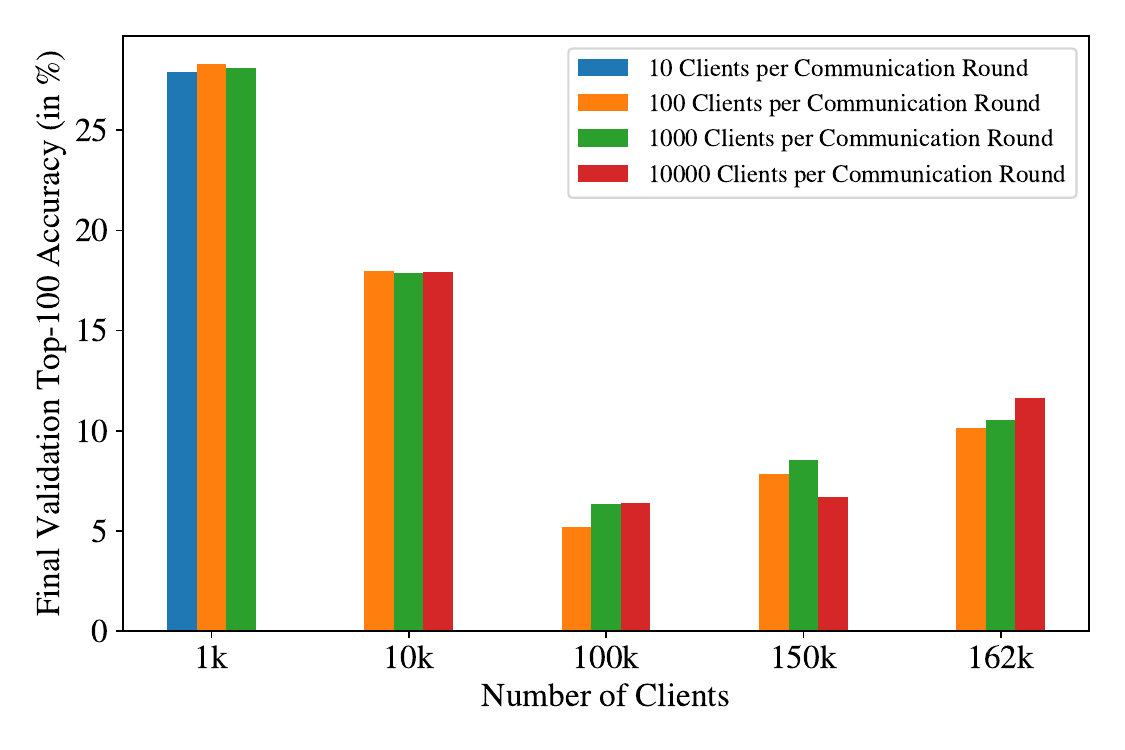}
        \caption{Candidate generator}
        \label{subfigure:federated-learning-experiment-results-candidate-generator-final-validation-top-100-accuracy-only}
    \end{subfigure}%
    \begin{subfigure}{0.66\textwidth}
        \centering
        \includegraphics[height=0.152\textheight]{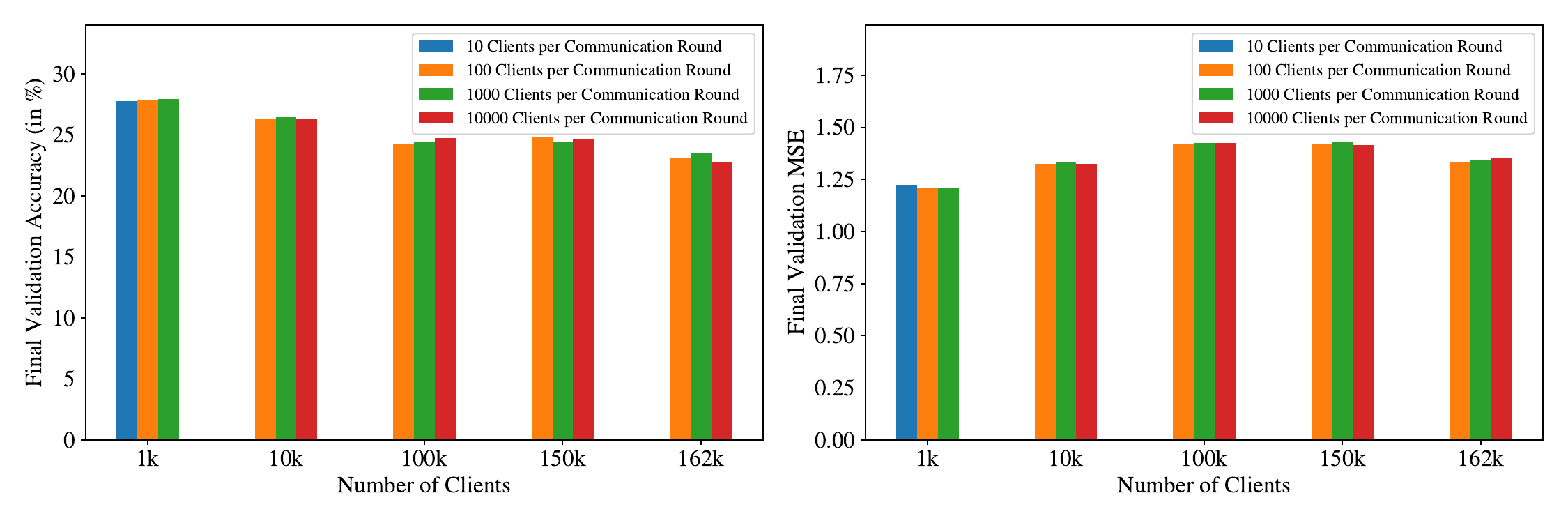}
        \caption{Ranker}
        \label{subfigure:federated-learning-experiment-results-ranker-final-validation-accuracy-and-mse-only}
    \end{subfigure}

    \caption{\Ac{fl} experiment results. For increased legibility, instead of showing the complete training graphs, only \textbf{(\subref{subfigure:federated-learning-experiment-results-candidate-generator-final-validation-top-100-accuracy-only})} the final validation top-100 accuracy, and \textbf{(\subref{subfigure:federated-learning-experiment-results-ranker-final-validation-accuracy-and-mse-only})} the final validation accuracy and \ac{mse} are shown. For reference, the full training graphs are depicted in Appendix~\ref{appendix:extended-federated-learning-and-fedq-experiment-results}.}
    \Description{\Ac{fl} experiment results. For increased legibility, instead of showing the complete training graphs, only \textbf{(\subref{subfigure:federated-learning-experiment-results-candidate-generator-final-validation-top-100-accuracy-only})} the final validation top-100 accuracy, and \textbf{(\subref{subfigure:federated-learning-experiment-results-ranker-final-validation-accuracy-and-mse-only})} the final validation accuracy and \ac{mse} are shown. For reference, the full training graphs are depicted in Appendix~\ref{appendix:extended-federated-learning-and-fedq-experiment-results}.}
    \label{figure:federated-learning-experiment-results-final-validation-metrics-only}
\end{figure}

We subsequently performed \ac{fl} experiments by simulating the \ac{fl} process. A detailed description of how this \ac{fl} simulator operates can be found in Appendix~\ref{appendix:federated-learning-simulator}. The \ac{fl} experiments use the same hyperparameter configuration as the baseline experiments, except for the learning rates of the candidate generator, which had to be decreased by one order of magnitude to stabilize the training. A broad range of different numbers of clients in the dataset were selected in order to simulate the impact of varied local data distributions on the performance of the global model. Different client sub-sampling rates were employed to determine the optimal number of clients per communication round for the individual scale of the experiments, ensuring accurate client updates to be aggregated by the central server. With a range of 1k to 150k clients, the underlying datasets were randomly split into equal-sized local datasets for the clients and randomly distributed among them, assuring that they have approximately the same local data-generating distribution, especially in the low scale experiments. As the number of clients grows, the sizes of the local datasets shrink, which in turn reduces the likelihood of receiving an \ac{iid} subset of the underlying dataset and gradually increases the non-\ac{iid}-ness of the client data. The 162k experiments split the underlying dataset using the user IDs provided by MovieLens, thereby ensuring that each \ac{fl} client receives the samples from exactly one real-world user. As a result, each client's local dataset has a unique data-generating distribution. Additionally, in the 162k setup, the local datasets are imbalanced, as the users of the MovieLens dataset have varying numbers of samples. Furthermore, with an increasing number of clients the local datasets become smaller, thus increasing the negative effects from noisy updates. This setup allows us to clearly identify the effects of small local datasets and non-\ac{iid}-ness to be compared to our \ac{fedq} method. The results of these experiments are shown in Figure~\ref{figure:federated-learning-experiment-results-final-validation-metrics-only}.

As can be seen in Figure~\ref{subfigure:federated-learning-experiment-results-candidate-generator-final-validation-top-100-accuracy-only}, the candidate generator is strongly affected by non-\ac{iid}-ness and small local datasets, as even the setup with only 1k clients already performs much worse compared to the non-\ac{fl} baseline, and increasing the number of clients decreases the performance significantly. The ranker, which can be seen in Figure~\ref{subfigure:federated-learning-experiment-results-ranker-final-validation-accuracy-and-mse-only}, is not as much affected by non-\ac{iid}-ness and small local datasets. Additionally, the performance drop from increasing the number of clients is not as pronounced as with the candidate generator. Still, the performance is significantly lower than the non-\ac{fl} baseline. The ranker performs better than the candidate generator, since the watch behavior varies more between users than rating behavior, e.g., two users with different watch histories may still rate the same movie similarly. Since the rating data is much more homogeneous, the data-generating distributions of the users do not differ as much as in the case of the watch history data.

Reasonably, one might expect that, as the number of clients grows and the sizes of the local datasets shrink, the performance of the candidate generator should gradually decline. The experiments, however, reveal that the performance declines between 1k and 100k clients, before increasing again with 150k and 162k clients. We believe this can be explained by viewing the performance penalty incurred by \ac{fl} compared to centralized training as a compound error. One of the components of this error arises from the non-\ac{iid}-ness of the clients, as the different local data generating distributions cause the clients to have disparate local objectives. This leads to contradicting client updates that, when averaged by the central server, can cancel out some of the training progress of other clients and result in an update to the global model that does not minimize the global objective. Another component of the error is caused by noisy client updates: The smaller the local dataset of a client is, the worse its estimation of the empirical loss becomes, which results in a noisy gradient and unstable training. Even with a homogeneous client population, this can lead to contradicting client updates, which cause the global model to not properly converge. In the setups with 1k and 10k clients, each client has a large local dataset, which results in stable local training and good client updates. This means that the compound error causing the decrease in performance is dominated by the increase in non-\ac{iid}-ness. As the number of clients increases, the sizes of the local datasets decrease, which in turn increases the heterogeneity of the clients. But this decrease in the size of the local datasets also causes the client updates to become noisier and the error induced by noisy client updates to become a significant component of the compound error. This increase in both error components causes the sharp decline in performance between the setups with 10k and 100k clients. Between the setups with 100k, 150k, and 162k clients, the non-\ac{iid}-ness and noisy client update error components do not significantly change, because the increase in the number of clients is not as large and the sizes of the local datasets do not change as much. At this point, a new error component comes into play: With the decrease of the local dataset sizes  also comes a decrease in the number of different MovieLens users represented in the local datasets. In the setup with 100k clients the number of different MovieLens users represented in the local datasets becomes small enough that the global model starts to be negatively impacted by the heterogeneity of the local dataset samples. The 1k and 10k setups do not suffer from this, as their local datasets have samples from so many MovieLens users that the effect averages out. In the 150k setup the number of MovieLens users decreases and in the 162k setup it is even guaranteed that each client only has samples from a single MovieLens user, thus gradually decreasing the negative impact of the heterogeneity of the local dataset samples. Again, the ranker model is not as affected by this, since the rating data is much more homogeneous than the watch history data, as described above.

\subsection{FedQ Experiments}
\label{section:fedq-experiments}

\begin{figure}[ht]
    \centering

    \begin{subfigure}{0.33\textwidth}
        \centering
        \includegraphics[height=0.152\textheight]{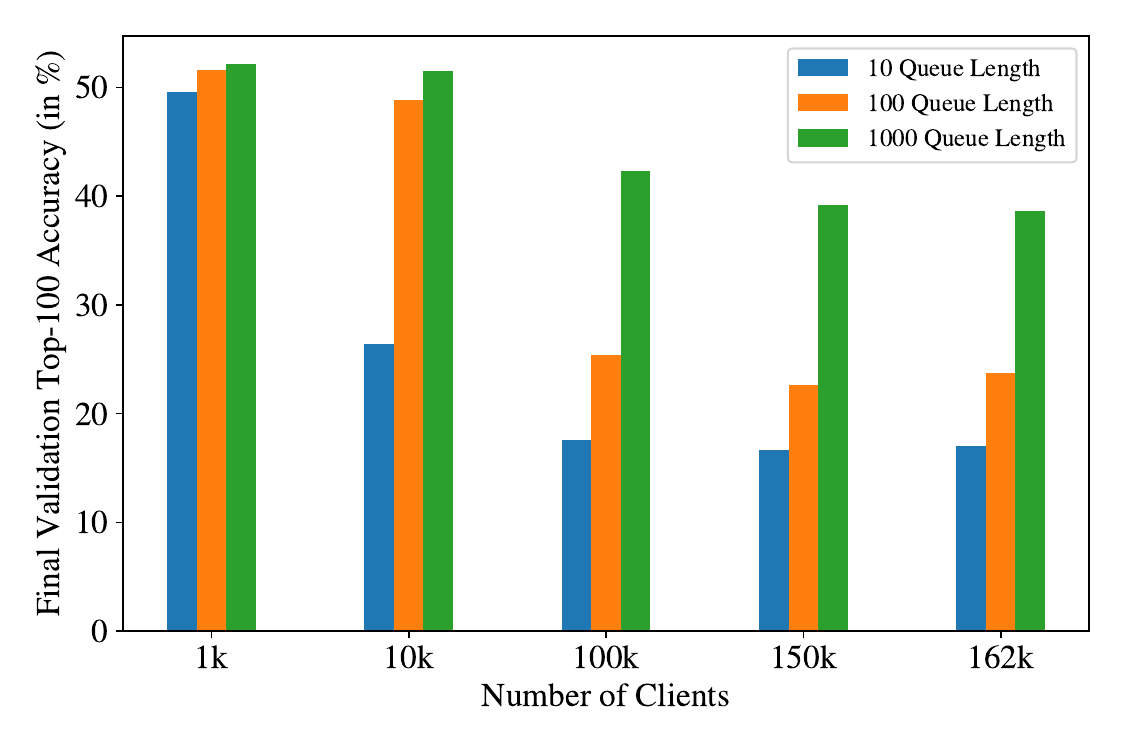}
        \caption{Candidate generator}
        \label{subfigure:fedq-experiment-results-candidate-generator-final-validation-top-100-accuracy-only}
    \end{subfigure}%
    \begin{subfigure}{0.66\textwidth}
        \centering
        \includegraphics[height=0.152\textheight]{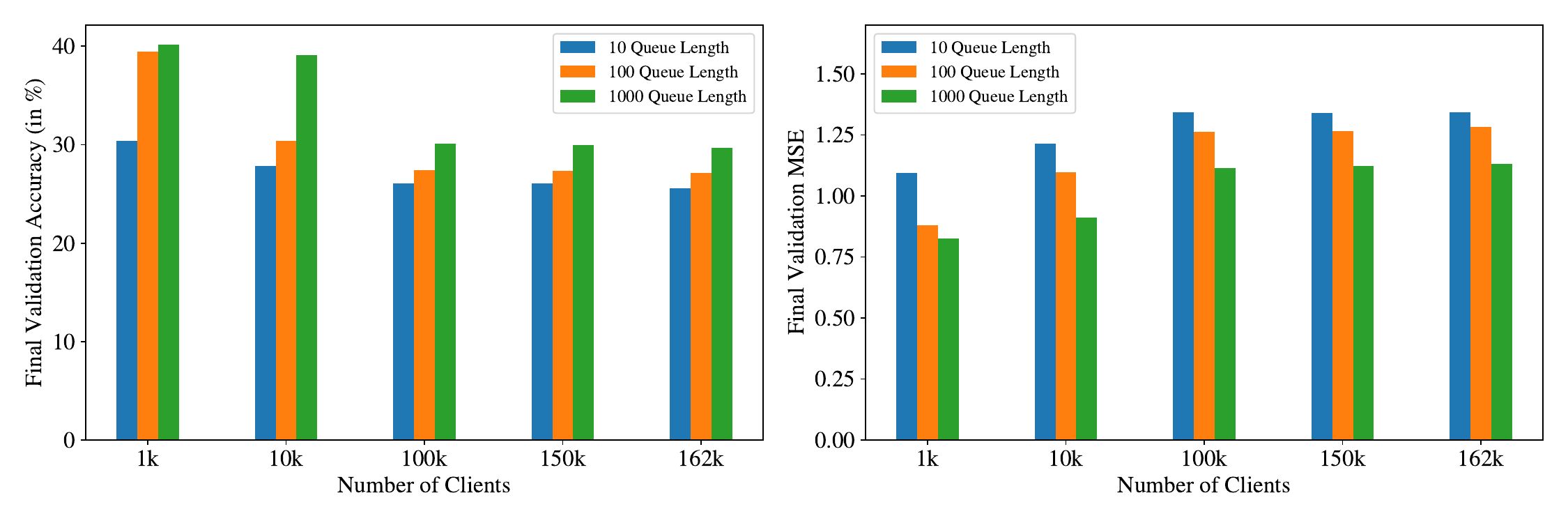}
        \caption{Ranker}
        \label{subfigure:fedq-experiment-results-ranker-final-validation-accuracy-and-mse-only}
    \end{subfigure}

    \caption{\Ac{fedq} experiment results for \textbf{(\subref{subfigure:fedq-experiment-results-candidate-generator-final-validation-top-100-accuracy-only})} the candidate generator and \textbf{(\subref{subfigure:fedq-experiment-results-ranker-final-validation-accuracy-and-mse-only})} the ranker. For increased legibility, instead of showing the complete training graphs, only the final validation top-100 accuracy, and the final validation accuracy and \ac{mse} are shown. For reference, the full training graphs are depicted in Appendix~\ref{appendix:extended-federated-learning-and-fedq-experiment-results}.}
    \Description{\Ac{fedq} experiment results for \textbf{(\subref{subfigure:fedq-experiment-results-candidate-generator-final-validation-top-100-accuracy-only})} the candidate generator and \textbf{(\subref{subfigure:fedq-experiment-results-ranker-final-validation-accuracy-and-mse-only})} the ranker. For increased legibility, instead of showing the complete training graphs, only the final validation top-100 accuracy, and the final validation accuracy and \ac{mse} are shown. For reference, the full training graphs are depicted in Appendix~\ref{appendix:extended-federated-learning-and-fedq-experiment-results}.}
    \label{figure:fedq-experiment-results-final-validation-metrics-only}
\end{figure}

As described in the previous section, the effects of non-\ac{iid}-ness and small local datasets result in a significant decrease in performance. Therefore, we employed the \ac{fedq} technique, described in Section~\ref{section:federated-recommender-systems-at-scale-using-queue-based-federated-learning}. We fixed the client sub-sampling rate at 1,000 clients per communication round and used varying queue lengths. In order to stabilize the training, the learning rate applied to the candidate generator was once again lowered in comparison to the non-\ac{fl} baseline experiments. The experiments were also conducted using the \ac{fl} simulator described in Appendix~\ref{appendix:federated-learning-simulator}, the results of which can be seen in Figure~\ref{figure:fedq-experiment-results-final-validation-metrics-only}.

As shown in Figure~\ref{subfigure:fedq-experiment-results-candidate-generator-final-validation-top-100-accuracy-only}, the candidate generator now performs much better as compared to standard \ac{fedavg} and in particular for the setups with 1k and 10k clients, \ac{fedq} even outperforms its baseline. The latter may be caused by a regularizing effect. In addition, the setups with large numbers of clients not only perform much better, but also in the expected way, as the performance slightly decreases with an increased number of clients. This provides evidence for our hypothesis that the candidate generator started to perform slightly better with increasing numbers of clients due to noisy updates induced by the decreased heterogeneity of the local dataset samples. Since the clients now train the global model sequentially, the number of samples adding to the local model update has drastically increased, thus resulting in much higher quality local updates. The ranker likewise shows a comparable improvement in performance and outperformed its non-\ac{fl} baseline, though with a smaller margin than the candidate generator, as shown in Figure~\ref{subfigure:fedq-experiment-results-ranker-final-validation-accuracy-and-mse-only}. Tables~\ref{table:candidate-generator-comparison-fl-and-fedq} and \ref{table:ranker-comparison-fl-and-fedq} compile the results of both the \ac{fl} and \ac{fedq} experiments and clearly show that \ac{fedq} outperforms \ac{fedavg} in every single experiment.

\begin{table}[ht]
    {\small
        \begin{tabular}{|r|c|c|c|c|c|}
            \cline{2-6}
            \multicolumn{1}{r|}{} & \textbf{1k Clients}      & \textbf{10k Clients}     & \textbf{100k Clients}    & \textbf{150k Clients}    & \textbf{162k Clients}    \\ [-2pt]
            \multicolumn{1}{r|}{} & {\footnotesize Accuracy} & {\footnotesize Accuracy} & {\footnotesize Accuracy} & {\footnotesize Accuracy} & {\footnotesize Accuracy} \\

            \hline
            \textsb{Sub-Sampling} & \multicolumn{5}{c|}{\textsb{\ac{fedavg}}}                                                                                            \\
            \hline

            10                    & 27.93\%                  & \textit{-}               & \textit{-}               & \textit{-}               & \textit{-}               \\
            100                   & 28.29\%                  & 17.98\%                  & 5.18\%                   & 7.83\%                   & 10.14\%                  \\
            1,000                 & 28.11\%                  & 17.86\%                  & 6.34\%                   & 8.53\%                   & 10.52\%                  \\
            10,000                & \textit{-}               & 17.91\%                  & 6.39\%                   & 6.68\%                   & 11.64\%                  \\

            \hline
            \textsb{Queue Length} & \multicolumn{5}{c|}{\textsb{\ac{fedq}}}                                                                                              \\
            \hline

            10                    & 49.61\%                  & 26.41\%                  & 17.59\%                  & 16.68\%                  & 17.01\%                  \\
            100                   & 51.57\%                  & 48.80\%                  & 25.35\%                  & 22.66\%                  & 23.72\%                  \\
            1,000                 & 52.12\%                  & 51.47\%                  & 42.34\%                  & 39.22\%                  & 38.67\%                  \\
            \hline
        \end{tabular}
    }

    \caption{Comparison of the candidate generator \ac{fl} and \ac{fedq} experiment results. The table reports the final validation top-100 accuracies after 300 communication rounds.}
    \Description{Comparison of the candidate generator \ac{fl} and \ac{fedq} experiment results. The table reports the final validation top-100 accuracies after 300 communication rounds.}
    \label{table:candidate-generator-comparison-fl-and-fedq}
\end{table}

\begin{table}[ht]
    {\small
        \begin{tabular}{|r|cc|cc|cc|cc|cc|}
            \cline{2-11}
            \multicolumn{1}{r|}{} & \multicolumn{2}{c|}{\textbf{1k Clients}} & \multicolumn{2}{c|}{\textbf{10k Clients}} & \multicolumn{2}{c|}{\textbf{100k Clients}} & \multicolumn{2}{c|}{\textbf{150k Clients}} & \multicolumn{2}{c|}{\textbf{162k Clients} \rule{0pt}{2.4ex}} \\ [-2pt]
            \multicolumn{1}{r|}{} & {\footnotesize Accuracy}   & {\footnotesize \ac{mse}}                    & {\footnotesize Accuracy}   & {\footnotesize \ac{mse}}                     & {\footnotesize Accuracy}   & {\footnotesize \ac{mse}}                      & {\footnotesize Accuracy}   & {\footnotesize \ac{mse}}                      & {\footnotesize Accuracy}   & {\footnotesize \ac{mse}}                                        \\

            \hline
            \textsb{Sub-Sampling} & \multicolumn{10}{c|}{\textsb{\ac{fedavg}}}                                                                                                                                                                                                    \\
            \hline

            10                    & 27.78\%    & 1.22                        & \textit{-} & \textit{-}                   & \textit{-} & \textit{-}                    & \textit{-} & \textit{-}                    & \textit{-} & \textit{-}                                      \\
            100                   & 27.91\%    & 1.22                        & 26.32\%    & 1.32                         & 24.27\%    & 1.42                          & 24.80\%    & 1.42                          & 23.14\%    & 1.33                                            \\
            1,000                 & 27.94\%    & 1.21                        & 26.47\%    & 1.33                         & 24.47\%    & 1.42                          & 24.43\%    & 1.43                          & 23.51\%    & 1.34                                            \\
            10,000                & \textit{-} & \textit{-}                  & 26.36\%    & 1.33                         & 24.74\%    & 1.42                          & 24.61\%    & 1.42                          & 22.77\%    & 1.36                                            \\

            \hline
            \textsb{Queue Length} & \multicolumn{10}{c|}{\textsb{\ac{fedq}}}                                                                                                                                                                                                      \\
            \hline

            10                    & 30.38\%    & 1.09                        & 27.85\%    & 1.22                         & 26.10\%    & 1.34                          & 26.10\%    & 1.34                          & 25.56\%    & 1.34                                            \\
            100                   & 39.45\%    & 0.88                        & 30.37\%    & 1.1                          & 27.43\%    & 1.26                          & 27.33\%    & 1.26                          & 27.15\%    & 1.28                                            \\
            1,000                 & 40.14\%    & 0.83                        & 39.07\%    & 0.91                         & 30.10\%    & 1.12                          & 29.98\%    & 1.12                          & 29.69\%    & 1.13                                            \\
            \hline
        \end{tabular}
    }

    \caption{Comparison of the ranker \ac{fl} and \ac{fedq} experiment results. The table reports the final validation accuracies and \acp{mse} after 300 communication rounds.}
    \Description{Comparison of the ranker \ac{fl} and \ac{fedq} experiment results. The table reports the final validation accuracies and \acp{mse} after 300 communication rounds.}
    \label{table:ranker-comparison-fl-and-fedq}
\end{table}

To further investigate the efficacy of \ac{fedq}, we have evaluated it using the LEAF benchmark~\cite{bibliography:leaf-a-benchmark-for-federated-settings}, which is a benchmark for testing \ac{fl} algorithms. The LEAF benchmark includes multiple different datasets that can naturally be partitioned into local datasets for \ac{fl} clients, as well as accompanying \ac{nn} models and metrics. We have benchmarked \ac{fedq} on four of LEAF's datasets and \ac{nn} architectures, which range from image classification using \acp{cnn}, to text classification and next word prediction using \acp{lstm}. The results of these experiments and a detailed evaluation can be found in Appendix~\ref{appendix:validation-of-fedq-on-the-leaf-federated-learning-benchmark}.

\subsection{Communication Compression Experiments}

As described in the Section~\ref{section:achieving-communication-efficiency}, \ac{fl} has, due to the continuous exchange of local updates between clients and central server, a significant communication overhead. We employed the recent \ac{nnc} standard to compress the \ac{nn} parametrizations communicated between the clients and the central server. The coding engine uses parameter quantization as a lossy preprocessing step and \ac{deepcabac} as arithmetic coder. The quantization of the parameters requires a hyperparameter called \ac{qp}, which controls the step size $\delta$ between quantization points and thus the rate-performance trade-off. A lower \ac{qp} results in a smaller step size and therefore in more quantization points and lower compression performance, while a higher \ac{qp} results in a larger step size and therefore in less quantization points and higher compression performance. To compute $\delta$ as demonstrated by Algorithm~\ref{algorithm:quantization}, it is necessary to provide an additional parameter $f_{QP}$, which incorporates the dependency between \acp{qp} and the quantization step sizes. Lower values of $f_{QP}$ result in larger neighboring quantization step sizes\footnote{\url{https://github.com/fraunhoferhhi/nncodec/wiki/usage}.}.

{\small
    \begin{algorithm}[ht]
        \caption{Quantization step size}
        \label{algorithm:quantization}

        \SetKwInOut{Input}{Input}
        \SetKwInOut{Output}{Output}

        \Input{$QP$ is the quantization parameter and $f_{QP}$ is the regulating parameter for mapping between \acp{qp} and quantization step sizes}
        \Output{Quantization step size $\delta$}
        $m \gets \left ( 1 << f_{QP} \right ) + \left ( QP + \left ( \left ( 1 << f_{QP} \right ) -1 \right ) \right ) $ \\
        $s \gets QP >>f_{QP} $ \\
        $\delta \gets m \cdot 2.0^{s - f_{QP}}$\\
        \KwRet $\delta$
    \end{algorithm}
}

\begin{figure}[ht]
    \centering

    \begin{subfigure}{0.5\textwidth}
        \centering
        \includegraphics[height=0.225\textheight]{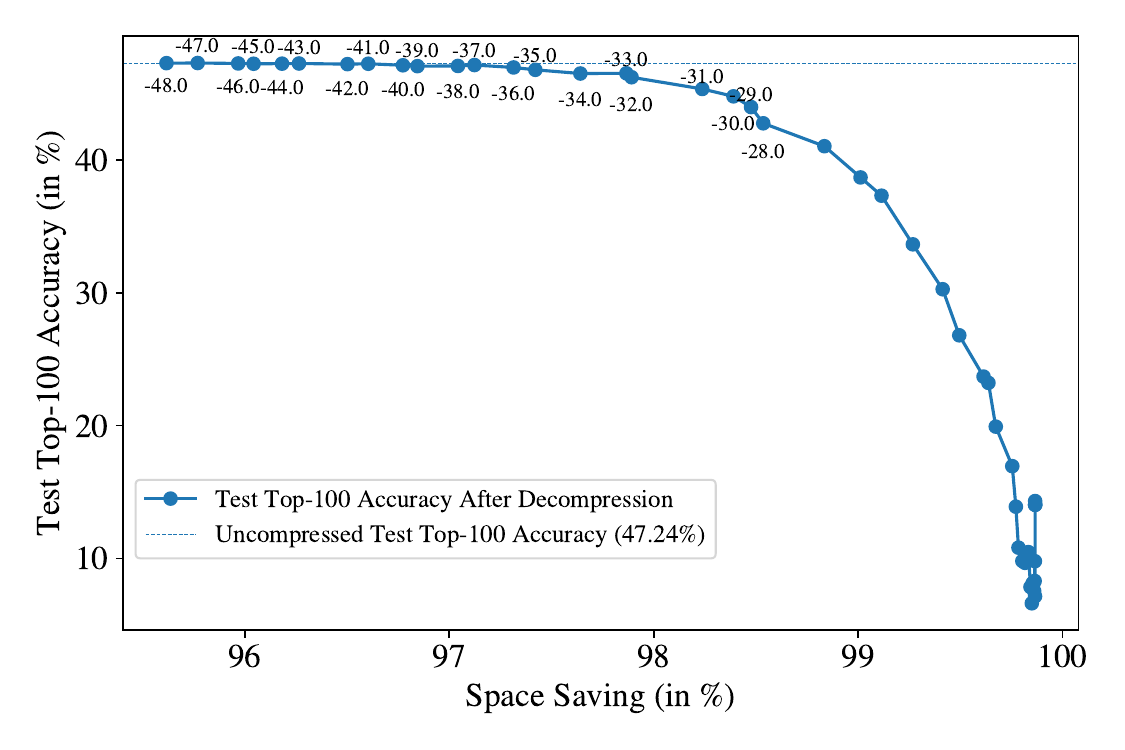}
        \caption{Candidate generator}
        \label{subfigure:compression-vs-accuracy-experiment-results-candidate-generator}
    \end{subfigure}%
    \hfill
    \begin{subfigure}{0.5\textwidth}
        \centering
        \includegraphics[height=0.225\textheight]{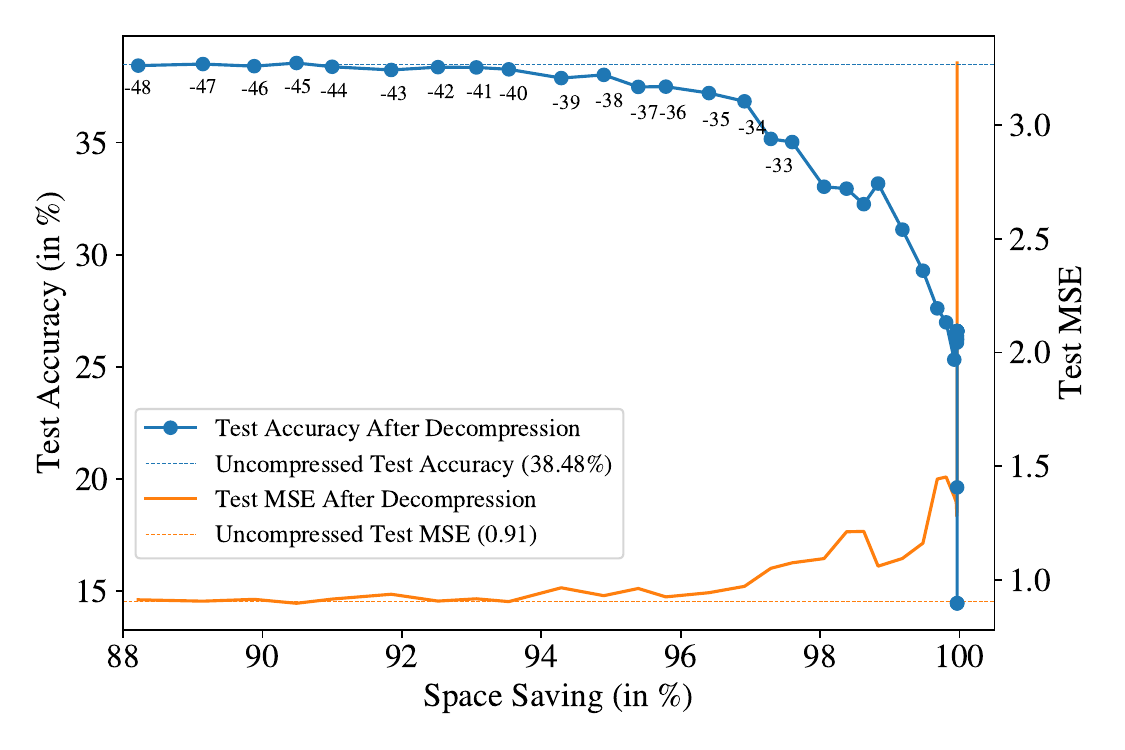}
        \caption{Ranker}
        \label{subfigure:compression-vs-accuracy-experiment-results-ranker}
    \end{subfigure}

    \caption{Compression vs. accuracy experiment results for \textbf{(\subref{subfigure:compression-vs-accuracy-experiment-results-candidate-generator})} the candidate generator and \textbf{(\subref{subfigure:compression-vs-accuracy-experiment-results-ranker})} the ranker.}
    \Description{Compression vs. accuracy experiment results for \textbf{(\subref{subfigure:compression-vs-accuracy-experiment-results-candidate-generator})} the candidate generator and \textbf{(\subref{subfigure:compression-vs-accuracy-experiment-results-ranker})} the ranker.}
    \label{figure:compression-vs-accuracy-experiment-results}
\end{figure}

Besides influencing the compression performance, the \ac{qp} also impacts the performance of the \ac{nn} model after decompression, i.e., if the \ac{qp} was chosen too large, the resulting performance is significantly decreased. In order to determine the optimal value of the \ac{qp} for the candidate generator and the ranker models, we performed an experiment, testing \ac{qp} values between -48 and 0. The results are shown in Figure~\ref{figure:compression-vs-accuracy-experiment-results}.

\begin{figure}[ht]
    \centering

    \begin{subfigure}{\textwidth}
        \centering
        \includegraphics[height=0.225\textheight]{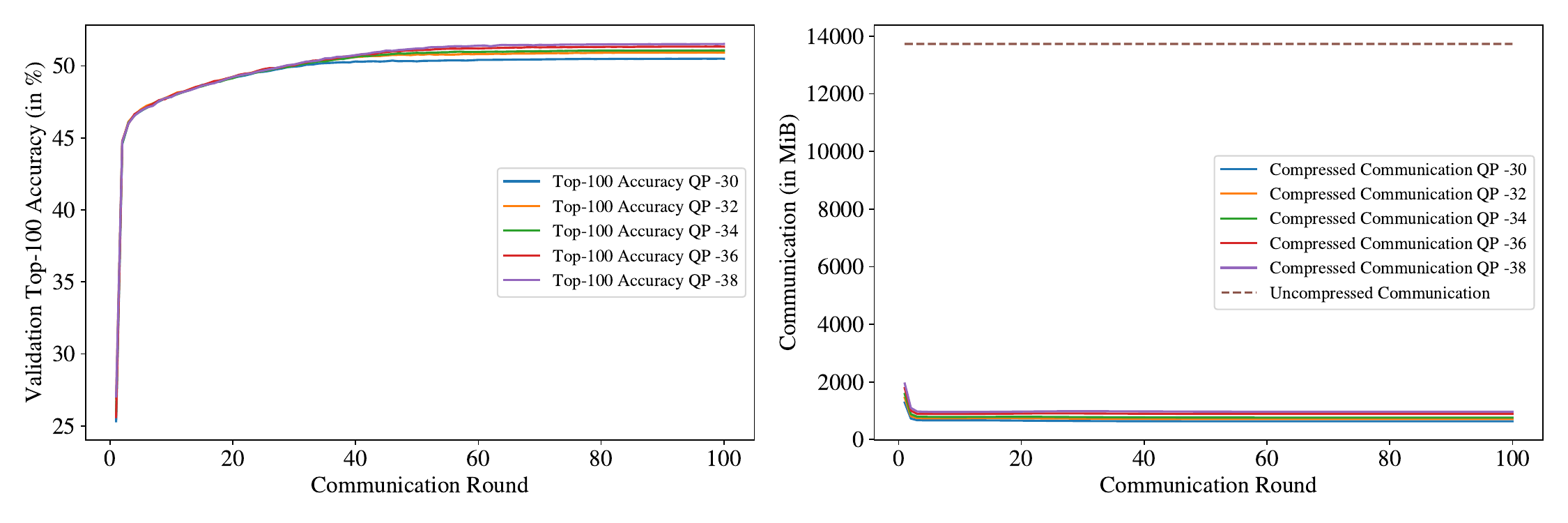}
        \caption{Candidate generator}
        \label{subfigure:federated-learning-compression-experiment-results-candidate-generator}
    \end{subfigure}%
    \hfill
    \begin{subfigure}{\textwidth}
        \centering
        \includegraphics[height=0.225\textheight]{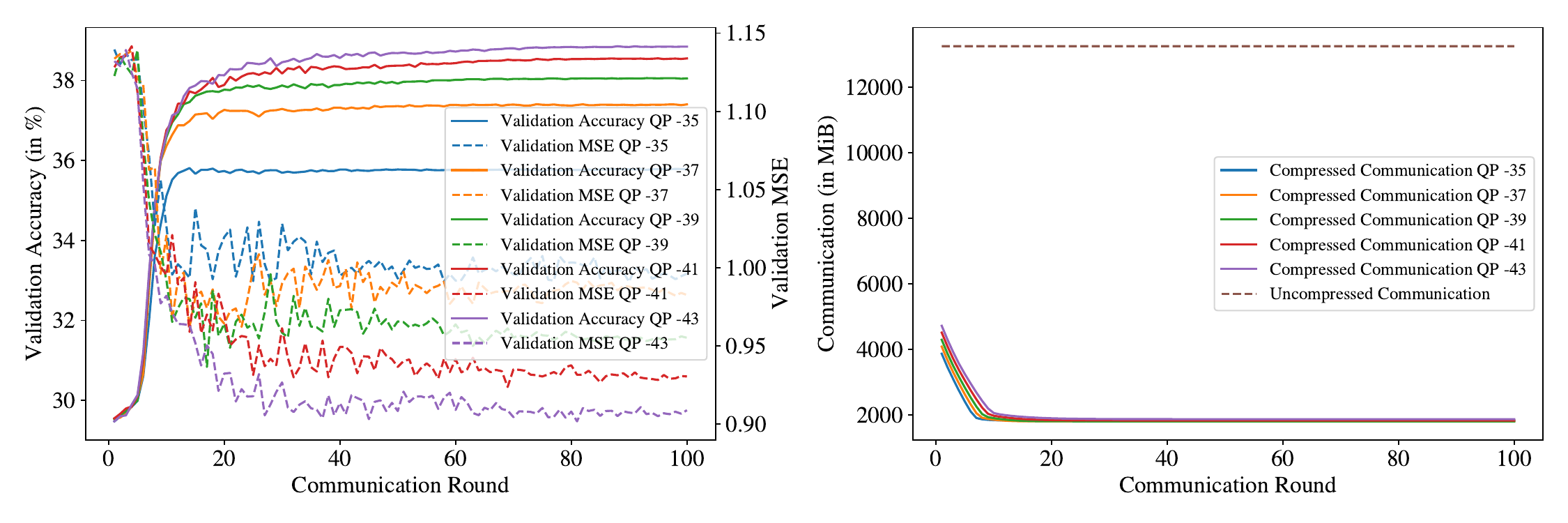}
        \caption{Ranker}
        \label{subfigure:federated-learning-compression-experiment-results-ranker}
    \end{subfigure}

    \caption{\Ac{fl} with communication compression experiment results for \textbf{(\subref{subfigure:federated-learning-compression-experiment-results-candidate-generator})} the candidate generator and \textbf{(\subref{subfigure:federated-learning-compression-experiment-results-ranker})} the ranker.}
    \Description{\Ac{fl} with communication compression experiment results for \textbf{(\subref{subfigure:federated-learning-compression-experiment-results-candidate-generator})} the candidate generator and \textbf{(\subref{subfigure:federated-learning-compression-experiment-results-ranker})} the ranker.}
    \label{figure:federated-learning-compression-experiment-results}
\end{figure}

For \ac{fl}, the \ac{qp} value should be chosen in a way to optimize the rate-distortion trade-off. As can be seen in Figure~\ref{figure:compression-vs-accuracy-experiment-results}, a \ac{qp} range of \textbf{-38} and \textbf{-30} for the candidate generator, and \textbf{-43} and \textbf{-35} for the ranker results in compression rates with no or marginal performance degradation. Since the compression performance (per client) in our setting is independent from the number of clients, we only performed experiments with 100 clients and no client sub-sampling, i.e., all clients were included in every communication round. The experiments also perform \ac{fedq} with a queue length of 10. Besides the number of clients, the client sub-sampling rate, and the \ac{fedq} queue length, the other hyperparameters of the experimental setup are identical to the \ac{fl} experiments in Section~\ref{section:federated-learning-experiments}. The results of this experiment are depicted in Figure~\ref{figure:federated-learning-compression-experiment-results}.

The plots on the left of each sub-figure show the model performance, while the plots on the right of each sub-figure demonstrate the compression performance for different \ac{qp} values. According to the compression performance plots, the initial number of communicated MiB is slightly higher, as in the beginning weights are initialized with random values. During the course of training, the entropy of the weights decreases, resulting in better compression performance. After a few communication rounds, the compression performance saturates at an almost constant value. For the candidate generator, the space saving, as compared to uncompressed communication, varies between \textbf{92.97\%} for \ac{qp} -38 and \textbf{95.37\%} for \ac{qp} -30. For the ranker, the space saving, as compared to uncompressed communication, varies between \textbf{85.88\%} for \ac{qp} -43 and \textbf{86.17\%} for \ac{qp} -35. The space savings are lower in comparison to the non-\ac{fl} baseline, where the candidate generator achieved \textbf{97.04\%} for \ac{qp} -38 and \textbf{98.39\%} for \ac{qp} -30, and the ranker \textbf{91.85\%} for \ac{qp} -43 and \textbf{96.4\%} for \ac{qp} -35. This seems to be an effect that is inherent to \ac{fl}. In centralized training, regularization methods produce small magnitude weights, which results in higher sparsity when applying quantization. The exact weights that are going towards zero can, however, differ between several training runs, which means that in an \ac{fl} setting each client can have different weights of small magnitude. Due to the averaging of the weights in \ac{fedavg}, the produced global model will most likely be less sparse than its constituent local models. For example, the overall entropy of the candidate generator that was trained using \ac{fl} with compression is \textbf{4.41 bits}, while the overall entropy of the candidate generator that was trained using \ac{fl} without compression is \textbf{7.5 bits}, and the baseline candidate generator which was trained centrally without compression has an entropy of \textbf{1.26 bits}. The overall entropy of the ranker that was trained using \ac{fl} with compression is \textbf{11.44 bits}, while the overall entropy of the ranker that was trained using \ac{fl} without compression is \textbf{12.59 bits} and the baseline ranker, which was trained centrally without compression, has an entropy of \textbf{3.67 bits}. This shows that models that are trained using \ac{fl} do, in fact, have a higher entropy and are thus less amenable to compression. Quantization on the other hand seems to induce more sparsity, thus lowering the resulting entropy for models trained with compression. Furthermore, this also shows why the ranker performs much worse in terms of space saving, as it has much higher entropy in general.

The candidate generator has excellent model performance, even for higher \acp{qp}, well outperforming the non-\ac{fl} baseline and showing the same performance characteristics as the \ac{fedq} experiments presented in Section~\ref{section:fedq-experiments}. As the loss in performance and the increase in compression performance are very small, any of the tested \acp{qp} are well-suited to be used, so we selected a \ac{qp} of -30, which offers the best overall space saving of \textbf{95.37\%} and a top-100 accuracy of \textbf{50.5\%}, which is only one percentage point smaller than the best accuracy and well above the non-\ac{fl} baseline of 47.15\%.

The training of the ranker model, however, seems to be much more affected by the compression as compared to the candidate generator, although the increase in compression performance is exceedingly small with increasing \acp{qp}. Only \acp{qp} -43 and -41 manage to meet the non-\ac{fl} baseline and none of the \acp{qp} achieve a performance that is in line with the results of the \ac{fedq} experiments. This is, however, to be expected since the lossy compression of \ac{nnc} may hurt the performance of the models. In this case, the difference between the best performance reached with compression is only slightly lower than the best performance of \ac{fedq} without compression. Therefore, \ac{qp} -43 was selected as it slightly outperforms the non-\ac{fl} baseline with \textbf{38.85\%} accuracy and an \ac{mse} of \textbf{0.91} but reaches almost the same space savings as the smallest \ac{qp} with \textbf{85.88\%} as compared to \textbf{86.17\%}.

\section{Conclusion \& Outlook}

Modern \acp{recsys}, especially the ones based on \ac{dl}, benefit from increasing amounts of personal information about its users. This has resulted in the collection of substantial amounts of personal data on many platforms in recent years, leading to a data privacy problem. Here, \Ac{fl} has emerged as a technique that intrinsically provides privacy and is therefore used in many scenarios where data privacy is of high priority. Consequently, we presented a movie \ac{recsys}, which is being trained end-to-end using \ac{fl} and scales well to exceptionally large numbers of users. We have identified major problems in such systems and proposed solutions to them. In particular, we have shown that the non-\ac{iid}-ness of the clients' local datasets, as well as small local datasets can significantly degrade the federated training of a \ac{recsys} and developed a novel technique, called \ac{fedq}, which satisfactorily counteracts this problem. Furthermore, the substantial overhead of constantly communicating \ac{nn} parametrizations between server and clients in \ac{fl} poses a problem, especially when clients are connected via mobile internet connections. For this, we have shown that the most recent \ac{nnc} compression technology can considerably reduce this communication overhead to a fraction of the uncompressed communication.

Beyond the proposed significant improvements to the overall \ac{recsys}, additional improvements can be achieved through further research. In the area of data privacy, differential privacy methods could be further investigated and combined with the quantization-induced privacy by \ac{nnc} communication compression. Another topic of interest is the learning of embeddings in an \ac{fl} setting, which is known to be problematic. Solutions proposed in the literature, seem to all depend on the partial disclosure of client data. Here, future work could investigate possibilities of learning embeddings in an \ac{fl} setting without disclosing private information. The space savings of the compression can be further improved by differential compression, i.e., only the difference between the global model and the updated local model is compressed, which is sparser and is thus more amenable to compression. Finally, the non-\ac{iid}-ness in the \ac{fedrec} scenario originates from different user preferences. The local datasets within user groups of similar preference should be much more homogeneous, leading the way to further model performance improvements for \acp{fedrec}.

\clearpage

\bibliographystyle{ACM-Reference-Format}
\bibliography{acm-tors-paper}


\begin{thebibliography}{134}


\ifx \showCODEN    \undefined \def \showCODEN     #1{\unskip}     \fi
\ifx \showDOI      \undefined \def \showDOI       #1{#1}\fi
\ifx \showISBNx    \undefined \def \showISBNx     #1{\unskip}     \fi
\ifx \showISBNxiii \undefined \def \showISBNxiii  #1{\unskip}     \fi
\ifx \showISSN     \undefined \def \showISSN      #1{\unskip}     \fi
\ifx \showLCCN     \undefined \def \showLCCN      #1{\unskip}     \fi
\ifx \shownote     \undefined \def \shownote      #1{#1}          \fi
\ifx \showarticletitle \undefined \def \showarticletitle #1{#1}   \fi
\ifx \showURL      \undefined \def \showURL       {\relax}        \fi
\providecommand\bibfield[2]{#2}
\providecommand\bibinfo[2]{#2}
\providecommand\natexlab[1]{#1}
\providecommand\showeprint[2][]{arXiv:#2}

\bibitem[Alam et~al\mbox{.}(2016)]%
        {bibliography:joint-multi-grain-topic-sentiment-modelling-semantic-aspects-for-online-reviews}
\bibfield{author}{\bibinfo{person}{Md.~Hijbul Alam}, \bibinfo{person}{Woo-Jong Ryu}, {and} \bibinfo{person}{SangKeun Lee}.} \bibinfo{year}{2016}\natexlab{}.
\newblock \showarticletitle{Joint Multi-Grain Topic Sentiment}.
\newblock \bibinfo{journal}{\emph{Information Sciences}} \bibinfo{volume}{339}, \bibinfo{number}{C} (\bibinfo{date}{April} \bibinfo{year}{2016}), \bibinfo{pages}{206--223}.
\newblock
\showISSN{0020-0255}
\urldef\tempurl%
\url{https://doi.org/10.1016/j.ins.2016.01.013}
\showDOI{\tempurl}


\bibitem[Alamgir et~al\mbox{.}(2022)]%
        {bibliography:federated-recommenders-methods-challenges-and-future}
\bibfield{author}{\bibinfo{person}{Zareen Alamgir}, \bibinfo{person}{Farwa~K. Khan}, {and} \bibinfo{person}{Saira Karim}.} \bibinfo{year}{2022}\natexlab{}.
\newblock \showarticletitle{Federated Recommenders: Methods, Challenges and Future}.
\newblock \bibinfo{journal}{\emph{Cluster Computing}} \bibinfo{volume}{25}, \bibinfo{number}{6} (\bibinfo{date}{June} \bibinfo{year}{2022}), \bibinfo{pages}{4075--4096}.
\newblock
\showISSN{1386-7857}
\urldef\tempurl%
\url{https://doi.org/10.1007/s10586-022-03644-w}
\showDOI{\tempurl}


\bibitem[{Ammad-ud-din} et~al\mbox{.}(2019)]%
        {bibliography:federated-collaborative-filtering-for-privacy-preserving-personalized-recommendation-system}
\bibfield{author}{\bibinfo{person}{Muhammad {Ammad-ud-din}}, \bibinfo{person}{Elena {Ivannikova}}, \bibinfo{person}{Suleiman~A. {Khan}}, \bibinfo{person}{Were {Oyomno}}, \bibinfo{person}{Qiang {Fu}}, \bibinfo{person}{Kuan {Eeik Tan}}, {and} \bibinfo{person}{Adrian {Flanagan}}.} \bibinfo{year}{2019}\natexlab{}.
\newblock \showarticletitle{{Federated Collaborative Filtering for Privacy-Preserving Personalized Recommendation System}}.
\newblock \bibinfo{journal}{\emph{arXiv e-prints}}  \bibinfo{volume}{abs/1901.09888} (\bibinfo{date}{Jan.} \bibinfo{year}{2019}), \bibinfo{numpages}{12}~pages.
\newblock
\showeprint[arxiv]{1901.09888}~[cs.IR]


\bibitem[Asad et~al\mbox{.}(2023)]%
        {bibliography:a-comprehensive-survey-on-privacy-preserving-techniques-in-federated-recommendation-systems}
\bibfield{author}{\bibinfo{person}{Muhammad Asad}, \bibinfo{person}{Saima Shaukat}, \bibinfo{person}{Ehsan Javanmardi}, \bibinfo{person}{Jin Nakazato}, {and} \bibinfo{person}{Manabu Tsukada}.} \bibinfo{year}{2023}\natexlab{}.
\newblock \showarticletitle{A Comprehensive Survey on Privacy-Preserving Techniques in Federated Recommendation Systems}.
\newblock \bibinfo{journal}{\emph{Applied Sciences}} \bibinfo{volume}{13}, \bibinfo{number}{10} (\bibinfo{year}{2023}), \bibinfo{numpages}{26}~pages.
\newblock
\showISSN{2076-3417}
\urldef\tempurl%
\url{https://doi.org/10.3390/app13106201}
\showDOI{\tempurl}


\bibitem[{Bird} et~al\mbox{.}(2009)]%
        {bibliography:natural-language-processing-with-python}
\bibfield{author}{\bibinfo{person}{Steven {Bird}}, \bibinfo{person}{Ewan {Klein}}, {and} \bibinfo{person}{Edward {Loper}}.} \bibinfo{year}{2009}\natexlab{}.
\newblock \bibinfo{booktitle}{\emph{Natural Language Processing with Python}}.
\newblock \bibinfo{publisher}{O'Reilly Media Inc.}, \bibinfo{address}{Sebastopol, California, United States of America}.
\newblock
\showISBNx{9780596516499}


\bibitem[{Brendan McMahan} et~al\mbox{.}(2018)]%
        {bibliography:learning-differentially-private-recurrent-language-models}
\bibfield{author}{\bibinfo{person}{H. {Brendan McMahan}}, \bibinfo{person}{Daniel {Ramage}}, \bibinfo{person}{Kunal {Talwar}}, {and} \bibinfo{person}{Li {Zhang}}.} \bibinfo{year}{2018}\natexlab{}.
\newblock \showarticletitle{{Learning Differentially Private Recurrent Language Models}}. In \bibinfo{booktitle}{\emph{International Conference on Learning Representations}}. \bibinfo{publisher}{OpenReview.net}, \bibinfo{address}{Vancouver, British Columbia, Canada}, \bibinfo{numpages}{14}~pages.
\newblock
\urldef\tempurl%
\url{https://openreview.net/forum?id=BJ0hF1Z0b}
\showURL{%
\tempurl}


\bibitem[{Caldas} et~al\mbox{.}(2019a)]%
        {bibliography:expanding-the-reach-of-federated-learning-by-reducing-client-resource-requirements}
\bibfield{author}{\bibinfo{person}{Sebastian {Caldas}}, \bibinfo{person}{J. Kone{\v{c}}n{\'{y}}}, \bibinfo{person}{H.~Brendan {McMahan}}, {and} \bibinfo{person}{Ameet {Talwalkar}}.} \bibinfo{year}{2019}\natexlab{a}.
\newblock \showarticletitle{Expanding the Reach of Federated Learning by Reducing Client Resource Requirements}.
\newblock \bibinfo{journal}{\emph{arXiv e-prints}}  \bibinfo{volume}{abs/1812.07210} (\bibinfo{date}{Jan.} \bibinfo{year}{2019}).
\newblock
\showeprint[arxiv]{1812.07210}~[cs.LG]


\bibitem[{Caldas} et~al\mbox{.}(2019b)]%
        {bibliography:leaf-a-benchmark-for-federated-settings}
\bibfield{author}{\bibinfo{person}{Sebastian {Caldas}}, \bibinfo{person}{Sai {Meher Karthik Duddu}}, \bibinfo{person}{Peter {Wu}}, \bibinfo{person}{Tian {Li}}, \bibinfo{person}{Jakub {Konečný}}, \bibinfo{person}{H.~Brendan {McMahan}}, \bibinfo{person}{Virginia {Smith}}, {and} \bibinfo{person}{Ameet {Talwalkar}}.} \bibinfo{year}{2019}\natexlab{b}.
\newblock \showarticletitle{{LEAF:} {A} Benchmark for Federated Settings}.
\newblock \bibinfo{journal}{\emph{CoRR}}  \bibinfo{volume}{abs/1812.01097} (\bibinfo{date}{Dec.} \bibinfo{year}{2019}).
\newblock
\urldef\tempurl%
\url{https://doi.org/10.48550/arXiv.1812.01097}
\showDOI{\tempurl}
\showeprint[arxiv]{1812.01097}~[cs.LG]


\bibitem[Cao et~al\mbox{.}(2022)]%
        {bibliography:class-aware-client-selection-for-effective-aggregation-in-federated-learning}
\bibfield{author}{\bibinfo{person}{Mei Cao}, \bibinfo{person}{Yujie Zhang}, \bibinfo{person}{Zezhong Ma}, {and} \bibinfo{person}{Mengying Zhao}.} \bibinfo{year}{2022}\natexlab{}.
\newblock \showarticletitle{C2S: Class-aware client selection for effective aggregation in federated learning}.
\newblock \bibinfo{journal}{\emph{High-Confidence Computing}} \bibinfo{volume}{2}, \bibinfo{number}{3} (\bibinfo{year}{2022}), \bibinfo{pages}{100068}.
\newblock
\showISSN{2667-2952}
\urldef\tempurl%
\url{https://doi.org/10.1016/j.hcc.2022.100068}
\showDOI{\tempurl}


\bibitem[Cao et~al\mbox{.}(2007)]%
        {bibliography:learning-to-rank}
\bibfield{author}{\bibinfo{person}{Zhe Cao}, \bibinfo{person}{Tao Qin}, \bibinfo{person}{Tie-Yan Liu}, \bibinfo{person}{Ming-Feng Tsai}, {and} \bibinfo{person}{Hang Li}.} \bibinfo{year}{2007}\natexlab{}.
\newblock \showarticletitle{Learning to Rank: From Pairwise Approach to Listwise Approach}. In \bibinfo{booktitle}{\emph{Proceedings of the 24th International Conference on Machine Learning}} (Corvalis, Oregon, USA) \emph{(\bibinfo{series}{ICML '07})}. \bibinfo{publisher}{Association for Computing Machinery}, \bibinfo{address}{New York, NY, USA}, \bibinfo{pages}{129--136}.
\newblock
\showISBNx{9781595937933}
\urldef\tempurl%
\url{https://doi.org/10.1145/1273496.1273513}
\showDOI{\tempurl}


\bibitem[Chai et~al\mbox{.}(2021)]%
        {bibliography:secure-federated-matrix-factorization}
\bibfield{author}{\bibinfo{person}{D. Chai}, \bibinfo{person}{L. Wang}, \bibinfo{person}{K. Chen}, {and} \bibinfo{person}{Q. Yang}.} \bibinfo{year}{2021}\natexlab{}.
\newblock \showarticletitle{Secure Federated Matrix Factorization}.
\newblock \bibinfo{journal}{\emph{IEEE Intelligent Systems}} \bibinfo{volume}{36}, \bibinfo{number}{05} (\bibinfo{date}{Sept.} \bibinfo{year}{2021}), \bibinfo{pages}{11--20}.
\newblock
\showISSN{1941-1294}
\urldef\tempurl%
\url{https://doi.org/10.1109/MIS.2020.3014880}
\showDOI{\tempurl}


\bibitem[{Chai} et~al\mbox{.}(2022)]%
        {bibliography:fedeval-a-benchmark}
\bibfield{author}{\bibinfo{person}{Di {Chai}}, \bibinfo{person}{Leye {Wang}}, \bibinfo{person}{Liu {Yang}}, \bibinfo{person}{Junxue {Zhang}}, \bibinfo{person}{Kai {Chen}}, {and} \bibinfo{person}{Qiang {Yang}}.} \bibinfo{year}{2022}\natexlab{}.
\newblock \showarticletitle{{FedEval: A Holistic Evaluation Framework for Federated Learning}}.
\newblock \bibinfo{journal}{\emph{arXiv e-prints}}  \bibinfo{volume}{abs/2011.09655} (\bibinfo{date}{Dec.} \bibinfo{year}{2022}), \bibinfo{numpages}{14}~pages.
\newblock
\urldef\tempurl%
\url{https://doi.org/10.48550/arXiv.2011.09655}
\showDOI{\tempurl}
\showeprint[arxiv]{2011.09655}~[cs.LG]


\bibitem[{Chen} et~al\mbox{.}(2019)]%
        {bibliography:federated-meta-learning-with-fast-convergence-and-efficient-communication}
\bibfield{author}{\bibinfo{person}{Fei {Chen}}, \bibinfo{person}{Mi {Luo}}, \bibinfo{person}{Zhenhua {Dong}}, \bibinfo{person}{Zhenguo {Li}}, {and} \bibinfo{person}{Xiuqiang {He}}.} \bibinfo{year}{2019}\natexlab{}.
\newblock \showarticletitle{{Federated Meta-Learning with Fast Convergence and Efficient Communication}}.
\newblock \bibinfo{journal}{\emph{arXiv e-prints}}  \bibinfo{volume}{1802.07876} (\bibinfo{date}{Dec.} \bibinfo{year}{2019}).
\newblock
\urldef\tempurl%
\url{https://doi.org/10.48550/arXiv.1802.07876}
\showDOI{\tempurl}
\showeprint[arxiv]{1802.07876}~[cs.LG]


\bibitem[{Chen} et~al\mbox{.}(2011)]%
        {bibliography:feature-based-matrix-factorization}
\bibfield{author}{\bibinfo{person}{Tianqi {Chen}}, \bibinfo{person}{Zhao {Zheng}}, \bibinfo{person}{Qiuxia {Lu}}, \bibinfo{person}{Weinan {Zhang}}, {and} \bibinfo{person}{Yong {Yu}}.} \bibinfo{year}{2011}\natexlab{}.
\newblock \showarticletitle{Feature-Based Matrix Factorization}.
\newblock \bibinfo{journal}{\emph{arXiv e-prints}}  \bibinfo{volume}{abs/1109.2271} (\bibinfo{date}{Dec.} \bibinfo{year}{2011}), \bibinfo{numpages}{12}~pages.
\newblock
\urldef\tempurl%
\url{https://doi.org/10.48550/arXiv.1109.2271}
\showDOI{\tempurl}
\showeprint[arxiv]{1109.2271}~[cs.AI]


\bibitem[Chen et~al\mbox{.}(2022)]%
        {bibliography:optimal-client-sampling-for-federated-learning}
\bibfield{author}{\bibinfo{person}{Wenlin Chen}, \bibinfo{person}{Samuel Horv{\'a}th}, {and} \bibinfo{person}{Peter Richt{\'a}rik}.} \bibinfo{year}{2022}\natexlab{}.
\newblock \showarticletitle{Optimal Client Sampling for Federated Learning}.
\newblock \bibinfo{journal}{\emph{Transactions on Machine Learning Research}} \bibinfo{volume}{2022}, \bibinfo{number}{08} (\bibinfo{year}{2022}), \bibinfo{numpages}{32}~pages.
\newblock
\showISSN{2835-8856}
\urldef\tempurl%
\url{https://openreview.net/forum?id=8GvRCWKHIL}
\showURL{%
\tempurl}


\bibitem[Cho et~al\mbox{.}(2014)]%
        {bibliography:on-the-properties-of-neural-machine-translation:}
\bibfield{author}{\bibinfo{person}{Kyunghyuna Cho}, \bibinfo{person}{Bart van Merriënboer}, \bibinfo{person}{Dzmitry Bahdanau}, {and} \bibinfo{person}{Yoshua Bengio}.} \bibinfo{year}{2014}\natexlab{}.
\newblock \showarticletitle{On the Properties of Neural Machine Translation: Encoder{-}Decoder Approaches}. In \bibinfo{booktitle}{\emph{Proceedings of {SSST}-8, Eighth Workshop on Syntax, Semantics and Structure in Statistical Translation}}. \bibinfo{publisher}{Association for Computational Linguistics}, \bibinfo{address}{Doha, Qatar}, \bibinfo{pages}{103--111}.
\newblock
\urldef\tempurl%
\url{https://doi.org/10.3115/v1/W14-4012}
\showDOI{\tempurl}


\bibitem[Choe et~al\mbox{.}(2021)]%
        {bibliography:recommendation-system-with-hierarchical-recurrent-neural-network-for-long-term-time-series}
\bibfield{author}{\bibinfo{person}{Byeongjin Choe}, \bibinfo{person}{Taegwan Kang}, {and} \bibinfo{person}{Kyomin Jung}.} \bibinfo{year}{2021}\natexlab{}.
\newblock \showarticletitle{Recommendation System With Hierarchical Recurrent Neural Network for Long-Term Time Series}.
\newblock \bibinfo{journal}{\emph{IEEE Access}} \bibinfo{volume}{9}, \bibinfo{number}{1} (\bibinfo{year}{2021}), \bibinfo{pages}{72033--72039}.
\newblock
\urldef\tempurl%
\url{https://doi.org/10.1109/ACCESS.2021.3079922}
\showDOI{\tempurl}


\bibitem[Choi et~al\mbox{.}(2017)]%
        {bibliography:towards-the-limit-of-network-quantization}
\bibfield{author}{\bibinfo{person}{Yoojin Choi}, \bibinfo{person}{Mostafa El-Khamy}, {and} \bibinfo{person}{Jungwon Lee}.} \bibinfo{year}{2017}\natexlab{}.
\newblock \showarticletitle{Towards the Limit of Network Quantization}. In \bibinfo{booktitle}{\emph{International Conference on Learning Representations}}. \bibinfo{publisher}{OpenReview.net}, \bibinfo{address}{Toulon, France}, \bibinfo{numpages}{14}~pages.
\newblock
\urldef\tempurl%
\url{https://openreview.net/forum?id=rJ8uNptgl}
\showURL{%
\tempurl}


\bibitem[Cohen et~al\mbox{.}(2017)]%
        {bibliography:emnist-extending-mnist-to-handwritten-letters}
\bibfield{author}{\bibinfo{person}{Gregory Cohen}, \bibinfo{person}{Saeed Afshar}, \bibinfo{person}{Jonathan Tapson}, {and} \bibinfo{person}{André van Schaik}.} \bibinfo{year}{2017}\natexlab{}.
\newblock \showarticletitle{EMNIST: Extending MNIST to handwritten letters}. In \bibinfo{booktitle}{\emph{2017 International Joint Conference on Neural Networks (IJCNN)}} (Anchorage, Alaska, United States of America). \bibinfo{publisher}{Institute of Electrical and Electronics Engineers ({IEEE})}, \bibinfo{address}{3 Park Avenue, 17th Floor, New York, NY 10016-5997 USA}, \bibinfo{pages}{2921--2926}.
\newblock
\urldef\tempurl%
\url{https://doi.org/10.1109/IJCNN.2017.7966217}
\showDOI{\tempurl}


\bibitem[Covington et~al\mbox{.}(2016)]%
        {bibliography:dnns-for-youtube-recommendations}
\bibfield{author}{\bibinfo{person}{Paul Covington}, \bibinfo{person}{Jay Adams}, {and} \bibinfo{person}{Emre Sargin}.} \bibinfo{year}{2016}\natexlab{}.
\newblock \showarticletitle{Deep Neural Networks for YouTube Recommendations}. In \bibinfo{booktitle}{\emph{Proceedings of the 10th ACM Conference on Recommender Systems}} (Boston, Massachusetts, USA) \emph{(\bibinfo{series}{RecSys '16})}. \bibinfo{publisher}{ACM (Association for Computer Machinery)}, \bibinfo{address}{New York, NY, USA}, \bibinfo{pages}{191--198}.
\newblock
\showISBNx{9781450340359}
\urldef\tempurl%
\url{https://doi.org/10.1145/2959100.2959190}
\showDOI{\tempurl}


\bibitem[Cramer et~al\mbox{.}(2015)]%
        {bibliography:secure-multiparty-computation-and-secret-sharing}
\bibfield{author}{\bibinfo{person}{Ronald Cramer}, \bibinfo{person}{Ivan~Bjerre Damg{\aa}rd}, {and} \bibinfo{person}{Jesper~Buus Nielsen}.} \bibinfo{year}{2015}\natexlab{}.
\newblock \bibinfo{booktitle}{\emph{Secure Multiparty Computation and Secret Sharing}}.
\newblock \bibinfo{publisher}{Cambridge University Press}, \bibinfo{address}{Cambridge, United Kingdom}.
\newblock
\showISBNx{9781107337756}
\urldef\tempurl%
\url{https://doi.org/10.1017/CBO9781107337756}
\showDOI{\tempurl}


\bibitem[Dimitrov et~al\mbox{.}(2022)]%
        {bibliography:data-leakage-in-federated-averaging}
\bibfield{author}{\bibinfo{person}{Dimitar~I. Dimitrov}, \bibinfo{person}{Mislav Balunović}, \bibinfo{person}{Nikola Konstantinov}, {and} \bibinfo{person}{Martin Vechev}.} \bibinfo{year}{2022}\natexlab{}.
\newblock \showarticletitle{Data Leakage in Federated Averaging}.
\newblock \bibinfo{journal}{\emph{arXiv e-prints}}  \bibinfo{volume}{abs/2206.12395} (\bibinfo{year}{2022}).
\newblock
\urldef\tempurl%
\url{https://doi.org/10.48550/ARXIV.2206.12395}
\showDOI{\tempurl}


\bibitem[Dwork(2008)]%
        {bibliography:differential-privacy-a-survey-of-results}
\bibfield{author}{\bibinfo{person}{Cynthia Dwork}.} \bibinfo{year}{2008}\natexlab{}.
\newblock \showarticletitle{Differential Privacy: A Survey of Results}. In \bibinfo{booktitle}{\emph{Theory and Applications of Models of Computation}}, \bibfield{editor}{\bibinfo{person}{Manindra Agrawal}, \bibinfo{person}{Dingzhu Du}, \bibinfo{person}{Zhenhua Duan}, {and} \bibinfo{person}{Angsheng Li}} (Eds.). \bibinfo{publisher}{Springer Berlin Heidelberg}, \bibinfo{address}{Berlin, Heidelberg}, \bibinfo{pages}{1--19}.
\newblock
\showISBNx{978-3-540-79228-4}


\bibitem[Dwork and Roth(2014)]%
        {bibliography:the-algorithmic-foundations-of-differential-privacy}
\bibfield{author}{\bibinfo{person}{Cynthia Dwork} {and} \bibinfo{person}{Aaron Roth}.} \bibinfo{year}{2014}\natexlab{}.
\newblock \showarticletitle{The Algorithmic Foundations of Differential Privacy}.
\newblock \bibinfo{journal}{\emph{Found. Trends Theor. Comput. Sci.}} \bibinfo{volume}{9}, \bibinfo{number}{3-4} (\bibinfo{date}{Aug.} \bibinfo{year}{2014}), \bibinfo{pages}{211--407}.
\newblock
\showISSN{1551-305X}
\urldef\tempurl%
\url{https://doi.org/10.1561/0400000042}
\showDOI{\tempurl}


\bibitem[{European Parliament}(2016)]%
        {bibliography:eu-gdpr}
\bibfield{author}{\bibinfo{person}{{European Parliament}}.} \bibinfo{year}{2016}\natexlab{}.
\newblock \bibinfo{booktitle}{\emph{Regulation (EU) 2016/679 of the European Parliament and of the Council of 27 April 2016 on the protection of natural persons with regard to the processing of personal data and on the free movement of such data, and repealing Directive 95/46/EC (General Data Protection Regulation)}}.
\newblock {European Union}.
\newblock
\urldef\tempurl%
\url{https://eur-lex.europa.eu/legal-content/EN/TXT/PDF/?uri=CELEX:32016R0679}
\showURL{%
\tempurl}


\bibitem[Fang and Quan(2021)]%
        {bibliography:privacy-preserving-machine-learning-with-homomorphic-encryption-and-federated-learning}
\bibfield{author}{\bibinfo{person}{Haokun Fang} {and} \bibinfo{person}{Qian Quan}.} \bibinfo{year}{2021}\natexlab{}.
\newblock \showarticletitle{Privacy Preserving Machine Learning with Homomorphic Encryption and Federated Learning}.
\newblock \bibinfo{journal}{\emph{Future Internet}} \bibinfo{volume}{13}, \bibinfo{number}{4} (\bibinfo{year}{2021}), \bibinfo{pages}{94}.
\newblock
\urldef\tempurl%
\url{https://doi.org/10.3390/fi13040094}
\showDOI{\tempurl}


\bibitem[Finn et~al\mbox{.}(2017)]%
        {bibliography:model-agnostic-meta-learning-for-fast-adaptation-of-deep-networks}
\bibfield{author}{\bibinfo{person}{Chelsea Finn}, \bibinfo{person}{Pieter Abbeel}, {and} \bibinfo{person}{Sergey Levine}.} \bibinfo{year}{2017}\natexlab{}.
\newblock \showarticletitle{Model-Agnostic Meta-Learning for Fast Adaptation of Deep Networks}. In \bibinfo{booktitle}{\emph{Proceedings of the 34th International Conference on Machine Learning - Volume 70}} (Sydney, NSW, Australia) \emph{(\bibinfo{series}{ICML'17})}. \bibinfo{publisher}{JMLR.org}, \bibinfo{address}{1269 Law Street, San Diego, CA 92109}, \bibinfo{pages}{1126--1135}.
\newblock


\bibitem[Flanagan et~al\mbox{.}(2021)]%
        {bibliography:federated-multi-view-matrix-factorization-for-personalized-recommendations}
\bibfield{author}{\bibinfo{person}{Adrian Flanagan}, \bibinfo{person}{Were Oyomno}, \bibinfo{person}{Alexander Grigorievskiy}, \bibinfo{person}{Kuan~E. Tan}, \bibinfo{person}{Suleiman~A. Khan}, {and} \bibinfo{person}{Muhammad Ammad-Ud-Din}.} \bibinfo{year}{2021}\natexlab{}.
\newblock \showarticletitle{Federated Multi-view Matrix Factorization for Personalized Recommendations}. In \bibinfo{booktitle}{\emph{Machine Learning and Knowledge Discovery in Databases}}, \bibfield{editor}{\bibinfo{person}{Frank Hutter}, \bibinfo{person}{Kristian Kersting}, \bibinfo{person}{Jefrey Lijffijt}, {and} \bibinfo{person}{Isabel Valera}} (Eds.). \bibinfo{publisher}{Springer International Publishing}, \bibinfo{address}{Ghent, Belgium}, \bibinfo{pages}{324--347}.
\newblock
\showISBNx{978-3-030-67661-2}
\urldef\tempurl%
\url{https://doi.org/10.1007/978-3-030-67661-2_20}
\showDOI{\tempurl}


\bibitem[Fraboni et~al\mbox{.}(2023)]%
        {bibliography:a-general-theory-for-client-sampling-in-federated-learning}
\bibfield{author}{\bibinfo{person}{Yann Fraboni}, \bibinfo{person}{Richard Vidal}, \bibinfo{person}{Laetitia Kameni}, {and} \bibinfo{person}{Marco Lorenzi}.} \bibinfo{year}{2023}\natexlab{}.
\newblock \showarticletitle{A General Theory for Client Sampling in Federated Learning}. In \bibinfo{booktitle}{\emph{Trustworthy Federated Learning: First International Workshop, FL 2022, Held in Conjunction with IJCAI 2022, Vienna, Austria, July 23, 2022, Revised Selected Papers}} (Vienna, Austria). \bibinfo{publisher}{Springer-Verlag}, \bibinfo{address}{Berlin, Heidelberg}, \bibinfo{pages}{46–58}.
\newblock
\showISBNx{978-3-031-28995-8}
\urldef\tempurl%
\url{https://doi.org/10.1007/978-3-031-28996-5_4}
\showDOI{\tempurl}


\bibitem[Geiping et~al\mbox{.}(2020)]%
        {bibliography:inverting-gradients}
\bibfield{author}{\bibinfo{person}{Jonas Geiping}, \bibinfo{person}{Hartmut Bauermeister}, \bibinfo{person}{Hannah Dröge}, {and} \bibinfo{person}{Michael Moeller}.} \bibinfo{year}{2020}\natexlab{}.
\newblock \showarticletitle{Inverting Gradients - How Easy is It to Break Privacy in Federated Learning?}. In \bibinfo{booktitle}{\emph{Proceedings of the 34th International Conference on Neural Information Processing Systems}} (Vancouver, British Columbia, Canada) \emph{(\bibinfo{series}{NIPS'20})}. \bibinfo{publisher}{Curran Associates Inc.}, \bibinfo{address}{Red Hook, NY, USA}, Article \bibinfo{articleno}{1421}, \bibinfo{numpages}{11}~pages.
\newblock
\showISBNx{9781713829546}


\bibitem[{Gholami} et~al\mbox{.}(2022)]%
        {bibliography:a-survey-of-quantization-methods-for-efficient-neural-network-inference}
\bibfield{author}{\bibinfo{person}{Amir {Gholami}}, \bibinfo{person}{Sehoon {Kim}}, \bibinfo{person}{Zhen {Dong}}, \bibinfo{person}{Zhewei {Yao}}, \bibinfo{person}{Michael~W. {Mahoney}}, {and} \bibinfo{person}{Kurt {Keutzer}}.} \bibinfo{year}{2022}\natexlab{}.
\newblock \bibinfo{booktitle}{\emph{Low-Power Computer Vision} (\bibinfo{edition}{1st} ed.)}.
\newblock \bibinfo{publisher}{Chapman and Hall/CRC}, \bibinfo{address}{New York, United States of America}, Chapter A Survey of Quantization Methods for Efficient Neural Network Inference, \bibinfo{pages}{288--324}.
\newblock
\showISBNx{9781003162810}
\urldef\tempurl%
\url{https://doi.org/10.1201/9781003162810}
\showDOI{\tempurl}


\bibitem[Go et~al\mbox{.}(2009)]%
        {bibliography:twitter-sentiment-classification-using-distant-supervision}
\bibfield{author}{\bibinfo{person}{Alec Go}, \bibinfo{person}{Richa Bhayani}, {and} \bibinfo{person}{Lei Huang}.} \bibinfo{year}{2009}\natexlab{}.
\newblock \bibinfo{booktitle}{\emph{Twitter Sentiment Classification using Distant Supervision}}.
\newblock \bibinfo{type}{CS224N Project Report}. \bibinfo{institution}{Stanford}.
\newblock


\bibitem[Golbeck(2016)]%
        {bibliography:user-privacy-concerns-with-common-data-used-in-recommender-systems}
\bibfield{author}{\bibinfo{person}{Jennifer Golbeck}.} \bibinfo{year}{2016}\natexlab{}.
\newblock \showarticletitle{User Privacy Concerns with Common Data Used in Recommender Systems}. In \bibinfo{booktitle}{\emph{Social Informatics}}, \bibfield{editor}{\bibinfo{person}{Emma Spiro} {and} \bibinfo{person}{Yong-Yeol Ahn}} (Eds.). \bibinfo{publisher}{Springer International Publishing}, \bibinfo{address}{Cham}, \bibinfo{pages}{468--480}.
\newblock


\bibitem[Gomez-Uribe and Hunt(2016)]%
        {bibliography:the-netflix-recommender-system}
\bibfield{author}{\bibinfo{person}{Carlos~A. Gomez-Uribe} {and} \bibinfo{person}{Neil Hunt}.} \bibinfo{year}{2016}\natexlab{}.
\newblock \showarticletitle{The Netflix Recommender System: Algorithms, Business Value, and Innovation}.
\newblock \bibinfo{journal}{\emph{ACM Trans. Manage. Inf. Syst.}} \bibinfo{volume}{6}, \bibinfo{number}{4}, Article \bibinfo{articleno}{13} (\bibinfo{date}{Dec.} \bibinfo{year}{2016}), \bibinfo{numpages}{19}~pages.
\newblock
\showISSN{2158-656X}
\urldef\tempurl%
\url{https://doi.org/10.1145/2843948}
\showDOI{\tempurl}


\bibitem[Grbovic and Cheng(2018)]%
        {bibliography:real-time-personalization-using-embeddings-for-search-ranking-at-airbnb}
\bibfield{author}{\bibinfo{person}{Mihajlo Grbovic} {and} \bibinfo{person}{Haibin Cheng}.} \bibinfo{year}{2018}\natexlab{}.
\newblock \showarticletitle{Real-Time Personalization Using Embeddings for Search Ranking at Airbnb}. In \bibinfo{booktitle}{\emph{Proceedings of the 24th ACM SIGKDD International Conference on Knowledge Discovery \& Data Mining}} (London, United Kingdom) \emph{(\bibinfo{series}{KDD '18})}. \bibinfo{publisher}{Association for Computing Machinery}, \bibinfo{address}{New York, NY, USA}, \bibinfo{pages}{311--320}.
\newblock
\showISBNx{9781450355520}
\urldef\tempurl%
\url{https://doi.org/10.1145/3219819.3219885}
\showDOI{\tempurl}


\bibitem[Grother and Hanaoka(1995)]%
        {bibliography:nist-special-database-19-handprinted-forms-and-characters-database}
\bibfield{author}{\bibinfo{person}{Patrick~J. Grother} {and} \bibinfo{person}{Kayee~K. Hanaoka}.} \bibinfo{year}{1995}\natexlab{}.
\newblock \bibinfo{booktitle}{\emph{NIST special database 19 handprinted forms and characters database}}.
\newblock \bibinfo{type}{{T}echnical {R}eport}. \bibinfo{institution}{National Institute of Standards and Technology}.
\newblock
\urldef\tempurl%
\url{https://doi.org/10.18434/T4H01C}
\showDOI{\tempurl}


\bibitem[Haase et~al\mbox{.}(2021)]%
        {bibliography:encoder-optimization-for-the-nnr-standard}
\bibfield{author}{\bibinfo{person}{Paul Haase}, \bibinfo{person}{Daniel Becking}, \bibinfo{person}{Heiner Kirchhoffer}, \bibinfo{person}{Karsten Müller}, \bibinfo{person}{Heiko Schwarz}, \bibinfo{person}{Wojciech Samek}, \bibinfo{person}{Detlev Marpe}, {and} \bibinfo{person}{Thomas Wiegand}.} \bibinfo{year}{2021}\natexlab{}.
\newblock \showarticletitle{Encoder Optimizations For The NNR Standard On Neural Network Compression}. In \bibinfo{booktitle}{\emph{2021 IEEE International Conference on Image Processing (ICIP)}} (Anchorage, Alaska, USA). \bibinfo{publisher}{Institute of Electrical and Electronics Engineers ({IEEE})}, \bibinfo{address}{3 Park Avenue, 17th Floor, New York, NY 10016-5997 USA}, \bibinfo{pages}{3522--3526}.
\newblock
\urldef\tempurl%
\url{https://doi.org/10.1109/ICIP42928.2021.9506655}
\showDOI{\tempurl}


\bibitem[Han et~al\mbox{.}(2016)]%
        {bibliography:deep-compression}
\bibfield{author}{\bibinfo{person}{Song Han}, \bibinfo{person}{Huizi Mao}, {and} \bibinfo{person}{William~J. Dally}.} \bibinfo{year}{2016}\natexlab{}.
\newblock \showarticletitle{Deep Compression: Compressing Deep Neural Network with Pruning, Trained Quantization and Huffman Coding}. In \bibinfo{booktitle}{\emph{4th International Conference on Learning Representations, {ICLR}, May 2-4, 2016, Conference Track Proceedings}}, \bibfield{editor}{\bibinfo{person}{Yoshua Bengio} {and} \bibinfo{person}{Yann LeCun}} (Eds.). \bibinfo{publisher}{ICLR}, \bibinfo{address}{San Juan, Puerto Rico}.
\newblock
\urldef\tempurl%
\url{http://arxiv.org/abs/1510.00149}
\showURL{%
\tempurl}


\bibitem[{Hard} et~al\mbox{.}(2019)]%
        {bibliography:federated-learning-for-mobile-keyboard-prediction}
\bibfield{author}{\bibinfo{person}{Andrew {Hard}}, \bibinfo{person}{Kanishka {Rao}}, \bibinfo{person}{Rajiv {Mathews}}, \bibinfo{person}{Swaroop {Ramaswamy}}, \bibinfo{person}{Fran{\c{c}}oise {Beaufays}}, \bibinfo{person}{Sean {Augenstein}}, \bibinfo{person}{Hubert {Eichner}}, \bibinfo{person}{Chlo{\'e} {Kiddon}}, {and} \bibinfo{person}{Daniel {Ramage}}.} \bibinfo{year}{2019}\natexlab{}.
\newblock \showarticletitle{{Federated Learning for Mobile Keyboard Prediction}}.
\newblock \bibinfo{journal}{\emph{arXiv e-prints}}  \bibinfo{volume}{abs/1811.03604} (\bibinfo{date}{Feb.} \bibinfo{year}{2019}), \bibinfo{numpages}{7}~pages.
\newblock
\showeprint[arxiv]{1811.03604}~[cs.CL]


\bibitem[Harper and Konstan(2015)]%
        {bibliography:the-movielens-datasets-history-and-context}
\bibfield{author}{\bibinfo{person}{F.~Maxwell Harper} {and} \bibinfo{person}{Joseph~A. Konstan}.} \bibinfo{year}{2015}\natexlab{}.
\newblock \showarticletitle{The MovieLens Datasets: History and Context}.
\newblock \bibinfo{journal}{\emph{ACM Trans. Interact. Intell. Syst.}} \bibinfo{volume}{5}, \bibinfo{number}{4}, Article \bibinfo{articleno}{19} (\bibinfo{date}{Dec.} \bibinfo{year}{2015}), \bibinfo{numpages}{19}~pages.
\newblock
\showISSN{2160-6455}
\urldef\tempurl%
\url{https://doi.org/10.1145/2827872}
\showDOI{\tempurl}


\bibitem[{He} et~al\mbox{.}(2021)]%
        {bibliography:fedgraphnn}
\bibfield{author}{\bibinfo{person}{Chaoyang {He}}, \bibinfo{person}{Keshav {Balasubramanian}}, \bibinfo{person}{Emir {Ceyani}}, \bibinfo{person}{Carl {Yang}}, \bibinfo{person}{Han {Xie}}, \bibinfo{person}{Lichao {Sun}}, \bibinfo{person}{Lifang {He}}, \bibinfo{person}{Liangwei {Yang}}, \bibinfo{person}{Philip~S. {Yu}}, \bibinfo{person}{Yu {Rong}}, \bibinfo{person}{Peilin {Zhao}}, \bibinfo{person}{Junzhou {Huang}}, \bibinfo{person}{Murali {Annavaram}}, {and} \bibinfo{person}{Salman {Avestimehr}}.} \bibinfo{year}{2021}\natexlab{}.
\newblock \showarticletitle{{FedGraphNN: A Federated Learning System and Benchmark for Graph Neural Networks}}. In \bibinfo{booktitle}{\emph{9th International Conference on Learning Representations}}. \bibinfo{publisher}{OpenReview.net}, \bibinfo{address}{Virtual Only}, \bibinfo{numpages}{17}~pages.
\newblock


\bibitem[He et~al\mbox{.}(2017)]%
        {bibliography:neural-collaborative-filtering}
\bibfield{author}{\bibinfo{person}{Xiangnan He}, \bibinfo{person}{Lizi Liao}, \bibinfo{person}{Hanwang Zhang}, \bibinfo{person}{Liqiang Nie}, \bibinfo{person}{Xia Hu}, {and} \bibinfo{person}{Tat-Seng Chua}.} \bibinfo{year}{2017}\natexlab{}.
\newblock \showarticletitle{Neural Collaborative Filtering}. In \bibinfo{booktitle}{\emph{Proceedings of the 26th International Conference on World Wide Web}} (Perth, Australia) \emph{(\bibinfo{series}{WWW '17})}. \bibinfo{publisher}{International World Wide Web Conferences Steering Committee}, \bibinfo{address}{Republic and Canton of Geneva, CHE}, \bibinfo{pages}{173--182}.
\newblock
\showISBNx{9781450349130}
\urldef\tempurl%
\url{https://doi.org/10.1145/3038912.3052569}
\showDOI{\tempurl}


\bibitem[Hermann(2022)]%
        {bibliography:artificial-intelligence-and-mass-personalization-of-communication-content}
\bibfield{author}{\bibinfo{person}{Erik Hermann}.} \bibinfo{year}{2022}\natexlab{}.
\newblock \showarticletitle{Artificial intelligence and mass personalization of communication content—An ethical and literacy perspective}.
\newblock \bibinfo{journal}{\emph{New Media \& Society}} \bibinfo{volume}{24}, \bibinfo{number}{5} (\bibinfo{year}{2022}), \bibinfo{pages}{1258--1277}.
\newblock
\urldef\tempurl%
\url{https://doi.org/10.1177/14614448211022702}
\showDOI{\tempurl}
\showeprint{https://doi.org/10.1177/14614448211022702}


\bibitem[Hinton et~al\mbox{.}(2015)]%
        {bibliography:distilling-the-knowledge-in-a-neural-network}
\bibfield{author}{\bibinfo{person}{Geoffrey Hinton}, \bibinfo{person}{Oriol Vinyals}, {and} \bibinfo{person}{Jeffrey Dean}.} \bibinfo{year}{2015}\natexlab{}.
\newblock \showarticletitle{Distilling the Knowledge in a Neural Network}. In \bibinfo{booktitle}{\emph{NIPS Deep Learning and Representation Learning Workshop}}. \bibinfo{publisher}{Morgan-Kaufmann}, \bibinfo{address}{Montréal, Québec, Canada}.
\newblock
\urldef\tempurl%
\url{http://arxiv.org/abs/1503.02531}
\showURL{%
\tempurl}


\bibitem[Hochreiter and Schmidhuber(1997)]%
        {bibliography:long-short-term-memory}
\bibfield{author}{\bibinfo{person}{Sepp Hochreiter} {and} \bibinfo{person}{Jürgen Schmidhuber}.} \bibinfo{year}{1997}\natexlab{}.
\newblock \showarticletitle{Long Short-Term Memory}.
\newblock \bibinfo{journal}{\emph{Neural Comput.}} \bibinfo{volume}{9}, \bibinfo{number}{8} (\bibinfo{date}{Nov.} \bibinfo{year}{1997}), \bibinfo{pages}{1735--1780}.
\newblock
\showISSN{0899-7667}
\urldef\tempurl%
\url{https://doi.org/10.1162/neco.1997.9.8.1735}
\showDOI{\tempurl}


\bibitem[{Hu} et~al\mbox{.}(2022)]%
        {bibliography:fedcat}
\bibfield{author}{\bibinfo{person}{Ming {Hu}}, \bibinfo{person}{Tian {Liu}}, \bibinfo{person}{Zhiwei {Ling}}, \bibinfo{person}{Zhihao {Yue}}, {and} \bibinfo{person}{Mingsong {Chen}}.} \bibinfo{year}{2022}\natexlab{}.
\newblock \showarticletitle{{FedCAT: Towards Accurate Federated Learning via Device Concatenation}}.
\newblock \bibinfo{journal}{\emph{arXiv e-prints}}  \bibinfo{volume}{abs/2202.12751} (\bibinfo{date}{Feb.} \bibinfo{year}{2022}), \bibinfo{numpages}{12}~pages.
\newblock
\showeprint[arxiv]{2202.12751}~[cs.LG]


\bibitem[{International Organization for Standardization (ISO)}(2022)]%
        {bibliography:iso-standard-compression-of-neural-networks-for-multimedia-content-description-and-analysis}
\bibfield{author}{\bibinfo{person}{{International Organization for Standardization (ISO)}}.} \bibinfo{year}{2022}\natexlab{}.
\newblock \bibinfo{booktitle}{\emph{Information technology - Multimedia content description interface — Part 17: Compression of neural networks for multimedia content description and analysis}}.
\newblock \bibinfo{type}{Standard}. \bibinfo{institution}{{International Organization for Standardization (ISO)}}, \bibinfo{address}{Geneva, Switzerland}.
\newblock


\bibitem[Ioffe and Szegedy(2015)]%
        {bibliography:batch-normalization-accelerating-deep-network-training-by-reducing-internal-covariate-shift}
\bibfield{author}{\bibinfo{person}{Sergey Ioffe} {and} \bibinfo{person}{Christian Szegedy}.} \bibinfo{year}{2015}\natexlab{}.
\newblock \showarticletitle{Batch Normalization: Accelerating Deep Network Training by Reducing Internal Covariate Shift}. In \bibinfo{booktitle}{\emph{Proceedings of the 32nd International Conference on International Conference on Machine Learning - Volume 37}} (Lille, France) \emph{(\bibinfo{series}{ICML'15})}. \bibinfo{publisher}{JMLR.org}, \bibinfo{address}{1269 Law Street, San Diego, CA 92109}, \bibinfo{pages}{448–456}.
\newblock


\bibitem[{Jeong} et~al\mbox{.}(2023)]%
        {bibliography:communication-efficient-on-device-machine-learning}
\bibfield{author}{\bibinfo{person}{Eunjeong {Jeong}}, \bibinfo{person}{Seungeun {Oh}}, \bibinfo{person}{Hyesung {Kim}}, \bibinfo{person}{Jihong {Park}}, \bibinfo{person}{Mehdi {Bennis}}, {and} \bibinfo{person}{Seong-Lyun {Kim}}.} \bibinfo{year}{2023}\natexlab{}.
\newblock \showarticletitle{{Communication-Efficient On-Device Machine Learning: Federated Distillation and Augmentation under Non-IID Private Data}}.
\newblock \bibinfo{journal}{\emph{arXiv e-prints}}  \bibinfo{volume}{abs/1811.11479} (\bibinfo{date}{Oct.} \bibinfo{year}{2023}), \bibinfo{numpages}{6}~pages.
\newblock
\showeprint[arxiv]{1811.11479}~[cs.LG]


\bibitem[Jia and Lei(2021)]%
        {bibliography:personalized-recommendation-algorithm-for-mobile-based-on-federated-matrix-factorization}
\bibfield{author}{\bibinfo{person}{Junjie Jia} {and} \bibinfo{person}{Zhipeng Lei}.} \bibinfo{year}{2021}\natexlab{}.
\newblock \showarticletitle{Personalized Recommendation Algorithm for Mobile Based on Federated Matrix Factorization}.
\newblock \bibinfo{journal}{\emph{Journal of Physics: Conference Series}} \bibinfo{volume}{1802}, \bibinfo{number}{3} (\bibinfo{date}{March} \bibinfo{year}{2021}), \bibinfo{pages}{032021}.
\newblock
\urldef\tempurl%
\url{https://doi.org/10.1088/1742-6596/1802/3/032021}
\showDOI{\tempurl}


\bibitem[Jie et~al\mbox{.}(2022)]%
        {bibliography:personalized-federated-recommendation-system-with-historical-parameter-clustering}
\bibfield{author}{\bibinfo{person}{Zhiyong Jie}, \bibinfo{person}{Shuhong Chen}, \bibinfo{person}{Junqiu Lai}, \bibinfo{person}{Muhammad Arif}, {and} \bibinfo{person}{Zongyuan He}.} \bibinfo{year}{2022}\natexlab{}.
\newblock \showarticletitle{Personalized federated recommendation system with historical parameter clustering}.
\newblock \bibinfo{journal}{\emph{Journal of Ambient Intelligence and Humanized Computing}} \bibinfo{volume}{14}, \bibinfo{number}{8} (\bibinfo{date}{02} \bibinfo{year}{2022}), \bibinfo{pages}{10555--10565}.
\newblock
\urldef\tempurl%
\url{https://doi.org/10.1007/s12652-022-03709-z}
\showDOI{\tempurl}


\bibitem[Kamp et~al\mbox{.}(2023)]%
        {bibliography:federated-learning-from-small-datasets}
\bibfield{author}{\bibinfo{person}{Michael Kamp}, \bibinfo{person}{Jonas Fischer}, {and} \bibinfo{person}{Jilles Vreeken}.} \bibinfo{year}{2023}\natexlab{}.
\newblock \showarticletitle{Federated Learning from Small Datasets}. In \bibinfo{booktitle}{\emph{The Eleventh International Conference on Learning Representations}}. \bibinfo{publisher}{OpenReview.net}, \bibinfo{address}{Kigali, Rwanda}, \bibinfo{numpages}{13}~pages.
\newblock
\urldef\tempurl%
\url{https://openreview.net/forum?id=hDDV1lsRV8}
\showURL{%
\tempurl}


\bibitem[Karimireddy et~al\mbox{.}(2020)]%
        {bilbiography:stochastic-controlled-averaging-for-on-device-federated-learning}
\bibfield{author}{\bibinfo{person}{Sai~Praneeth Karimireddy}, \bibinfo{person}{Satyen Kale}, \bibinfo{person}{Mehryar Mohri}, \bibinfo{person}{Sashank Reddi}, \bibinfo{person}{Sebastian Stich}, {and} \bibinfo{person}{Ananda~Theertha Suresh}.} \bibinfo{year}{2020}\natexlab{}.
\newblock \showarticletitle{{SCAFFOLD}: Stochastic Controlled Averaging for Federated Learning}. In \bibinfo{booktitle}{\emph{Proceedings of the 37th International Conference on Machine Learning}} \emph{(\bibinfo{series}{Proceedings of Machine Learning Research}, Vol.~\bibinfo{volume}{119})}, \bibfield{editor}{\bibinfo{person}{Hal~Daumé III} {and} \bibinfo{person}{Aarti Singh}} (Eds.). \bibinfo{publisher}{PMLR}, \bibinfo{address}{virtual}, \bibinfo{pages}{5132--5143}.
\newblock
\urldef\tempurl%
\url{https://proceedings.mlr.press/v119/karimireddy20a.html}
\showURL{%
\tempurl}


\bibitem[Kiefer and Wolfowitz(1952)]%
        {bibliography:stochastic-estimation-of-the-maximum-of-a-regression-function}
\bibfield{author}{\bibinfo{person}{J. Kiefer} {and} \bibinfo{person}{J. Wolfowitz}.} \bibinfo{year}{1952}\natexlab{}.
\newblock \showarticletitle{Stochastic Estimation of the Maximum of a Regression Function}.
\newblock \bibinfo{journal}{\emph{The Annals of Mathematical Statistics}} \bibinfo{volume}{23}, \bibinfo{number}{3} (\bibinfo{year}{1952}), \bibinfo{pages}{462--466}.
\newblock
\showISSN{00034851}
\urldef\tempurl%
\url{http://www.jstor.org/stable/2236690}
\showURL{%
\tempurl}


\bibitem[Kim et~al\mbox{.}(2018)]%
        {bibliography:efficient-privacy-preserving-matrix-factorization-for-recommendation-via-fully-homomorphic-encryption}
\bibfield{author}{\bibinfo{person}{Jinsu Kim}, \bibinfo{person}{Dongyoung Koo}, \bibinfo{person}{Yuna Kim}, \bibinfo{person}{Hyunsoo Yoon}, \bibinfo{person}{Junbum Shin}, {and} \bibinfo{person}{Sungwook Kim}.} \bibinfo{year}{2018}\natexlab{}.
\newblock \showarticletitle{Efficient Privacy-Preserving Matrix Factorization for Recommendation via Fully Homomorphic Encryption}.
\newblock \bibinfo{journal}{\emph{ACM Trans. Priv. Secur.}} \bibinfo{volume}{21}, \bibinfo{number}{4}, Article \bibinfo{articleno}{17} (\bibinfo{date}{jun} \bibinfo{year}{2018}), \bibinfo{numpages}{30}~pages.
\newblock
\showISSN{2471-2566}
\urldef\tempurl%
\url{https://doi.org/10.1145/3212509}
\showDOI{\tempurl}


\bibitem[Kingma and Ba(2015)]%
        {bibliography:adam-a-method-for-stochastic-optimization}
\bibfield{author}{\bibinfo{person}{Diederik~P. Kingma} {and} \bibinfo{person}{Jimmy Ba}.} \bibinfo{year}{2015}\natexlab{}.
\newblock \showarticletitle{Adam: {A} Method for Stochastic Optimization}. In \bibinfo{booktitle}{\emph{3rd International Conference on Learning Representations, {ICLR} 2015, San Diego, CA, USA, May 7-9, 2015, Conference Track Proceedings}}, \bibfield{editor}{\bibinfo{person}{Yoshua Bengio} {and} \bibinfo{person}{Yann LeCun}} (Eds.). \bibinfo{publisher}{International Conference on Learning Representations}, \bibinfo{address}{2710 E Corridor Drive, Appleton, WI 54913}.
\newblock
\urldef\tempurl%
\url{http://arxiv.org/abs/1412.6980}
\showURL{%
\tempurl}


\bibitem[Kirchhoffer et~al\mbox{.}(2022)]%
        {bibliography:overview-of-the-neural-network-compression-and-representation-standard}
\bibfield{author}{\bibinfo{person}{Heiner Kirchhoffer}, \bibinfo{person}{Paul Haase}, \bibinfo{person}{Wojciech Samek}, \bibinfo{person}{Karsten Müller}, \bibinfo{person}{Hamed Rezazadegan-Tavakoli}, \bibinfo{person}{Francesco Cricri}, \bibinfo{person}{Emre~B. Aksu}, \bibinfo{person}{Miska~M. Hannuksela}, \bibinfo{person}{Wei Jiang}, \bibinfo{person}{Wei Wang}, \bibinfo{person}{Shan Liu}, \bibinfo{person}{Swayambhoo Jain}, \bibinfo{person}{Shahab Hamidi-Rad}, \bibinfo{person}{Fabien Racapé}, {and} \bibinfo{person}{Werner Bailer}.} \bibinfo{year}{2022}\natexlab{}.
\newblock \showarticletitle{Overview of the Neural Network Compression and Representation (NNR) Standard}.
\newblock \bibinfo{journal}{\emph{IEEE Transactions on Circuits and Systems for Video Technology}} \bibinfo{volume}{32}, \bibinfo{number}{5} (\bibinfo{year}{2022}), \bibinfo{pages}{3203--3216}.
\newblock
\urldef\tempurl%
\url{https://doi.org/10.1109/TCSVT.2021.3095970}
\showDOI{\tempurl}


\bibitem[Konečný et~al\mbox{.}(2016)]%
        {bibliography:federated-optimization}
\bibfield{author}{\bibinfo{person}{Jakub Konečný}, \bibinfo{person}{Hugh~Brendan McMahan}, \bibinfo{person}{Daniel Ramage}, {and} \bibinfo{person}{Peter Richtárik}.} \bibinfo{year}{2016}\natexlab{}.
\newblock \showarticletitle{Federated Optimization: Distributed Machine Learning for On-Device Intelligence}.
\newblock \bibinfo{journal}{\emph{CoRR}}  \bibinfo{volume}{abs/1610.02527} (\bibinfo{date}{Oct.} \bibinfo{year}{2016}), \bibinfo{numpages}{38}~pages.
\newblock
\showeprint[arxiv]{1610.02527}
\urldef\tempurl%
\url{http://arxiv.org/abs/1610.02527}
\showURL{%
\tempurl}


\bibitem[Konečný et~al\mbox{.}(2018)]%
        {bibliography:federated-learning-strategies-for-improving-communication-efficiency}
\bibfield{author}{\bibinfo{person}{Jakub Konečný}, \bibinfo{person}{H.~Brendan McMahan}, \bibinfo{person}{Felix~X. Yu}, \bibinfo{person}{Ananda~Theertha Suresh}, \bibinfo{person}{Dave Bacon}, {and} \bibinfo{person}{Peter Richtárik}.} \bibinfo{year}{2018}\natexlab{}.
\newblock \showarticletitle{Federated Learning: Strategies for Improving Communication Efficiency}. In \bibinfo{booktitle}{\emph{6th International Conference on Learning Representations}}. \bibinfo{publisher}{OpenReview.net}, \bibinfo{address}{Vancouver, British Columbia, Canada}, \bibinfo{numpages}{10}~pages.
\newblock
\urldef\tempurl%
\url{https://openreview.net/forum?id=B1EPYJ-C-}
\showURL{%
\tempurl}


\bibitem[Koren(2008)]%
        {bibliography:factorization-meets-the-neighborhood}
\bibfield{author}{\bibinfo{person}{Yehuda Koren}.} \bibinfo{year}{2008}\natexlab{}.
\newblock \showarticletitle{Factorization Meets the Neighborhood: A Multifaceted Collaborative Filtering Model}. In \bibinfo{booktitle}{\emph{Proceedings of the 14th ACM SIGKDD International Conference on Knowledge Discovery and Data Mining}} (Las Vegas, Nevada, USA) \emph{(\bibinfo{series}{KDD '08})}. \bibinfo{publisher}{Association for Computing Machinery}, \bibinfo{address}{New York, NY, USA}, \bibinfo{pages}{426--434}.
\newblock
\showISBNx{9781605581934}
\urldef\tempurl%
\url{https://doi.org/10.1145/1401890.1401944}
\showDOI{\tempurl}


\bibitem[Koren et~al\mbox{.}(2009)]%
        {bibliography:matrix-factorization-techniques-for-recommender-systems}
\bibfield{author}{\bibinfo{person}{Yehuda Koren}, \bibinfo{person}{Robert Bell}, {and} \bibinfo{person}{Chris Volinsky}.} \bibinfo{year}{2009}\natexlab{}.
\newblock \showarticletitle{Matrix Factorization Techniques for Recommender Systems}.
\newblock \bibinfo{journal}{\emph{Computer}} \bibinfo{volume}{42}, \bibinfo{number}{8} (\bibinfo{date}{Aug.} \bibinfo{year}{2009}), \bibinfo{pages}{30--37}.
\newblock
\showISSN{1558-0814}
\urldef\tempurl%
\url{https://doi.org/10.1109/MC.2009.263}
\showDOI{\tempurl}


\bibitem[Kozyreva et~al\mbox{.}(2021)]%
        {bilbiography:public-attitudes-towards-algorithmic-personalization-and-use-of-personal-data-online}
\bibfield{author}{\bibinfo{person}{Anastasia Kozyreva}, \bibinfo{person}{Philipp Lorenz-Spreen}, \bibinfo{person}{Ralph Hertwig}, \bibinfo{person}{Stephan Lewandowsky}, {and} \bibinfo{person}{Stefan~M Herzog}.} \bibinfo{year}{2021}\natexlab{}.
\newblock \showarticletitle{Public attitudes towards algorithmic personalization and use of personal data online: Evidence from Germany, Great Britain, and the United States}.
\newblock \bibinfo{journal}{\emph{Humanities and Social Sciences Communications}} \bibinfo{volume}{8}, \bibinfo{number}{1} (\bibinfo{year}{2021}), \bibinfo{pages}{1--11}.
\newblock


\bibitem[Lam et~al\mbox{.}(2006)]%
        {bibliography:do-you-trust-your-recommendations}
\bibfield{author}{\bibinfo{person}{Shyong K.~``Tony'' Lam}, \bibinfo{person}{Dan Frankowski}, {and} \bibinfo{person}{John Riedl}.} \bibinfo{year}{2006}\natexlab{}.
\newblock \showarticletitle{Do You Trust Your Recommendations? An Exploration of Security and Privacy Issues in Recommender Systems}. In \bibinfo{booktitle}{\emph{Emerging Trends in Information and Communication Security}}, \bibfield{editor}{\bibinfo{person}{Günter Müller}} (Ed.). \bibinfo{publisher}{Springer Berlin Heidelberg}, \bibinfo{address}{Berlin, Heidelberg}, \bibinfo{pages}{14--29}.
\newblock
\showISBNx{978-3-540-34642-5}


\bibitem[Lang and Shlezinger(2022)]%
        {bibliography:joint-privacy-enhancement-and-quantization-in-federated-learning}
\bibfield{author}{\bibinfo{person}{Natalie Lang} {and} \bibinfo{person}{Nir Shlezinger}.} \bibinfo{year}{2022}\natexlab{}.
\newblock \showarticletitle{Joint Privacy Enhancement and Quantization in Federated Learning}. In \bibinfo{booktitle}{\emph{2022 IEEE International Symposium on Information Theory (ISIT)}} (Aalto University, Espoo, Finland). \bibinfo{publisher}{Institute of Electrical and Electronics Engineers ({IEEE})}, \bibinfo{address}{3 Park Avenue, 17th Floor, New York, NY 10016-5997 USA}, \bibinfo{pages}{2040--2045}.
\newblock
\showISSN{2157-8117}
\urldef\tempurl%
\url{https://doi.org/10.1109/ISIT50566.2022.9834551}
\showDOI{\tempurl}


\bibitem[Lecun et~al\mbox{.}(1998)]%
        {bibliography:gradient-based-learning-applied-to-document-recognition}
\bibfield{author}{\bibinfo{person}{Y. Lecun}, \bibinfo{person}{L. Bottou}, \bibinfo{person}{Y. Bengio}, {and} \bibinfo{person}{P. Haffner}.} \bibinfo{year}{1998}\natexlab{}.
\newblock \showarticletitle{Gradient-based learning applied to document recognition}.
\newblock \bibinfo{journal}{\emph{Proc. IEEE}} \bibinfo{volume}{86}, \bibinfo{number}{11} (\bibinfo{year}{1998}), \bibinfo{pages}{2278--2324}.
\newblock
\urldef\tempurl%
\url{https://doi.org/10.1109/5.726791}
\showDOI{\tempurl}


\bibitem[LeCun et~al\mbox{.}(1990)]%
        {bibliography:optimal-brain-damage}
\bibfield{author}{\bibinfo{person}{Yann LeCun}, \bibinfo{person}{John Denker}, {and} \bibinfo{person}{Sara Solla}.} \bibinfo{year}{1990}\natexlab{}.
\newblock \showarticletitle{Optimal Brain Damage}. In \bibinfo{booktitle}{\emph{Advances in Neural Information Processing Systems}}, \bibfield{editor}{\bibinfo{person}{D.~Touretzky}} (Ed.), Vol.~\bibinfo{volume}{2}. \bibinfo{publisher}{Morgan-Kaufmann}, \bibinfo{address}{Denver, Colorado, USA}.
\newblock
\urldef\tempurl%
\url{https://proceedings.neurips.cc/paper/1989/file/6c9882bbac1c7093bd25041881277658-Paper.pdf}
\showURL{%
\tempurl}


\bibitem[Leroy et~al\mbox{.}(2019)]%
        {bibliography:federated-learning-for-keyword-spotting}
\bibfield{author}{\bibinfo{person}{David Leroy}, \bibinfo{person}{Alice Coucke}, \bibinfo{person}{Thibaut Lavril}, \bibinfo{person}{Thibault Gisselbrecht}, {and} \bibinfo{person}{Joseph Dureau}.} \bibinfo{year}{2019}\natexlab{}.
\newblock \showarticletitle{Federated Learning for Keyword Spotting}. In \bibinfo{booktitle}{\emph{ICASSP 2019 - 2019 IEEE International Conference on Acoustics, Speech and Signal Processing (ICASSP)}} (Brighton, United Kingdom). \bibinfo{publisher}{Institute of Electrical and Electronics Engineers ({IEEE})}, \bibinfo{address}{3 Park Avenue, 17th Floor, New York, NY 10016-5997 USA}, \bibinfo{pages}{6341--6345}.
\newblock
\urldef\tempurl%
\url{https://doi.org/10.1109/ICASSP.2019.8683546}
\showDOI{\tempurl}


\bibitem[Li et~al\mbox{.}(2019)]%
        {bibliography:federated-optimization-in-heterogeneous-networks}
\bibfield{author}{\bibinfo{person}{Tian Li}, \bibinfo{person}{Anit~Kumar Sahu}, \bibinfo{person}{Manzil Zaheer}, \bibinfo{person}{Maziar Sanjabi}, \bibinfo{person}{Ameet Talwalkar}, {and} \bibinfo{person}{Virginia Smith}.} \bibinfo{year}{2019}\natexlab{}.
\newblock \showarticletitle{Federated Optimization for Heterogeneous Networks}. In \bibinfo{booktitle}{\emph{ICML Workshop on Adaptive {\&} Multitask Learning: Algorithms {\&} Systems}}. \bibinfo{publisher}{OpenReview.net}, \bibinfo{address}{Long Beach, California, United States of America}, \bibinfo{numpages}{16}~pages.
\newblock
\urldef\tempurl%
\url{https://openreview.net/forum?id=SkgwE5Ss3N}
\showURL{%
\tempurl}


\bibitem[Li et~al\mbox{.}(2020b)]%
        {bibliography:fair-ressource-allocation-in-federated-learning}
\bibfield{author}{\bibinfo{person}{Tian Li}, \bibinfo{person}{Maziar Sanjabi}, \bibinfo{person}{Ahmad Beirami}, {and} \bibinfo{person}{Virginia Smith}.} \bibinfo{year}{2020}\natexlab{b}.
\newblock \showarticletitle{Fair Resource Allocation in Federated Learning}. In \bibinfo{booktitle}{\emph{8th International Conference on Learning Representations, {ICLR} 2020, Addis Ababa, Ethiopia, April 26-30, 2020}}. \bibinfo{publisher}{OpenReview.net}, \bibinfo{address}{Addis Ababa, Ethiopia}.
\newblock
\urldef\tempurl%
\url{https://openreview.net/forum?id=ByexElSYDr}
\showURL{%
\tempurl}


\bibitem[Li et~al\mbox{.}(2020a)]%
        {bibliography:on-the-convergence-of-fedavg-on-non-iid-data}
\bibfield{author}{\bibinfo{person}{Xiang Li}, \bibinfo{person}{Kaixuan Huang}, \bibinfo{person}{Wenhao Yang}, \bibinfo{person}{Shusen Wang}, {and} \bibinfo{person}{Zhihua Zhang}.} \bibinfo{year}{2020}\natexlab{a}.
\newblock \showarticletitle{On the Convergence of FedAvg on Non-IID Data}. In \bibinfo{booktitle}{\emph{International Conference on Learning Representations}}. \bibinfo{publisher}{OpenReview.net}, \bibinfo{address}{Addis Ababa, Ethiopia}, \bibinfo{numpages}{26}~pages.
\newblock
\urldef\tempurl%
\url{https://openreview.net/forum?id=HJxNAnVtDS}
\showURL{%
\tempurl}


\bibitem[Li et~al\mbox{.}(2021)]%
        {bilbiography:federated-learning-on-non-iid-features-via-local-batch-normalization}
\bibfield{author}{\bibinfo{person}{Xiaoxiao Li}, \bibinfo{person}{Meirui JIANG}, \bibinfo{person}{Xiaofei Zhang}, \bibinfo{person}{Michael Kamp}, {and} \bibinfo{person}{Qi Dou}.} \bibinfo{year}{2021}\natexlab{}.
\newblock \showarticletitle{Fed{BN}: Federated Learning on Non-{IID} Features via Local Batch Normalization}. In \bibinfo{booktitle}{\emph{International Conference on Learning Representations (ICLR) 2021}}. \bibinfo{publisher}{OpenReview.net}, \bibinfo{address}{Vienna, Austria}, \bibinfo{numpages}{27}~pages.
\newblock
\urldef\tempurl%
\url{https://openreview.net/forum?id=6YEQUn0QICG}
\showURL{%
\tempurl}


\bibitem[Liang et~al\mbox{.}(2021)]%
        {bibliography:lossless-federated-recommendation-with-explicit-feedback}
\bibfield{author}{\bibinfo{person}{Feng Liang}, \bibinfo{person}{Weike Pan}, {and} \bibinfo{person}{Zhong Ming}.} \bibinfo{year}{2021}\natexlab{}.
\newblock \showarticletitle{Fedrec++: Lossless federated recommendation with explicit feedback}. In \bibinfo{booktitle}{\emph{Proceedings of the AAAI conference on artificial intelligence}}, Vol.~\bibinfo{volume}{35}. \bibinfo{publisher}{AAAI Press}, \bibinfo{address}{Washington, DC, USA}, \bibinfo{pages}{4224--4231}.
\newblock


\bibitem[Lin et~al\mbox{.}(2022)]%
        {bibliography:fednlp}
\bibfield{author}{\bibinfo{person}{Bill~Yuchen Lin}, \bibinfo{person}{Chaoyang He}, \bibinfo{person}{Zihang Ze}, \bibinfo{person}{Hulin Wang}, \bibinfo{person}{Yufen Hua}, \bibinfo{person}{Christophe Dupuy}, \bibinfo{person}{Rahul Gupta}, \bibinfo{person}{Mahdi Soltanolkotabi}, \bibinfo{person}{Xiang Ren}, {and} \bibinfo{person}{Salman Avestimehr}.} \bibinfo{year}{2022}\natexlab{}.
\newblock \showarticletitle{{F}ed{NLP}: Benchmarking Federated Learning Methods for Natural Language Processing Tasks}. In \bibinfo{booktitle}{\emph{Findings of the Association for Computational Linguistics: NAACL 2022}}, \bibfield{editor}{\bibinfo{person}{Marine Carpuat}, \bibinfo{person}{Marie-Catherine de~Marneffe}, {and} \bibinfo{person}{Ivan Vladimir~Meza Ruiz}} (Eds.). \bibinfo{publisher}{Association for Computational Linguistics}, \bibinfo{address}{Seattle, United States of America}, \bibinfo{pages}{157--175}.
\newblock
\urldef\tempurl%
\url{https://doi.org/10.18653/v1/2022.findings-naacl.13}
\showDOI{\tempurl}


\bibitem[Lin et~al\mbox{.}(2016)]%
        {bibliography:fixed-point-quantization-of-deep-convolutional-networks}
\bibfield{author}{\bibinfo{person}{Darryl~D. Lin}, \bibinfo{person}{Sachin~S. Talathi}, {and} \bibinfo{person}{V.~Sreekanth Annapureddy}.} \bibinfo{year}{2016}\natexlab{}.
\newblock \showarticletitle{Fixed Point Quantization of Deep Convolutional Networks}. In \bibinfo{booktitle}{\emph{Proceedings of the 33rd International Conference on International Conference on Machine Learning - Volume 48}} (New York, NY, USA) \emph{(\bibinfo{series}{ICML'16})}. \bibinfo{publisher}{JMLR.org}, \bibinfo{address}{1269 Law Street, San Diego, CA 92109}, \bibinfo{pages}{2849--2858}.
\newblock


\bibitem[Lin et~al\mbox{.}(2021a)]%
        {bibliography:fedrec-federated-recommendation-with-explicit-feedback}
\bibfield{author}{\bibinfo{person}{Guanyu Lin}, \bibinfo{person}{Feng Liang}, \bibinfo{person}{Weike Pan}, {and} \bibinfo{person}{Zhong Ming}.} \bibinfo{year}{2021}\natexlab{a}.
\newblock \showarticletitle{FedRec: Federated Recommendation With Explicit Feedback}.
\newblock \bibinfo{journal}{\emph{IEEE Intelligent Systems}} \bibinfo{volume}{36}, \bibinfo{number}{5} (\bibinfo{date}{Sept.} \bibinfo{year}{2021}), \bibinfo{pages}{21--30}.
\newblock
\showISSN{1941-1294}
\urldef\tempurl%
\url{https://doi.org/10.1109/MIS.2020.3017205}
\showDOI{\tempurl}


\bibitem[Lin et~al\mbox{.}(2021b)]%
        {bibliography:federated-recommendation-via-fake-marks-and-secret-sharing}
\bibfield{author}{\bibinfo{person}{Zhaohao Lin}, \bibinfo{person}{Weike Pan}, {and} \bibinfo{person}{Zhong Ming}.} \bibinfo{year}{2021}\natexlab{b}.
\newblock \showarticletitle{FR-FMSS: Federated Recommendation via Fake Marks and Secret Sharing}. In \bibinfo{booktitle}{\emph{Proceedings of the 15th ACM Conference on Recommender Systems}} (Amsterdam, Netherlands) \emph{(\bibinfo{series}{RecSys '21})}. \bibinfo{publisher}{Association for Computing Machinery}, \bibinfo{address}{New York, NY, USA}, \bibinfo{pages}{668--673}.
\newblock
\showISBNx{9781450384582}
\urldef\tempurl%
\url{https://doi.org/10.1145/3460231.3478855}
\showDOI{\tempurl}


\bibitem[Liu(2009)]%
        {bibliography:learning-to-rank-for-information-retrieval}
\bibfield{author}{\bibinfo{person}{Tie-Yan Liu}.} \bibinfo{year}{2009}\natexlab{}.
\newblock \showarticletitle{Learning to Rank for Information Retrieval}.
\newblock \bibinfo{journal}{\emph{Foundations and Trends in Information Retrieval}} \bibinfo{volume}{3}, \bibinfo{number}{3} (\bibinfo{date}{March} \bibinfo{year}{2009}), \bibinfo{pages}{225--331}.
\newblock
\showISSN{1554-0669}
\urldef\tempurl%
\url{https://doi.org/10.1561/1500000016}
\showDOI{\tempurl}


\bibitem[Liu et~al\mbox{.}(2020)]%
        {bibliography:a-secure-federated-transfer-learning-framework}
\bibfield{author}{\bibinfo{person}{Yang Liu}, \bibinfo{person}{Yan Kang}, \bibinfo{person}{Chaoping Xing}, \bibinfo{person}{Tianjian Chen}, {and} \bibinfo{person}{Qiang Yang}.} \bibinfo{year}{2020}\natexlab{}.
\newblock \showarticletitle{A Secure Federated Transfer Learning Framework}.
\newblock \bibinfo{journal}{\emph{IEEE Intelligent Systems}} \bibinfo{volume}{35}, \bibinfo{number}{4} (\bibinfo{year}{2020}), \bibinfo{pages}{70--82}.
\newblock
\urldef\tempurl%
\url{https://doi.org/10.1109/MIS.2020.2988525}
\showDOI{\tempurl}


\bibitem[Liu et~al\mbox{.}(2015)]%
        {bibliography:deeplearning-face-attributes-in-the-wild}
\bibfield{author}{\bibinfo{person}{Ziwei Liu}, \bibinfo{person}{Ping Luo}, \bibinfo{person}{Xiaogang Wang}, {and} \bibinfo{person}{Xiaoou Tang}.} \bibinfo{year}{2015}\natexlab{}.
\newblock \showarticletitle{Deep Learning Face Attributes in the Wild}. In \bibinfo{booktitle}{\emph{2015 IEEE International Conference on Computer Vision (ICCV)}} (Santiago, Chile). \bibinfo{publisher}{Institute of Electrical and Electronics Engineers ({IEEE})}, \bibinfo{address}{3 Park Avenue, 17th Floor, New York, NY 10016-5997 USA}, \bibinfo{pages}{3730--3738}.
\newblock
\urldef\tempurl%
\url{https://doi.org/10.1109/ICCV.2015.425}
\showDOI{\tempurl}


\bibitem[Lloyd(1982)]%
        {bibliography:least-squares-quantization-in-pcm}
\bibfield{author}{\bibinfo{person}{S. Lloyd}.} \bibinfo{year}{1982}\natexlab{}.
\newblock \showarticletitle{Least squares quantization in PCM}.
\newblock \bibinfo{journal}{\emph{IEEE Transactions on Information Theory}} \bibinfo{volume}{28}, \bibinfo{number}{2} (\bibinfo{year}{1982}), \bibinfo{pages}{129--137}.
\newblock
\urldef\tempurl%
\url{https://doi.org/10.1109/TIT.1982.1056489}
\showDOI{\tempurl}


\bibitem[{Luo} et~al\mbox{.}(2021)]%
        {bibliography:real-world-image-datasets-for-federated-learning}
\bibfield{author}{\bibinfo{person}{Jiahuan {Luo}}, \bibinfo{person}{Xueyang {Wu}}, \bibinfo{person}{Yun {Luo}}, \bibinfo{person}{Anbu {Huang}}, \bibinfo{person}{Yunfeng {Huang}}, \bibinfo{person}{Yang {Liu}}, {and} \bibinfo{person}{Qiang {Yang}}.} \bibinfo{year}{2021}\natexlab{}.
\newblock \showarticletitle{{Real-World Image Datasets for Federated Learning}}.
\newblock \bibinfo{journal}{\emph{arXiv e-prints}}  \bibinfo{volume}{abs/1910.11089} (\bibinfo{date}{Jan.} \bibinfo{year}{2021}), \bibinfo{numpages}{8}~pages.
\newblock
\showeprint[arxiv]{1910.11089}~[cs.CV]


\bibitem[MacKenzie et~al\mbox{.}(2013)]%
        {bibliography:how-retailers-can-keep-up-with-customers}
\bibfield{author}{\bibinfo{person}{Ian MacKenzie}, \bibinfo{person}{Chris Meyer}, {and} \bibinfo{person}{Steve Noble}.} \bibinfo{year}{2013}\natexlab{}.
\newblock \bibinfo{booktitle}{\emph{How retailers can keep up with consumers}}.
\newblock McKinsey \& Company.
\newblock
\urldef\tempurl%
\url{https://www.mckinsey.com/industries/retail/our-insights/how-retailers-can-keep-up-with-consumers}
\showURL{%
\tempurl}


\bibitem[McMahan et~al\mbox{.}(2017)]%
        {bibliography:communication-efficient-learning-of-deep-networks-from-decentralized-data}
\bibfield{author}{\bibinfo{person}{Hugh~Brendan McMahan}, \bibinfo{person}{Eider Moore}, \bibinfo{person}{Daniel Ramage}, \bibinfo{person}{Seth Hampson}, {and} \bibinfo{person}{Blaise~Agüera y Arcas}.} \bibinfo{year}{2017}\natexlab{}.
\newblock \showarticletitle{{Communication-Efficient Learning of Deep Networks from Decentralized Data}}. In \bibinfo{booktitle}{\emph{Proceedings of the 20th International Conference on Artificial Intelligence and Statistics (AISTATS) 2017}}, \bibfield{editor}{\bibinfo{person}{Aarti Singh} {and} \bibinfo{person}{Jerry Zhu}} (Eds.), Vol.~\bibinfo{volume}{54}. \bibinfo{publisher}{JMLR, Inc. and Microtome Publishing}, \bibinfo{address}{Fort Lauderdale, Florida, USA}, \bibinfo{pages}{1273--1282}.
\newblock
\showISSN{2640-3498}


\bibitem[Minto et~al\mbox{.}(2021)]%
        {bibliography:stronger-privacy-for-federated-collaborative-filtering-with-implicit-feedback}
\bibfield{author}{\bibinfo{person}{Lorenzo Minto}, \bibinfo{person}{Moritz Haller}, \bibinfo{person}{Benjamin Livshits}, {and} \bibinfo{person}{Hamed Haddadi}.} \bibinfo{year}{2021}\natexlab{}.
\newblock \showarticletitle{Stronger privacy for federated collaborative filtering with implicit feedback}. In \bibinfo{booktitle}{\emph{Proceedings of the 15th ACM Conference on Recommender Systems}}. \bibinfo{publisher}{ACM (Association for Computer Machinery)}, \bibinfo{address}{New York, NY, USA}, \bibinfo{pages}{342--350}.
\newblock


\bibitem[{Moving Picture Experts Group (MPEG) working group of ISO/IEC}(2021)]%
        {bibliography:mpeg-7-compression-of-neural-networks-for-multimedia-content-description-and-analysis}
\bibfield{author}{\bibinfo{person}{{Moving Picture Experts Group (MPEG) working group of ISO/IEC}}.} \bibinfo{year}{2021}\natexlab{}.
\newblock \bibinfo{booktitle}{\emph{MPEG-7: Compression of Neural Networks for Multimedia Content Description and analysis}}.
\newblock \bibinfo{type}{Standard}. \bibinfo{institution}{Moving Picture Experts Group (MPEG) working group of ISO/IEC}, \bibinfo{address}{Hannover, DE}.
\newblock


\bibitem[Muhammad et~al\mbox{.}(2020)]%
        {bibliography:fedfast-going-beyond-average-for-faster-training-of-federated-recommender-systems}
\bibfield{author}{\bibinfo{person}{Khalil Muhammad}, \bibinfo{person}{Qinqin Wang}, \bibinfo{person}{Diarmuid O'Reilly-Morgan}, \bibinfo{person}{Elias Tragos}, \bibinfo{person}{Barry Smyth}, \bibinfo{person}{Neil Hurley}, \bibinfo{person}{James Geraci}, {and} \bibinfo{person}{Aonghus Lawlor}.} \bibinfo{year}{2020}\natexlab{}.
\newblock \showarticletitle{FedFast: Going Beyond Average for Faster Training of Federated Recommender Systems}. In \bibinfo{booktitle}{\emph{KDD '20: Proceedings of the 26th ACM SIGKDD International Conference on Knowledge Discovery \& Data Mining}} (Virtual Event, CA, USA) \emph{(\bibinfo{series}{KDD '20})}. \bibinfo{publisher}{Association for Computing Machinery}, \bibinfo{address}{New York, New York, USA}, \bibinfo{pages}{1234--1242}.
\newblock
\showISBNx{9781450379984}
\urldef\tempurl%
\url{https://doi.org/10.1145/3394486.3403176}
\showDOI{\tempurl}


\bibitem[Neumann et~al\mbox{.}(2020)]%
        {bibliography:deepcabac-plug-and-play-compression-of-neural-network-weights-and-weight-updates}
\bibfield{author}{\bibinfo{person}{David Neumann}, \bibinfo{person}{Felix Sattler}, \bibinfo{person}{Heiner Kirchhoffer}, \bibinfo{person}{Simon Wiedemann}, \bibinfo{person}{Karsten Müller}, \bibinfo{person}{Heiko Schwarz}, \bibinfo{person}{Thomas Wiegand}, \bibinfo{person}{Detlev Marpe}, {and} \bibinfo{person}{Wojciech Samek}.} \bibinfo{year}{2020}\natexlab{}.
\newblock \showarticletitle{DeepCABAC: Plug\&Play Compression of Neural Network Weights and Weight Updates}. In \bibinfo{booktitle}{\emph{{IEEE} International Conference on Image Processing, {ICIP} 2020, October 25-28, 2020}}. \bibinfo{publisher}{{IEEE}}, \bibinfo{address}{Abu Dhabi, United Arab Emirates}, \bibinfo{pages}{21--25}.
\newblock
\urldef\tempurl%
\url{https://doi.org/10.1109/ICIP40778.2020.9190821}
\showDOI{\tempurl}


\bibitem[Ovi et~al\mbox{.}(2023)]%
        {bibliography:mixed-quantization-enabled-federated-learning-to-tackle-gradient-inversion-attacks}
\bibfield{author}{\bibinfo{person}{Pretom~Roy Ovi}, \bibinfo{person}{Emon Dey}, \bibinfo{person}{Nirmalya Roy}, {and} \bibinfo{person}{Aryya Gangopadhyay}.} \bibinfo{year}{2023}\natexlab{}.
\newblock \showarticletitle{Mixed Quantization Enabled Federated Learning to Tackle Gradient Inversion Attacks}. In \bibinfo{booktitle}{\emph{2023 IEEE/CVF Conference on Computer Vision and Pattern Recognition Workshops (CVPRW)}}. \bibinfo{publisher}{Institute of Electrical and Electronics Engineers ({IEEE})}, \bibinfo{address}{Vancouver, British Columbia, Canada}, \bibinfo{pages}{5046--5054}.
\newblock
\urldef\tempurl%
\url{https://doi.org/10.1109/CVPRW59228.2023.00533}
\showDOI{\tempurl}


\bibitem[Paszke et~al\mbox{.}(2019)]%
        {bibliography:pytorch}
\bibfield{author}{\bibinfo{person}{Adam Paszke}, \bibinfo{person}{Sam Gross}, \bibinfo{person}{Francisco Massa}, \bibinfo{person}{Adam Lerer}, \bibinfo{person}{James Bradbury}, \bibinfo{person}{Gregory Chanan}, \bibinfo{person}{Trevor Killeen}, \bibinfo{person}{Zeming Lin}, \bibinfo{person}{Natalia Gimelshein}, \bibinfo{person}{Luca Antiga}, \bibinfo{person}{Alban Desmaison}, \bibinfo{person}{Andreas Kopf}, \bibinfo{person}{Edward Yang}, \bibinfo{person}{Zachary DeVito}, \bibinfo{person}{Martin Raison}, \bibinfo{person}{Alykhan Tejani}, \bibinfo{person}{Sasank Chilamkurthy}, \bibinfo{person}{Benoit Steiner}, \bibinfo{person}{Lu Fang}, \bibinfo{person}{Junjie Bai}, {and} \bibinfo{person}{Soumith Chintala}.} \bibinfo{year}{2019}\natexlab{}.
\newblock \showarticletitle{PyTorch: An Imperative Style, High-Performance Deep Learning Library}.
\newblock In \bibinfo{booktitle}{\emph{Advances in Neural Information Processing Systems 32}}, \bibfield{editor}{\bibinfo{person}{H.~Wallach}, \bibinfo{person}{H.~Larochelle}, \bibinfo{person}{A.~Beygelzimer}, \bibinfo{person}{F.~d\textquotesingle Alché-Buc}, \bibinfo{person}{E.~Fox}, {and} \bibinfo{person}{R.~Garnett}} (Eds.). \bibinfo{publisher}{Curran Associates, Inc.}, \bibinfo{address}{Vancouver, British Columbia, Canada}, \bibinfo{pages}{8024--8035}.
\newblock
\urldef\tempurl%
\url{http://papers.neurips.cc/paper/9015-pytorch-an-imperative-style-high-performance-deep-learning-library.pdf}
\showURL{%
\tempurl}


\bibitem[Perifanis and Efraimidis(2022)]%
        {bibliography:federated-neural-collaborative-filtering}
\bibfield{author}{\bibinfo{person}{Vasileios Perifanis} {and} \bibinfo{person}{Pavlos~S. Efraimidis}.} \bibinfo{year}{2022}\natexlab{}.
\newblock \showarticletitle{Federated Neural Collaborative Filtering}.
\newblock \bibinfo{journal}{\emph{Know.-Based Syst.}} \bibinfo{volume}{242}, \bibinfo{number}{C} (\bibinfo{date}{April} \bibinfo{year}{2022}), \bibinfo{numpages}{16}~pages.
\newblock
\showISSN{0950-7051}
\urldef\tempurl%
\url{https://doi.org/10.1016/j.knosys.2022.108441}
\showDOI{\tempurl}


\bibitem[Phong et~al\mbox{.}(2018)]%
        {bibliography:privacy-preserving-deep-learning-via-additively-homomorphic-encryption}
\bibfield{author}{\bibinfo{person}{Le~Trieu Phong}, \bibinfo{person}{Yoshinori Aono}, \bibinfo{person}{Takuya Hayashi}, \bibinfo{person}{Lihua Wang}, {and} \bibinfo{person}{Shiho Moriai}.} \bibinfo{year}{2018}\natexlab{}.
\newblock \showarticletitle{Privacy-Preserving Deep Learning via Additively Homomorphic Encryption}.
\newblock \bibinfo{journal}{\emph{IEEE Transactions on Information Forensics and Security}} \bibinfo{volume}{13}, \bibinfo{number}{5} (\bibinfo{year}{2018}), \bibinfo{pages}{1333--1345}.
\newblock
\urldef\tempurl%
\url{https://doi.org/10.1109/TIFS.2017.2787987}
\showDOI{\tempurl}


\bibitem[Reisizadeh et~al\mbox{.}(2020)]%
        {bibliography:fedpaq-a-communication-efficient-federated-learning-method-with-periodic-averaging-and-Quantization}
\bibfield{author}{\bibinfo{person}{Amirhossein Reisizadeh}, \bibinfo{person}{Aryan Mokhtari}, \bibinfo{person}{Hamed Hassani}, \bibinfo{person}{Ali Jadbabaie}, {and} \bibinfo{person}{Ramtin Pedarsani}.} \bibinfo{year}{2020}\natexlab{}.
\newblock \showarticletitle{FedPAQ: A Communication-Efficient Federated Learning Method with Periodic Averaging and Quantization}. In \bibinfo{booktitle}{\emph{Proceedings of the Twenty Third International Conference on Artificial Intelligence and Statistics}} \emph{(\bibinfo{series}{Proceedings of Machine Learning Research}, Vol.~\bibinfo{volume}{108})}, \bibfield{editor}{\bibinfo{person}{Silvia Chiappa} {and} \bibinfo{person}{Roberto Calandra}} (Eds.). \bibinfo{publisher}{PMLR}, \bibinfo{address}{Online}, \bibinfo{pages}{2021--2031}.
\newblock
\urldef\tempurl%
\url{https://proceedings.mlr.press/v108/reisizadeh20a.html}
\showURL{%
\tempurl}


\bibitem[Ribero et~al\mbox{.}(2022)]%
        {bibliography:federating-recommendations-using-differentially-private-prototypes}
\bibfield{author}{\bibinfo{person}{M\'{o}nica Ribero}, \bibinfo{person}{Jette Henderson}, \bibinfo{person}{Sinead Williamson}, {and} \bibinfo{person}{Haris Vikalo}.} \bibinfo{year}{2022}\natexlab{}.
\newblock \showarticletitle{Federating Recommendations Using Differentially Private Prototypes}.
\newblock \bibinfo{journal}{\emph{Pattern Recogn.}} \bibinfo{volume}{129}, \bibinfo{number}{C} (\bibinfo{date}{Sept.} \bibinfo{year}{2022}), \bibinfo{numpages}{14}~pages.
\newblock
\showISSN{0031-3203}
\urldef\tempurl%
\url{https://doi.org/10.1016/j.patcog.2022.108746}
\showDOI{\tempurl}


\bibitem[Robbins and Monro(1951)]%
        {bibliography:a-stochastic-optimization-method}
\bibfield{author}{\bibinfo{person}{Herbert Robbins} {and} \bibinfo{person}{Sutton Monro}.} \bibinfo{year}{1951}\natexlab{}.
\newblock \showarticletitle{A Stochastic Approximation Method}.
\newblock \bibinfo{journal}{\emph{The Annals of Mathematical Statistics}} \bibinfo{volume}{22}, \bibinfo{number}{3} (\bibinfo{year}{1951}), \bibinfo{pages}{400--407}.
\newblock
\showISSN{00034851}
\urldef\tempurl%
\url{http://www.jstor.org/stable/2236626}
\showURL{%
\tempurl}


\bibitem[{Rønn Hansen} et~al\mbox{.}(2022)]%
        {bibliography:larynx-cancer-survival-model-developed-through-open-source-federated-learning}
\bibfield{author}{\bibinfo{person}{Christian {Rønn Hansen}}, \bibinfo{person}{Gareth Price}, \bibinfo{person}{Matthew Field}, \bibinfo{person}{Nis Sarup}, \bibinfo{person}{Ruta Zukauskaite}, \bibinfo{person}{Jørgen Johansen}, \bibinfo{person}{Jesper~Grau Eriksen}, \bibinfo{person}{Farhannah Aly}, \bibinfo{person}{Andrew McPartlin}, \bibinfo{person}{Lois Holloway}, \bibinfo{person}{David Thwaites}, {and} \bibinfo{person}{Carsten Brink}.} \bibinfo{year}{2022}\natexlab{}.
\newblock \showarticletitle{Larynx cancer survival model developed through open-source federated learning}.
\newblock \bibinfo{journal}{\emph{Radiotherapy and Oncology}} \bibinfo{volume}{176}, \bibinfo{number}{1} (\bibinfo{date}{Nov.} \bibinfo{year}{2022}), \bibinfo{pages}{179--186}.
\newblock
\showISSN{0167-8140}
\urldef\tempurl%
\url{https://doi.org/10.1016/j.radonc.2022.09.023}
\showDOI{\tempurl}


\bibitem[Sattler et~al\mbox{.}(2019)]%
        {bibliography:sparse-binary-compression}
\bibfield{author}{\bibinfo{person}{Felix Sattler}, \bibinfo{person}{Simon Wiedemann}, \bibinfo{person}{Klaus~Robert Müller}, {and} \bibinfo{person}{Wojciech Samek}.} \bibinfo{year}{2019}\natexlab{}.
\newblock \showarticletitle{Sparse Binary Compression: Towards Distributed Deep Learning with minimal Communication}. In \bibinfo{booktitle}{\emph{2019 International Joint Conference on Neural Networks, IJCNN 2019}} \emph{(\bibinfo{series}{Proceedings of the International Joint Conference on Neural Networks})}. \bibinfo{publisher}{Institute of Electrical and Electronics Engineers Inc.}, \bibinfo{address}{Budapest, Hungary}.
\newblock
\urldef\tempurl%
\url{https://doi.org/10.1109/IJCNN.2019.8852172}
\showDOI{\tempurl}


\bibitem[Sattler et~al\mbox{.}(2020)]%
        {bibliography:robust-and-communication-efficient-federated-learning-from-non-iid-data}
\bibfield{author}{\bibinfo{person}{Felix Sattler}, \bibinfo{person}{Simon Wiedemann}, \bibinfo{person}{Klaus-Robert Müller}, {and} \bibinfo{person}{Wojciech Samek}.} \bibinfo{year}{2020}\natexlab{}.
\newblock \showarticletitle{Robust and Communication-Efficient Federated Learning From Non-i.i.d. Data}.
\newblock \bibinfo{journal}{\emph{IEEE Transactions on Neural Networks and Learning Systems}} \bibinfo{volume}{31}, \bibinfo{number}{9} (\bibinfo{year}{2020}), \bibinfo{pages}{3400--3413}.
\newblock
\urldef\tempurl%
\url{https://doi.org/10.1109/TNNLS.2019.2944481}
\showDOI{\tempurl}


\bibitem[Schrage(2017)]%
        {bibliography:great-digital-companies-build-great-recommendation-engines}
\bibfield{author}{\bibinfo{person}{Michael Schrage}.} \bibinfo{year}{2017}\natexlab{}.
\newblock \bibinfo{booktitle}{\emph{Great Digital Companies Build Great Recommendation Engines}}.
\newblock Harvard Business Review.
\newblock
\urldef\tempurl%
\url{https://hbr.org/2017/08/great-digital-companies-build-great-recommendation-engines}
\showURL{%
\tempurl}


\bibitem[Schwartz(2004)]%
        {bibliography:the-tyranny-of-choice}
\bibfield{author}{\bibinfo{person}{Barry Schwartz}.} \bibinfo{year}{2004}\natexlab{}.
\newblock \showarticletitle{The Tyranny of Choice}.
\newblock \bibinfo{journal}{\emph{Scientific American}} \bibinfo{volume}{290}, \bibinfo{number}{4} (\bibinfo{date}{April} \bibinfo{year}{2004}), \bibinfo{pages}{70--75}.
\newblock
\showCODEN{SCAMAC}
\showISSN{0036-8733 (print), 1946-7087 (electronic)}
\urldef\tempurl%
\url{https://doi.org/10.1038/scientificamerican0404-70}
\showDOI{\tempurl}


\bibitem[Sedhain et~al\mbox{.}(2015)]%
        {bibliography:autoRec-autoencoders-meet-collaborative-filtering}
\bibfield{author}{\bibinfo{person}{Suvash Sedhain}, \bibinfo{person}{Aditya~Krishna Menon}, \bibinfo{person}{Scott Sanner}, {and} \bibinfo{person}{Lexing Xie}.} \bibinfo{year}{2015}\natexlab{}.
\newblock \showarticletitle{AutoRec: Autoencoders Meet Collaborative Filtering}. In \bibinfo{booktitle}{\emph{Proceedings of the 24th International Conference on World Wide Web}} (Florence, Italy) \emph{(\bibinfo{series}{WWW '15 Companion})}. \bibinfo{publisher}{Association for Computing Machinery}, \bibinfo{address}{New York, NY, USA}, \bibinfo{pages}{111--112}.
\newblock
\showISBNx{9781450334730}
\urldef\tempurl%
\url{https://doi.org/10.1145/2740908.2742726}
\showDOI{\tempurl}


\bibitem[Seol and Kim(2023)]%
        {bibliography:performance-enhancement-in-federated-learning-by-reducing-class-imbalance-of-non-iid-data}
\bibfield{author}{\bibinfo{person}{Mihye Seol} {and} \bibinfo{person}{Taejoon Kim}.} \bibinfo{year}{2023}\natexlab{}.
\newblock \showarticletitle{Performance Enhancement in Federated Learning by Reducing Class Imbalance of Non-IID Data}.
\newblock \bibinfo{journal}{\emph{Sensors}} \bibinfo{volume}{23}, \bibinfo{number}{3} (\bibinfo{year}{2023}), \bibinfo{numpages}{16}~pages.
\newblock
\showISSN{1424-8220}
\urldef\tempurl%
\url{https://doi.org/10.3390/s23031152}
\showDOI{\tempurl}


\bibitem[Shakespeare(1994)]%
        {bibliography:the-complete-works-of-william-shakespeare}
\bibfield{author}{\bibinfo{person}{William Shakespeare}.} \bibinfo{year}{1994}\natexlab{}.
\newblock \bibinfo{booktitle}{\emph{The Complete Works of William Shakespeare}}. \bibinfo{series}{Project Gutenberg}, Vol.~\bibinfo{volume}{100}.
\newblock \bibinfo{publisher}{Project Gutenberg}, \bibinfo{address}{P.O. Box 2782, Champaign, IL 61825-2782, USA}.
\newblock
\urldef\tempurl%
\url{https://www.gutenberg.org/ebooks/100}
\showURL{%
\tempurl}


\bibitem[Shamir(1979)]%
        {bibliography:how-to-share-a-secret}
\bibfield{author}{\bibinfo{person}{Adi Shamir}.} \bibinfo{year}{1979}\natexlab{}.
\newblock \showarticletitle{How to share a secret}.
\newblock \bibinfo{journal}{\emph{Commun. ACM}} \bibinfo{volume}{22}, \bibinfo{number}{11} (\bibinfo{year}{1979}), \bibinfo{pages}{612--613}.
\newblock


\bibitem[Sherstinsky(2020)]%
        {bibliography:fundamentals-of-recurrent-neural-network}
\bibfield{author}{\bibinfo{person}{Alex Sherstinsky}.} \bibinfo{year}{2020}\natexlab{}.
\newblock \showarticletitle{Fundamentals of Recurrent Neural Network (RNN) and Long Short-Term Memory (LSTM) network}.
\newblock \bibinfo{journal}{\emph{Physica D: Nonlinear Phenomena}} \bibinfo{volume}{404}, \bibinfo{number}{1} (\bibinfo{date}{March} \bibinfo{year}{2020}), \bibinfo{pages}{132306}.
\newblock
\showISSN{0167-2789}
\urldef\tempurl%
\url{https://doi.org/10.1016/j.physd.2019.132306}
\showDOI{\tempurl}


\bibitem[Shokri and Shmatikov(2015)]%
        {bibliography:privacy-preserving-deep-learning}
\bibfield{author}{\bibinfo{person}{Reza Shokri} {and} \bibinfo{person}{Vitaly Shmatikov}.} \bibinfo{year}{2015}\natexlab{}.
\newblock \showarticletitle{Privacy-Preserving Deep Learning}. In \bibinfo{booktitle}{\emph{Proceedings of the 22nd ACM SIGSAC Conference on Computer and Communications Security}} (Denver, Colorado, USA) \emph{(\bibinfo{series}{CCS '15})}. \bibinfo{publisher}{Association for Computing Machinery}, \bibinfo{address}{New York, NY, USA}, \bibinfo{pages}{1310--1321}.
\newblock
\showISBNx{9781450338325}
\urldef\tempurl%
\url{https://doi.org/10.1145/2810103.2813687}
\showDOI{\tempurl}


\bibitem[Smith et~al\mbox{.}(2022)]%
        {bibliography:recommender-systems-and-algorithmic-hate}
\bibfield{author}{\bibinfo{person}{Jessie~J. Smith}, \bibinfo{person}{Lucia Jayne}, {and} \bibinfo{person}{Robin Burke}.} \bibinfo{year}{2022}\natexlab{}.
\newblock \showarticletitle{Recommender Systems and Algorithmic Hate}. In \bibinfo{booktitle}{\emph{Proceedings of the 16th ACM Conference on Recommender Systems}} (Seattle, WA, USA) \emph{(\bibinfo{series}{RecSys '22})}. \bibinfo{publisher}{Association for Computing Machinery}, \bibinfo{address}{New York, NY, USA}, \bibinfo{pages}{592--597}.
\newblock
\showISBNx{9781450392785}
\urldef\tempurl%
\url{https://doi.org/10.1145/3523227.3551480}
\showDOI{\tempurl}


\bibitem[Stoll(2022)]%
        {bibliography:devices-used-to-watch-online-video-on-demand-worldwide}
\bibfield{author}{\bibinfo{person}{Julia Stoll}.} \bibinfo{year}{2022}\natexlab{}.
\newblock \bibinfo{booktitle}{\emph{Devices used to watch online video on demand (VOD) worldwide in 1st quarter 2022 and 2nd quarter 2022}}.
\newblock {Statista}.
\newblock
\urldef\tempurl%
\url{https://www.statista.com/statistics/1329449/vod-device-usage-share-worldwide/}
\showURL{%
\tempurl}


\bibitem[Sun et~al\mbox{.}(2022)]%
        {bibliography:adaptive-random-walk-gradient-descent-for-decentralized-optimization}
\bibfield{author}{\bibinfo{person}{Tao Sun}, \bibinfo{person}{Dongsheng Li}, {and} \bibinfo{person}{Bao Wang}.} \bibinfo{year}{2022}\natexlab{}.
\newblock \showarticletitle{Adaptive Random Walk Gradient Descent for Decentralized Optimization}. In \bibinfo{booktitle}{\emph{Proceedings of the 39th International Conference on Machine Learning}} \emph{(\bibinfo{series}{Proceedings of Machine Learning Research}, Vol.~\bibinfo{volume}{162})}, \bibfield{editor}{\bibinfo{person}{Kamalika Chaudhuri}, \bibinfo{person}{Stefanie Jegelka}, \bibinfo{person}{Le~Song}, \bibinfo{person}{Csaba Szepesvari}, \bibinfo{person}{Gang Niu}, {and} \bibinfo{person}{Sivan Sabato}} (Eds.). \bibinfo{publisher}{PMLR}, \bibinfo{address}{Baltimore, Maryland, USA}, \bibinfo{pages}{20790--20809}.
\newblock
\urldef\tempurl%
\url{https://proceedings.mlr.press/v162/sun22b.html}
\showURL{%
\tempurl}


\bibitem[{Sun} et~al\mbox{.}(2023)]%
        {bibliography:a-survey-on-federated-recommendation-systems}
\bibfield{author}{\bibinfo{person}{Zehua {Sun}}, \bibinfo{person}{Yonghui {Xu}}, \bibinfo{person}{Yong {Liu}}, \bibinfo{person}{Wei {He}}, \bibinfo{person}{Lanju {Kong}}, \bibinfo{person}{Fangzhao {Wu}}, \bibinfo{person}{Yali {Jiang}}, {and} \bibinfo{person}{Lizhen {Cui}}.} \bibinfo{year}{2023}\natexlab{}.
\newblock \showarticletitle{{A Survey on Federated Recommendation Systems}}.
\newblock \bibinfo{journal}{\emph{arXiv e-prints}}  \bibinfo{volume}{2301.00767} (\bibinfo{date}{March} \bibinfo{year}{2023}), \bibinfo{numpages}{15}~pages.
\newblock
\urldef\tempurl%
\url{https://doi.org/10.48550/arXiv.2301.00767}
\showDOI{\tempurl}
\showeprint[arxiv]{2301.00767}~[cs.IR]


\bibitem[Tang and Wang(2018)]%
        {bibliography:personalized-top-n-sequential-recommendation-via-convolutional-sequence-embedding}
\bibfield{author}{\bibinfo{person}{Jiaxi Tang} {and} \bibinfo{person}{Ke Wang}.} \bibinfo{year}{2018}\natexlab{}.
\newblock \showarticletitle{Personalized Top-N Sequential Recommendation via Convolutional Sequence Embedding}. In \bibinfo{booktitle}{\emph{Proceedings of the Eleventh ACM International Conference on Web Search and Data Mining}} (Marina Del Rey, CA, USA) \emph{(\bibinfo{series}{WSDM '18})}. \bibinfo{publisher}{Association for Computing Machinery}, \bibinfo{address}{New York, NY, USA}, \bibinfo{pages}{565--573}.
\newblock
\showISBNx{9781450355810}
\urldef\tempurl%
\url{https://doi.org/10.1145/3159652.3159656}
\showDOI{\tempurl}


\bibitem[Triastcyn et~al\mbox{.}(2022)]%
        {bibliography:decentralized-learning-with-random-walks-and-communication-efficient-adaptive-optimization}
\bibfield{author}{\bibinfo{person}{Aleksei Triastcyn}, \bibinfo{person}{Matthias Reisser}, {and} \bibinfo{person}{Christos Louizos}.} \bibinfo{year}{2022}\natexlab{}.
\newblock \showarticletitle{Decentralized Learning with Random Walks and Communication-Efficient Adaptive Optimization}. In \bibinfo{booktitle}{\emph{Workshop on Federated Learning: Recent Advances and New Challenges (in Conjunction with NeurIPS 2022)}}. \bibinfo{publisher}{NeurIPS}, \bibinfo{address}{New Orleans, LA, USA}.
\newblock


\bibitem[Wainakh et~al\mbox{.}(2019)]%
        {bibliography:efficient-privacy-preserving-recommendations-based-on-social-graphs}
\bibfield{author}{\bibinfo{person}{Aidmar Wainakh}, \bibinfo{person}{Tim Grube}, \bibinfo{person}{Jörg Daubert}, {and} \bibinfo{person}{Max Mühlhäuser}.} \bibinfo{year}{2019}\natexlab{}.
\newblock \showarticletitle{Efficient Privacy-Preserving Recommendations Based on Social Graphs}. In \bibinfo{booktitle}{\emph{Proceedings of the 13th ACM Conference on Recommender Systems}} (Copenhagen, Denmark) \emph{(\bibinfo{series}{RecSys '19})}. \bibinfo{publisher}{Association for Computing Machinery}, \bibinfo{address}{New York, NY, USA}, \bibinfo{pages}{78--86}.
\newblock
\showISBNx{9781450362436}
\urldef\tempurl%
\url{https://doi.org/10.1145/3298689.3347013}
\showDOI{\tempurl}


\bibitem[{Wang} et~al\mbox{.}(2022)]%
        {bibliography:on-the-unreasonable-effectiveness-of-federated-averaging-with-heterogeneous-data}
\bibfield{author}{\bibinfo{person}{Jianyu {Wang}}, \bibinfo{person}{Rudrajit {Das}}, \bibinfo{person}{Gauri {Joshi}}, \bibinfo{person}{Satyen {Kale}}, \bibinfo{person}{Zheng {Xu}}, {and} \bibinfo{person}{Tong {Zhang}}.} \bibinfo{year}{2022}\natexlab{}.
\newblock \showarticletitle{On the Unreasonable Effectiveness of Federated Averaging with Heterogeneous Data}.
\newblock \bibinfo{journal}{\emph{arXiv e-prints}}  \bibinfo{volume}{abs/2206.04723} (\bibinfo{date}{June} \bibinfo{year}{2022}), \bibinfo{numpages}{21}~pages.
\newblock
\urldef\tempurl%
\url{https://doi.org/10.48550/arXiv.2206.04723}
\showDOI{\tempurl}
\showeprint[arxiv]{2206.04723}~[cs.LG]


\bibitem[Wang et~al\mbox{.}(2021)]%
        {bibliography:demystifying-model-averaging-for-communication-efficient-federated-matrix-factorization}
\bibfield{author}{\bibinfo{person}{Shuai Wang}, \bibinfo{person}{Richard~Cornelius Suwandi}, {and} \bibinfo{person}{Tsung-Hui Chang}.} \bibinfo{year}{2021}\natexlab{}.
\newblock \showarticletitle{Demystifying Model Averaging for Communication-Efficient Federated Matrix Factorization}. In \bibinfo{booktitle}{\emph{ICASSP 2021 - 2021 IEEE International Conference on Acoustics, Speech and Signal Processing (ICASSP)}} (Toronto, Ontario, Canada). \bibinfo{publisher}{Institute of Electrical and Electronics Engineers ({IEEE})}, \bibinfo{address}{3 Park Avenue, 17th Floor, New York, NY 10016-5997 USA}, \bibinfo{pages}{3680--3684}.
\newblock
\showISSN{2379-190X}
\urldef\tempurl%
\url{https://doi.org/10.1109/ICASSP39728.2021.9413927}
\showDOI{\tempurl}


\bibitem[Wang et~al\mbox{.}(2023)]%
        {bibliography:batch-normalization-damages-federated-learning-on-non-iid-data-analysis-and-remedy}
\bibfield{author}{\bibinfo{person}{Yanmeng Wang}, \bibinfo{person}{Qingjiang Shi}, {and} \bibinfo{person}{Tsung-Hui Chang}.} \bibinfo{year}{2023}\natexlab{}.
\newblock \showarticletitle{Batch Normalization Damages Federated Learning on NON-IID Data: Analysis and Remedy}. In \bibinfo{booktitle}{\emph{ICASSP 2023 - 2023 IEEE International Conference on Acoustics, Speech and Signal Processing (ICASSP)}}. \bibinfo{publisher}{Institute of Electrical and Electronics Engineers ({IEEE})}, \bibinfo{address}{3 Park Avenue, 17th Floor, New York, NY 10016-5997 USA}, \bibinfo{pages}{1--5}.
\newblock
\urldef\tempurl%
\url{https://doi.org/10.1109/ICASSP49357.2023.10095399}
\showDOI{\tempurl}


\bibitem[Wei et~al\mbox{.}(2020a)]%
        {bibliography:federated-learning-with-differential-privacy}
\bibfield{author}{\bibinfo{person}{Kang Wei}, \bibinfo{person}{Jun Li}, \bibinfo{person}{Ming Ding}, \bibinfo{person}{Chuan Ma}, \bibinfo{person}{Howard~H. Yang}, \bibinfo{person}{Farhad Farokhi}, \bibinfo{person}{Shi Jin}, \bibinfo{person}{Tony Q.~S. Quek}, {and} \bibinfo{person}{H. Vincent~Poor}.} \bibinfo{year}{2020}\natexlab{a}.
\newblock \showarticletitle{Federated Learning With Differential Privacy: Algorithms and Performance Analysis}.
\newblock \bibinfo{journal}{\emph{Trans. Info. For. Sec.}} \bibinfo{volume}{15}, \bibinfo{number}{1} (\bibinfo{date}{Jan.} \bibinfo{year}{2020}), \bibinfo{pages}{3454--3469}.
\newblock
\showISSN{1556-6013}
\urldef\tempurl%
\url{https://doi.org/10.1109/TIFS.2020.2988575}
\showDOI{\tempurl}


\bibitem[Wei et~al\mbox{.}(2020b)]%
        {bibliography:a-framework-for-evaluating-privacy-leakages-in-federated-learning}
\bibfield{author}{\bibinfo{person}{Wenqi Wei}, \bibinfo{person}{Ling Liu}, \bibinfo{person}{Margaret Loper}, \bibinfo{person}{Ka-Ho Chow}, \bibinfo{person}{Mehmet~Emre Gursoy}, \bibinfo{person}{Stacey Truex}, {and} \bibinfo{person}{Yanzhao Wu}.} \bibinfo{year}{2020}\natexlab{b}.
\newblock \showarticletitle{A Framework for Evaluating Client Privacy Leakages in Federated Learning}. In \bibinfo{booktitle}{\emph{Computer Security -- ESORICS 2020}}, \bibfield{editor}{\bibinfo{person}{Liqun Chen}, \bibinfo{person}{Ninghui Li}, \bibinfo{person}{Kaitai Liang}, {and} \bibinfo{person}{Steve Schneider}} (Eds.). \bibinfo{publisher}{Springer International Publishing}, \bibinfo{address}{Cham}, \bibinfo{pages}{545--566}.
\newblock
\showISBNx{978-3-030-58951-6}


\bibitem[Weissenbacher et~al\mbox{.}(2018)]%
        {bibliography:overview-of-the-third-social-media-mining-for-health-workshop}
\bibfield{author}{\bibinfo{person}{Davy Weissenbacher}, \bibinfo{person}{Abeed Sarker}, \bibinfo{person}{Michael~J. Paul}, {and} \bibinfo{person}{Graciela Gonzalez-Hernandez}.} \bibinfo{year}{2018}\natexlab{}.
\newblock \showarticletitle{Overview of the Third Social Media Mining for Health ({SMM}4{H}) Shared Tasks at {EMNLP} 2018}. In \bibinfo{booktitle}{\emph{Proceedings of the 2018 {EMNLP} Workshop {SMM}4{H}: The 3rd Social Media Mining for Health Applications Workshop {\&} Shared Task}}. \bibinfo{publisher}{Association for Computational Linguistics}, \bibinfo{address}{Brussels, Belgium}, \bibinfo{pages}{13--16}.
\newblock
\urldef\tempurl%
\url{https://doi.org/10.18653/v1/W18-5904}
\showDOI{\tempurl}


\bibitem[Wiedemann et~al\mbox{.}(2020a)]%
        {bibliography:deepcabac-a-universal-compression-algorithm-for-deep-neural-networks}
\bibfield{author}{\bibinfo{person}{Simon Wiedemann}, \bibinfo{person}{Heiner Kirchhoffer}, \bibinfo{person}{Stefan Matlage}, \bibinfo{person}{Paul Haase}, \bibinfo{person}{Arturo Marban}, \bibinfo{person}{Talmaj Marinč}, \bibinfo{person}{David Neumann}, \bibinfo{person}{Tung Nguyen}, \bibinfo{person}{Heiko Schwarz}, \bibinfo{person}{Thomas Wiegand}, \bibinfo{person}{Detlev Marpe}, {and} \bibinfo{person}{Wojciech Samek}.} \bibinfo{year}{2020}\natexlab{a}.
\newblock \showarticletitle{DeepCABAC: A Universal Compression Algorithm for Deep Neural Networks}.
\newblock \bibinfo{journal}{\emph{IEEE Journal of Selected Topics in Signal Processing}} \bibinfo{volume}{14}, \bibinfo{number}{4} (\bibinfo{year}{2020}), \bibinfo{pages}{700--714}.
\newblock
\urldef\tempurl%
\url{https://doi.org/10.1109/JSTSP.2020.2969554}
\showDOI{\tempurl}


\bibitem[Wiedemann et~al\mbox{.}(2020b)]%
        {bibliography:a-universal-compression-algorithm-for-deep-neural-networks}
\bibfield{author}{\bibinfo{person}{Simon Wiedemann}, \bibinfo{person}{Heiner Kirchhoffer}, \bibinfo{person}{Stefan Matlage}, \bibinfo{person}{Paul Haase}, \bibinfo{person}{Arturo Marban}, \bibinfo{person}{Talmaj Marinč}, \bibinfo{person}{David Neumann}, \bibinfo{person}{Tung Nguyen}, \bibinfo{person}{Heiko Schwarz}, \bibinfo{person}{Thomas Wiegand}, \bibinfo{person}{Detlev Marpe}, {and} \bibinfo{person}{Wojciech Samek}.} \bibinfo{year}{2020}\natexlab{b}.
\newblock \showarticletitle{DeepCABAC: A Universal Compression Algorithm for Deep Neural Networks}.
\newblock \bibinfo{journal}{\emph{IEEE Journal of Selected Topics in Signal Processing}} \bibinfo{volume}{14}, \bibinfo{number}{4} (\bibinfo{year}{2020}), \bibinfo{pages}{700--714}.
\newblock
\urldef\tempurl%
\url{https://doi.org/10.1109/JSTSP.2020.2969554}
\showDOI{\tempurl}


\bibitem[Wu et~al\mbox{.}(2022)]%
        {bibliography:communication-efficient-federated-learning-via-knowledge-distillation}
\bibfield{author}{\bibinfo{person}{Chuhan Wu}, \bibinfo{person}{Fangzhao Wu}, \bibinfo{person}{Lingjuan Lyu}, \bibinfo{person}{Yongfeng Huang}, {and} \bibinfo{person}{Xing Xie}.} \bibinfo{year}{2022}\natexlab{}.
\newblock \showarticletitle{Communication-efficient federated learning via knowledge distillation}.
\newblock \bibinfo{journal}{\emph{Nature Communications}} \bibinfo{volume}{13}, \bibinfo{number}{1} (\bibinfo{date}{April} \bibinfo{year}{2022}), \bibinfo{numpages}{8}~pages.
\newblock
\urldef\tempurl%
\url{https://doi.org/10.1038/s41467-022-29763-x}
\showDOI{\tempurl}


\bibitem[Wu et~al\mbox{.}(2020)]%
        {bibliography:mind-a-large-scale-dataset-for-news-recommendation}
\bibfield{author}{\bibinfo{person}{Fangzhao Wu}, \bibinfo{person}{Ying Qiao}, \bibinfo{person}{Jiun-Hung Chen}, \bibinfo{person}{Chuhan Wu}, \bibinfo{person}{Tao Qi}, \bibinfo{person}{Jianxun Lian}, \bibinfo{person}{Danyang Liu}, \bibinfo{person}{Xing Xie}, \bibinfo{person}{Jianfeng Gao}, \bibinfo{person}{Winnie Wu}, {and} \bibinfo{person}{Ming Zhou}.} \bibinfo{year}{2020}\natexlab{}.
\newblock \showarticletitle{{MIND}: A Large-scale Dataset for News Recommendation}. In \bibinfo{booktitle}{\emph{Proceedings of the 58th Annual Meeting of the Association for Computational Linguistics}}. \bibinfo{publisher}{Association for Computational Linguistics}, \bibinfo{address}{Online}, \bibinfo{pages}{3597--3606}.
\newblock
\urldef\tempurl%
\url{https://doi.org/10.18653/v1/2020.acl-main.331}
\showDOI{\tempurl}


\bibitem[Wu et~al\mbox{.}(2016)]%
        {bibliography:collaborative-denoising-auto-encoders-for-top-N-recommender-systems}
\bibfield{author}{\bibinfo{person}{Yao Wu}, \bibinfo{person}{Christopher DuBois}, \bibinfo{person}{Alice~X. Zheng}, {and} \bibinfo{person}{Martin Ester}.} \bibinfo{year}{2016}\natexlab{}.
\newblock \showarticletitle{Collaborative Denoising Auto-Encoders for Top-N Recommender Systems}. In \bibinfo{booktitle}{\emph{Proceedings of the Ninth ACM International Conference on Web Search and Data Mining}} (San Francisco, California, USA) \emph{(\bibinfo{series}{WSDM '16})}. \bibinfo{publisher}{Association for Computing Machinery}, \bibinfo{address}{New York, NY, USA}, \bibinfo{pages}{153--162}.
\newblock
\showISBNx{9781450337168}
\urldef\tempurl%
\url{https://doi.org/10.1145/2835776.2835837}
\showDOI{\tempurl}


\bibitem[Wu and He(2020)]%
        {bibliography:group-normalization}
\bibfield{author}{\bibinfo{person}{Yuxin Wu} {and} \bibinfo{person}{Kaiming He}.} \bibinfo{year}{2020}\natexlab{}.
\newblock \showarticletitle{Group Normalization}.
\newblock \bibinfo{journal}{\emph{International Journal of Computer Vision}} \bibinfo{volume}{128}, \bibinfo{number}{3} (\bibinfo{date}{01 Mar} \bibinfo{year}{2020}), \bibinfo{pages}{742--755}.
\newblock
\showISSN{1573-1405}
\urldef\tempurl%
\url{https://doi.org/10.1007/s11263-019-01198-w}
\showDOI{\tempurl}


\bibitem[Yang et~al\mbox{.}(2021)]%
        {bibliography:federated-collective-matrix-factorization-for-heterogeneous-collaborative-filtering}
\bibfield{author}{\bibinfo{person}{Enyue Yang}, \bibinfo{person}{Yunfeng Huang}, \bibinfo{person}{Feng Liang}, \bibinfo{person}{Weike Pan}, {and} \bibinfo{person}{Zhong Ming}.} \bibinfo{year}{2021}\natexlab{}.
\newblock \showarticletitle{FCMF: Federated collective matrix factorization for heterogeneous collaborative filtering}.
\newblock \bibinfo{journal}{\emph{Knowledge-Based Systems}} \bibinfo{volume}{220}, \bibinfo{number}{1} (\bibinfo{date}{March} \bibinfo{year}{2021}), \bibinfo{pages}{106946}.
\newblock
\showISSN{0950-7051}
\urldef\tempurl%
\url{https://doi.org/10.1016/j.knosys.2021.106946}
\showDOI{\tempurl}


\bibitem[Yelp(2021)]%
        {bibliography:yelp-dataset}
\bibfield{author}{\bibinfo{person}{Yelp}.} \bibinfo{year}{2021}\natexlab{}.
\newblock \bibinfo{booktitle}{\emph{Yelp Dataset}}.
\newblock {Yelp Inc.}
\newblock
\urldef\tempurl%
\url{https://www.yelp.com/dataset}
\showURL{%
\tempurl}


\bibitem[Ying et~al\mbox{.}(2018)]%
        {bibliography:graph-convolutional-neural-networks-for-web-scale-recommender-systems}
\bibfield{author}{\bibinfo{person}{Rex Ying}, \bibinfo{person}{Ruining He}, \bibinfo{person}{Kaifeng Chen}, \bibinfo{person}{Pong Eksombatchai}, \bibinfo{person}{William~L. Hamilton}, {and} \bibinfo{person}{Jure Leskovec}.} \bibinfo{year}{2018}\natexlab{}.
\newblock \showarticletitle{Graph Convolutional Neural Networks for Web-Scale Recommender Systems}. In \bibinfo{booktitle}{\emph{Proceedings of the 24th ACM SIGKDD International Conference on Knowledge Discovery \& Data Mining}} (London, United Kingdom) \emph{(\bibinfo{series}{KDD '18})}. \bibinfo{publisher}{Association for Computing Machinery}, \bibinfo{address}{New York, NY, USA}, \bibinfo{pages}{974--983}.
\newblock
\showISBNx{9781450355520}
\urldef\tempurl%
\url{https://doi.org/10.1145/3219819.3219890}
\showDOI{\tempurl}


\bibitem[Yue et~al\mbox{.}(2023)]%
        {bibliography:gradient-obfuscation-gives-a-false-sense-of-security-in-federated-learning}
\bibfield{author}{\bibinfo{person}{Kai Yue}, \bibinfo{person}{Richeng Jin}, \bibinfo{person}{Chau-Wai Wong}, \bibinfo{person}{Dror Baron}, {and} \bibinfo{person}{Huaiyu Dai}.} \bibinfo{year}{2023}\natexlab{}.
\newblock \showarticletitle{Gradient Obfuscation Gives a False Sense of Security in Federated Learning}. In \bibinfo{booktitle}{\emph{Proceedings of the 32nd USENIX Conference on Security Symposium}} (Anaheim, California, United States of America) \emph{(\bibinfo{series}{SEC '23})}. \bibinfo{publisher}{USENIX Association}, \bibinfo{address}{USA}, Article \bibinfo{articleno}{357}, \bibinfo{numpages}{18}~pages.
\newblock
\showISBNx{978-1-939133-37-3}


\bibitem[Zaccone et~al\mbox{.}(2022)]%
        {bibliography:speeding-up-heterogeneous-federated-learning-with-sequentially-trained-superclients}
\bibfield{author}{\bibinfo{person}{Riccardo Zaccone}, \bibinfo{person}{Andrea Rizzardi}, \bibinfo{person}{Debora Caldarola}, \bibinfo{person}{Marco Ciccone}, {and} \bibinfo{person}{Barbara Caputo}.} \bibinfo{year}{2022}\natexlab{}.
\newblock \showarticletitle{Speeding up Heterogeneous Federated Learning with Sequentially Trained Superclients}. In \bibinfo{booktitle}{\emph{2022 26th International Conference on Pattern Recognition (ICPR)}} (Montréal, Québec, Canada). \bibinfo{publisher}{Institute of Electrical and Electronics Engineers ({IEEE})}, \bibinfo{address}{3 Park Avenue, 17th Floor, New York, NY 10016-5997 USA}, \bibinfo{pages}{3376--3382}.
\newblock
\urldef\tempurl%
\url{https://doi.org/10.1109/ICPR56361.2022.9956084}
\showDOI{\tempurl}


\bibitem[Zhang et~al\mbox{.}(2022)]%
        {bibliography:lightweight-federated-recommendation-with-privacy-preserving-matrix-factorization}
\bibfield{author}{\bibinfo{person}{Honglei Zhang}, \bibinfo{person}{Fangyuan Luo}, \bibinfo{person}{Jun Wu}, \bibinfo{person}{Xiangnan He}, {and} \bibinfo{person}{Yidong Li}.} \bibinfo{year}{2022}\natexlab{}.
\newblock \showarticletitle{LightFR: Lightweight Federated Recommendation with Privacy-Preserving Matrix Factorization}.
\newblock \bibinfo{journal}{\emph{ACM Trans. Inf. Syst.}} \bibinfo{volume}{41}, \bibinfo{number}{2} (\bibinfo{date}{Dec.} \bibinfo{year}{2022}), \bibinfo{pages}{1--28}.
\newblock
\showISSN{1046-8188}
\urldef\tempurl%
\url{https://doi.org/10.1145/3578361}
\showDOI{\tempurl}
\newblock
\shownote{Just Accepted}.


\bibitem[Zhang and Jiang(2021)]%
        {bibliography:a-vertical-federation-recommendation-method-based-on-clustering-and-latent-factor-model}
\bibfield{author}{\bibinfo{person}{JianFei Zhang} {and} \bibinfo{person}{YuChen Jiang}.} \bibinfo{year}{2021}\natexlab{}.
\newblock \showarticletitle{A vertical federation recommendation method based on clustering and latent factor model}. In \bibinfo{booktitle}{\emph{2021 International Conference on Electronic Information Engineering and Computer Science (EIECS)}}. \bibinfo{publisher}{Institute of Electrical and Electronics Engineers ({IEEE})}, \bibinfo{address}{3 Park Avenue, 17th Floor, New York, NY 10016-5997 USA}, \bibinfo{pages}{362--366}.
\newblock
\urldef\tempurl%
\url{https://doi.org/10.1109/EIECS53707.2021.9587935}
\showDOI{\tempurl}


\bibitem[{Zhao} et~al\mbox{.}(2020)]%
        {bibliography:improved-deep-leakage-from-gradients}
\bibfield{author}{\bibinfo{person}{Bo {Zhao}}, \bibinfo{person}{Konda {Reddy Mopuri}}, {and} \bibinfo{person}{Hakan {Bilen}}.} \bibinfo{year}{2020}\natexlab{}.
\newblock \showarticletitle{iDLG: Improved Deep Leakage from Gradients}.
\newblock \bibinfo{journal}{\emph{arXiv e-prints}}  \bibinfo{volume}{abs/2001.02610} (\bibinfo{date}{Jan.} \bibinfo{year}{2020}), \bibinfo{numpages}{5}~pages.
\newblock
\urldef\tempurl%
\url{https://doi.org/10.48550/arXiv.2001.02610}
\showDOI{\tempurl}
\showeprint[arxiv]{2001.02610}~[cs.LG]


\bibitem[Zhu et~al\mbox{.}(2021)]%
        {bibliography:federated-learning-on-non-iid-data-a-survey}
\bibfield{author}{\bibinfo{person}{Hangyu Zhu}, \bibinfo{person}{Jinjin Xu}, \bibinfo{person}{Shiqing Liu}, {and} \bibinfo{person}{Yaochu Jin}.} \bibinfo{year}{2021}\natexlab{}.
\newblock \showarticletitle{Federated Learning on Non-IID Data: A Survey}.
\newblock \bibinfo{journal}{\emph{Neurocomput.}} \bibinfo{volume}{465}, \bibinfo{number}{C} (\bibinfo{date}{Nov.} \bibinfo{year}{2021}), \bibinfo{pages}{371--390}.
\newblock
\showISSN{0925-2312}
\urldef\tempurl%
\url{https://doi.org/10.1016/j.neucom.2021.07.098}
\showDOI{\tempurl}


\bibitem[Zhu et~al\mbox{.}(2019)]%
        {bibliography:deep-leakage-from-gradients}
\bibfield{author}{\bibinfo{person}{Ligeng Zhu}, \bibinfo{person}{Zhijian Liu}, {and} \bibinfo{person}{Song Han}.} \bibinfo{year}{2019}\natexlab{}.
\newblock \showarticletitle{Deep Leakage from Gradients}. In \bibinfo{booktitle}{\emph{Advances in Neural Information Processing Systems}}, \bibfield{editor}{\bibinfo{person}{H.~Wallach}, \bibinfo{person}{H.~Larochelle}, \bibinfo{person}{A.~Beygelzimer}, \bibinfo{person}{F.~d\textquotesingle Alch\'{e}-Buc}, \bibinfo{person}{E.~Fox}, {and} \bibinfo{person}{R.~Garnett}} (Eds.), Vol.~\bibinfo{volume}{32}. \bibinfo{publisher}{Curran Associates, Inc.}, \bibinfo{address}{Vancouver, British Columbia, Canada}.
\newblock
\urldef\tempurl%
\url{https://proceedings.neurips.cc/paper/2019/file/60a6c4002cc7b29142def8871531281a-Paper.pdf}
\showURL{%
\tempurl}


\end{thebibliography}

\begin{acronym}
    \acro{ane}[ANE]{Apple Neural Engine}
    \acro{batchnorm}[BatchNorm]{batch normalization}
    \acro{blstm}[BLSTM]{bi-directional long short-term memory}
    \acro{brnn}[BRNN]{bi-directional recurrent neural network}
    \acro{celeba}[CelebA]{Large-scale CelebFaces Attributes Dataset}
    \acro{cnn}[CNN]{convolutional neural network}
    \acro{deepcabac}[DeepCABAC]{Deep Context-Adaptive Binary Arithmetic Coding}
    \acro{dl}[DL]{deep learning}
    \acro{dnn}[DNN]{deep neural network}
    \acro{emnist}[EMNIST]{Extended MNIST}
    \acro{eu}[EU]{European Union}
    \acroplural{eu}[EU's]{European Union's}
    \acro{fedavg}[FedAvg]{federated averaging}
    \acro{fedcat}[FedCat]{federated learning via device concatenation}
    \acro{feddc}[FedDC]{federated daisy-chaining}
    \acro{fedq}[FedQ]{federated learning with client queuing}
    \acro{fedrec}[FedRec]{federated recommender system}
    \acroplural{fedrec}[FedRec's]{federated recommender systems}
    \acro{fedseq}[FedSeq]{federated learning via sequential superclients training}
    \acro{fedsgd}[FedSGD]{federated stochastic gradient descent}
    \acro{femnist}[FEMNIST]{Federated EMNIST}
    \acro{fl}[FL]{federated learning}
    \acro{gcn}[GCN]{graph convolutional network}
    \acro{gdpr}[GDPR]{General Data Protection Regulation}
    \acro{glove}[GloVe]{Global Vectors for Word Representation}
    \acro{groupnorm}[GroupNorm]{group normalization}
    \acro{gru}[GRU]{gated recurrent unit}
    \acro{iid}[i.i.d.]{independent and identically distributed}
    \acro{lstm}[LSTM]{long short-term memory}
    \acroplural{lstm}[LSTMs]{long short-term memory networks}
    \acro{maml}[MAML]{model-agnostic meta-learning}
    \acro{ml}[ML]{machine learning}
    \acro{mnist}[MNIST]{Modified NIST Database}
    \acro{mse}[MSE]{mean squared error}
    \acro{ncf}[NCF]{neural collaborative filtering}
    \acro{nist}[NIST]{National Institute of Standards and Technology}
    \acro{nn}[NN]{neural network}
    \acro{nnc}[NNC]{neural network coding}
    \acro{psnr}[PSNR]{peak signal-to-noise ratio}
    \acro{qp}[QP]{quantization parameter}
    \acro{recsys}[RecSys]{recommender system}
    \acroplural{recsys}[RecSys's]{recommender systems}
    \acro{relu}[ReLU]{rectified linear unit}
    \acro{rnn}[RNN]{recurrent neural network}
    \acro{sgd}[SGD]{stochastic gradient descent}
    \acro{soc}[SoC]{System on a Chip}
    \acro{tmdb}[TMDb]{The Movie Database}
    \acro{uml}[UML]{Unified Modeling Language}
    \acro{vod}[VoD]{video-on-demand}
\end{acronym}

\clearpage

\appendix

\section{Federated Learning Simulator}
\label{appendix:federated-learning-simulator}

The unique requirements for the \ac{fedrec} presented in this work, especially the considerable number of \ac{fl} clients involved, render its experimental evaluation exceedingly difficult. Performing the experiments under real-world conditions, i.e., deploying real devices that communicate with the central server via a network connection, was infeasible. Simulating the \ac{fl} process is common in research, since most of the time the algorithmic and methodological underpinnings of the process are to be researched. Simulating an \ac{fl} system with the required number of clients proved to be challenging, however, as simultaneously keeping all clients, their local datasets, and their local models in main memory is impractical. This meant that some concessions had to be made in order to be able to perform the simulations.

The first and most obvious concession is that the clients must be strictly trained sequentially, which increases training times considerably, but allows for the training of the clients on limited computing resources. This enables the simulator to run on hardware whose processing capabilities allow for the training of at least one client.

The second concession is that the clients cannot remain in main memory at the same time. In fact, because of the potential high amount of data involved in the training process, intermediate results can also not be stored on hard disks. \Ac{fl} clients are usually comprised of the following data: a local dataset, a local model, and the data required by the optimizer. At the very least, an optimizer must store the gradient of the loss function with respect to the weights, whose size is equal to the size of the \ac{nn} model itself. Furthermore, some optimization algorithms require the storage of additional information. For example, the Adam optimizer~\cite{bibliography:adam-a-method-for-stochastic-optimization} also stores estimates of the first and second moments of the gradient, which are both equal in size to the gradient. Finally, the central server must store the global model, as well as all client updates, which are each equal in size to the global model. Although a single \ac{recsys} \ac{nn} model is only tens of Megabytes in size, this can add up to an inhibitive amount of data. For example, in the federated training of the candidate generator \ac{nn} model with more than 162.000 clients, the amount of storage required for all clients adds up to approximately 32 terabytes when using the regular \ac{sgd} optimizer~\cite{bibliography:a-stochastic-optimization-method, bibliography:stochastic-estimation-of-the-maximum-of-a-regression-function} and approximately 52 terabytes when using the Adam optimizer. Considering these memory requirements, it becomes obvious that this is not a viable option for the simulation.

The \ac{fl} simulator employs multiple simple improvements to circumvent the need to keep all data in the main memory simultaneously. Since the clients are trained sequentially, they do not have to keep their local datasets and models in memory simultaneously. At the beginning of each training round, the central server sends the parameters of the global model to the current client, which will load its local dataset and instantiate a local model using the parameters of the global model before starting training. After the local training has finished, the client sends the updated parameters of its local model back to the central server and frees up the memory resources occupied by its local dataset and model.

The second improvement is to not store all client updates on the central server and only aggregate them when all client updates have been received. Instead, a cumulative mean is kept in memory, which is updated each time the central server receives an update from a client. Since the client updates are weighted by the amount of data each client has trained on, this means that the central server must ask all participating clients at the beginning of each training round to reveal the size of their local dataset. In a real-world scenario, each client would send this information when it sends its training updates to the central server, but to be able to keep a cumulative mean, the server must know the local dataset sizes of all clients in advance. The central server then computes the percentage of training data every client contributes to the overall training, which it uses as the weights for the cumulative mean. When a client finishes its local training and sends the training updates back to the central server, the central server multiplies the received update by the weight of the respective client and adds it to the cumulative mean. After all clients have finished their training and the central server has aggregated all their contributions, the cumulative mean is equal to the actual mean of the client updates, which is then used as the updated parametrization of the global model.

Finally, the last improvement is to only use the \ac{sgd} optimizer, as it is stateless and requires no memory at all beyond having to store the gradient. However, since the gradient is volatile and only needs to be stored until it has been applied to the weights of the local model, its memory footprint is as low as possible. Other optimizers, such as Adam, need to store further information, which cannot be discarded and must be kept in memory for the entire duration of the federated training, thus rendering these optimizers unusable. Depending on the model and the training objective, the choice of optimizer can influence convergence time, as well as the final performance of the model. In the present case \ac{sgd} can train the \ac{recsys} models to the same level of performance in a similar amount of time as Adam.

Employed in unison, these improvements not only bring the computational requirements down to a manageable level, but they also reduce the memory footprint to a small fraction of the theoretical requirements. In fact, only the validation dataset and model of the central server, the cumulative mean of the client updates, the local training dataset and model of the current client, and the gradient calculated by the current client's optimizer will ever be in memory at the same time. This reduces the memory requirements of the \ac{fl} simulator from several tens of terabytes down to a few gigabytes.

The temporal complexity of the \ac{fl} simulator is still relatively high, however. Although a round of local training of a single client only requires a couple of minutes, this adds up to a substantial amount of training time considering that the federated training procedure must be repeated dozens or even hundreds of times. This is a problem that is also faced by real-world \ac{fl} systems and is usually solved by client sub-sampling~\cite{bibliography:a-general-theory-for-client-sampling-in-federated-learning, bibliography:optimal-client-sampling-for-federated-learning}, i.e., only selecting a small random subset of clients from the client population for each round of training. This is also the solution that we have chosen: For each round of federated training, we only select between 100 and 10,000 from the more than 162,000 clients. In conclusion, all these measures make it possible to train the \ac{recsys} models using \ac{fl} in a matter of a few days.

\section{Candidate Generator Experiments}
\label{appendix:candidate-generator}

In this appendix we will perform various experiments to determine the optimal model architecture for the candidate generator model described in Section~\ref{section:candidate-generation}, and then provide a detailed explanation of the chosen \ac{nn} architecture.

\subsection{Model Type Experiment}
\label{appendix:candidate-generator-model-type-experiment}

\begin{figure}[ht]
    \centering
    \includegraphics[height=0.225\textheight]{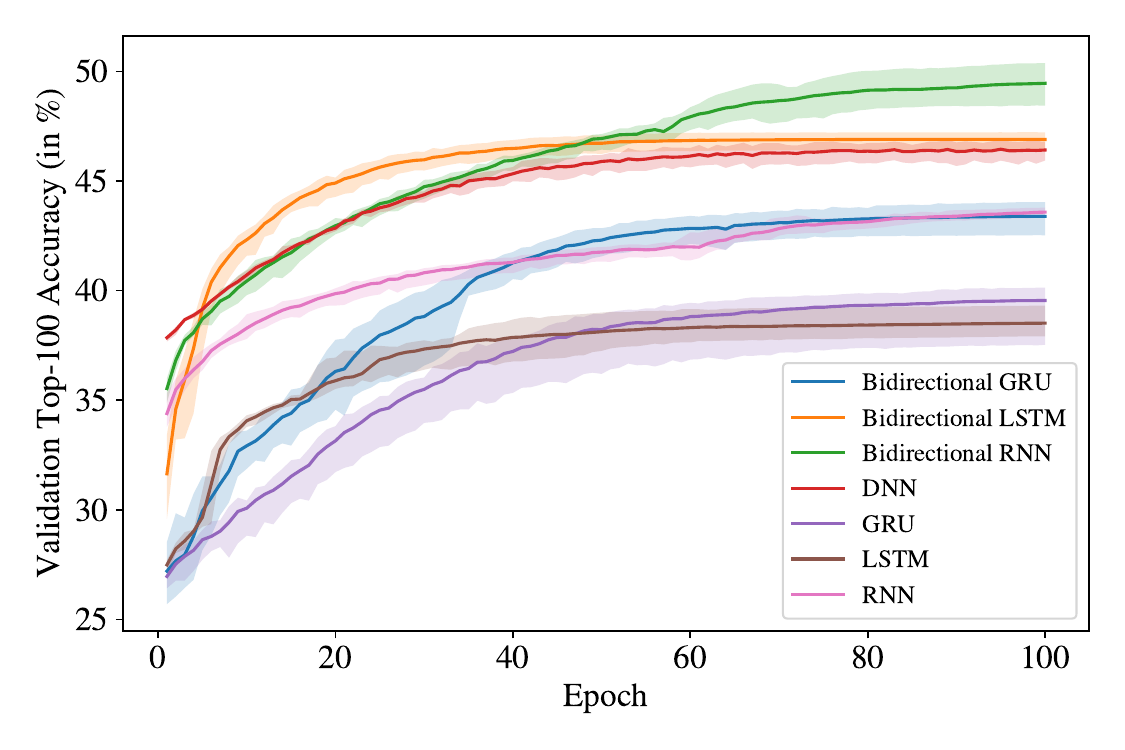}

    \caption{Validation top-100 accuracy results vs. number of epochs for different candidate generator model types.}
    \Description{Validation top-100 accuracy results vs. number of epochs for different candidate generator model types.}
    \label{figure:candidate-generator-model-type-experiment-results}
\end{figure}

There are many possible architectures for candidate generator models based on \acp{nn} ranging from simple \acp{dnn}~\cite{bibliography:dnns-for-youtube-recommendations} and \acp{rnn}~\cite{bibliography:recommendation-system-with-hierarchical-recurrent-neural-network-for-long-term-time-series} to more elaborate autoencoder architectures~\cite{bibliography:collaborative-denoising-auto-encoders-for-top-N-recommender-systems}. Since the \ac{nn} model will be trained using \ac{fl}, the size of the model is a crucial factor. Mobile devices, such as smartphones, are likely candidates for training the \ac{recsys}, as smartphones are the most used devices to watch online \ac{vod} content~\cite{bibliography:devices-used-to-watch-online-video-on-demand-worldwide}. Although some modern smartphones even have dedicated hardware for \ac{nn} training and inference\footnote{For example, the \ac{ane} introduced with the iPhone X's A11 \ac{soc} and Google's Tensor \ac{soc} introduced with the Pixel 6 line of smartphones.}, they are still very resource-constrained as compared to contemporary \ac{ml} hardware. Therefore, only the simplest architectures can be considered for the candidate generator. The most basic \ac{nn} architecture are feed-forward fully-connected \acp{dnn}. However, as the candidate generator will be trained on time-series data, \acp{rnn} would be a more appropriate choice. Therefore, an experiment with a simple feed-forward fully-connected architecture and multiple simple recurrent architectures, including plain \acp{rnn}~\cite{bibliography:fundamentals-of-recurrent-neural-network}, \ac{lstm}~\cite{bibliography:long-short-term-memory} networks, and \acp{gru}~\cite{bibliography:on-the-properties-of-neural-machine-translation:}, was conducted. The recurrent architectures were all trained as both unidirectional and bidirectional models. The results of this experiment are shown in Figure~\ref{figure:candidate-generator-model-type-experiment-results}.

The \ac{lstm} and the \ac{gru} have the worst average performance of the tested model architectures. They clearly show the pitfalls of recurrent architectures: Although the best-performing recurrent architectures reach best-in-class performances, they are tricky to train and show a large variance in training performance. Surprisingly, the \acp{rnn} are the highest performing among the recurrent architectures. Generally, the bi-directional versions of the recurrent architectures outperform their unidirectional counterpart. The feed-forward fully-connected model (denoted as \ac{dnn}) reaches an acceptable performance, which is almost as high as that of the \ac{brnn} or the \ac{blstm}. Just considering the performance of the tested architectures, the \ac{brnn} should be favored, but it also has its downsides: (1) it is the slowest to converge with an average wall clock time of roughly 200 hours as compared to an average wall clock time of roughly 55 hours for the \ac{dnn}, which is almost 4 times as long, and (2) the complexity of the two architectures differs significantly, while the \ac{dnn} only has 17,994,852 trainable parameters, the \ac{brnn} has 128,494,436 trainable parameters, which is more than 7 times as many. The same is true for the \ac{blstm}: It is much slower to converge in terms of wall clock time and is significantly larger. Especially considering that the model must be trained on resource-constrained devices, the simpler but also well-performing \ac{dnn} architecture was selected.

\subsection{Movie Embedding Layer Size Experiment}
\label{appendix:candidate-generator-number-of-movie-embedding-dimensions}

\begin{figure}[ht]
    \centering
    \includegraphics[height=0.225\textheight]{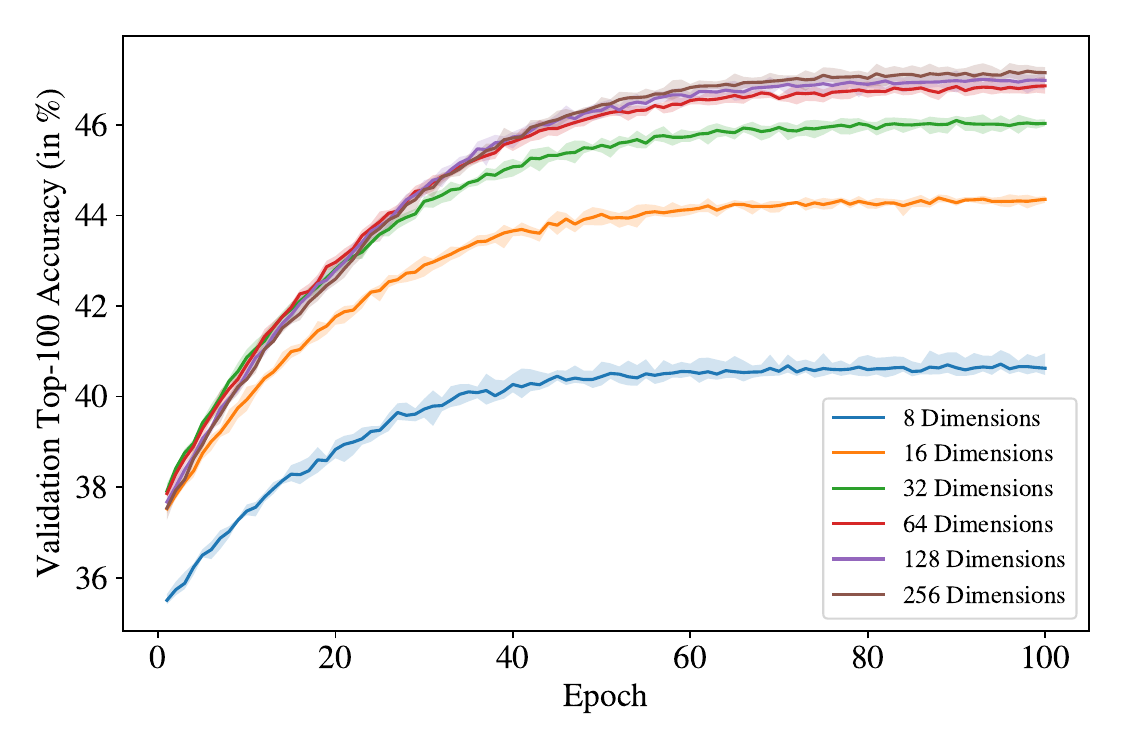}

    \caption{Validation top-100 accuracy results vs. number of epochs for different numbers of dimensions of the movie embedding layer in the candidate generator model.}
    \Description{Validation top-100 accuracy results vs. number of epochs for different numbers of dimensions of the movie embedding layer in the candidate generator model.}
    \label{figure:candidate-generator-number-of-movie-embedding-dimensions-experiment-results}
\end{figure}

The size of the embedding vectors has a substantial impact on the classification result: they cannot fully capture the latent information from the data when they are too small. Additionally, there is a computational cost and a risk of overfitting when they are too large, which means that more data (or regularization) is needed to properly train the model. We determined the optimal size of the embedding vectors experimentally by testing different sizes, as shown in Figure~\ref{figure:candidate-generator-number-of-movie-embedding-dimensions-experiment-results}. The results demonstrate, that increasing the size of the movie embedding vectors directly results in a performance gain, but the return on investment falls off quickly: While doubling from a size of 32 to a size of 64 results in a sizable performance increase of roughly 0.83 percentage points on average, doubling it again to 128 only yields a rise of roughly 0.12 percentage points on average. This means that a 64-dimensional embedding vector provides the best trade-off between performance and model size.

\subsection{Number of Hidden Layers Experiment}
\label{appendix:candidate-generator-number-of-hidden-layers}

\begin{figure}[ht]
    \centering
    \includegraphics[height=0.225\textheight]{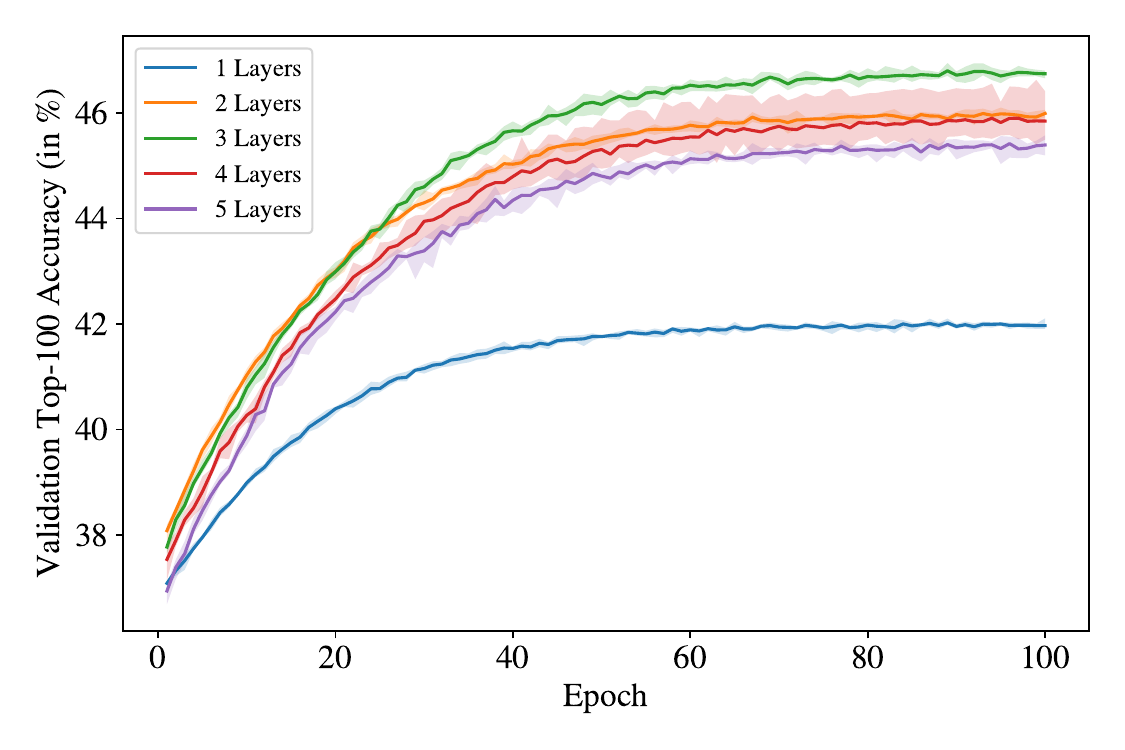}

    \caption{Validation top-100 accuracy results vs. number of epochs for different numbers of hidden layers of the candidate generator model.}
    \Description{Validation top-100 accuracy results vs. number of epochs for different numbers of hidden layers of the candidate generator model.}
    \label{figure:candidate-generator-number-of-hidden-layers-experiment-results}
\end{figure}

Likewise, the number of hidden layers in the candidate generator model also impacts both the performance, as well as the size of the resulting model. We performed an experiment with varying numbers of hidden layers. The results are shown in Figure~\ref{figure:candidate-generator-number-of-hidden-layers-experiment-results}, giving an optimum of a 3-layer configuration, as both increasing and decreasing the number of hidden layers results in inferior performance.

\subsection{Candidate Generator Model Architecture}
\label{appendix:candidate-generator-model-architecture}

The final \ac{nn} architecture that was chosen for the candidate generator has a 64-dimensional embedding layer for the movies in the watch history inputs, followed by three hidden fully-connected layers, which are each followed by a normalization layer and a \ac{relu} activation. The hidden layers with their normalization layers and \ac{relu} activations are then followed by an output fully-connected layer, which feeds its logits into a softmax.

Ever since its introduction, \ac{batchnorm}~\cite{bibliography:batch-normalization-accelerating-deep-network-training-by-reducing-internal-covariate-shift} has been a mainstay in deep learning. Today it is used in a wide variety of \ac{nn} architectures. Unfortunately, \ac{batchnorm} also comes with some drawbacks. First and foremost, \ac{batchnorm} normalizes along the batch dimension, which causes problems with small batch sizes as the estimation of the statistics of a batch become more error prone the smaller the batch size becomes, which can make the training unstable. Especially in \ac{fl} the clients tend to use small batch sizes because of the limited computing power. \Ac{groupnorm}~\cite{bibliography:group-normalization} was introduced to deal with this problem. Instead of estimating the mean and variance of the data based on the batches, it divides the data into groups and measures the statistics within these groups. This makes \ac{groupnorm} independent of the batch size.

Secondly, \ac{batchnorm} keeps, besides its two trainable parameters $\gamma$ and $\beta$, a running average of the mean and the variance of the batches. This makes it complicated to use in \ac{fl}, as the running averages of the mean and the variance cannot be simply averaged. \citet{bilbiography:federated-learning-on-non-iid-features-via-local-batch-normalization} propose to only communicate the trainable parameters of \ac{batchnorm}, i.e., $\gamma$ and $\beta$, to the central server for aggregation but keep the running average of the batch statistics local. \Ac{groupnorm}, however, has the edge over \ac{batchnorm} in this case, as it does not keep a running average of the data statistics and instead always estimates them from the current input.

\citeauthor{bibliography:batch-normalization-damages-federated-learning-on-non-iid-data-analysis-and-remedy} have performed a convergence analysis and were able to show that, although several schemes have been proposed to remedy the problems of \ac{batchnorm} in \ac{fl}, most of them still suffer a loss in performance due to the fact that a mismatch between the local and global statistics, incurred by non-\ac{iid} data distributions, causes a gradient deviation, which in turn leads the model to converge to a biased solution with a slower rate. To avoid all of the above-mentioned problems, we have decided to use \ac{groupnorm} for all \ac{fl} experiments and \ac{batchnorm} for all non-\ac{fl} experiments. A detailed breakdown of the layers that comprise the \ac{nn} architecture of the candidate generator model is presented in Table~\ref{table:candidate-generator-model-architecture}.

\begin{table}[ht]
    {\small
        \begin{tabular}{|llr|}
            \hline
            \textbf{Type}                                & \textbf{Shape}              & \textbf{Parameters}                                                   \\
            \hline

            Embedding Layer                              & $53,797 \times 64$          & 3,443,008                                                             \\

            \Xhline{0.1pt}

            \multirow{2}{*}{Fully-Connected Layer}       & Weights: $1,024 \times 64$  & \multirow{2}{*}{66,560}                                               \\
                                                         & Bias: $1,024$               &                                                                       \\

            \Xhline{0.1pt}

            \multirow{2}{*}{\ac{batchnorm} Layer}        & Gamma: $1,024$              & \multirow{2}{*}{2,048}                                                \\
                                                         & Beta: $1,024$               &                                                                       \\

            \multicolumn{3}{|c|}{\textsb{or}                                                                          \rule{0pt}{3ex} \rule[-1.5ex]{0pt}{0pt}} \\

            \multirow{2}{*}{GroupNorm Layer (32 Groups)} & Gamma: $1,024$              & \multirow{2}{*}{2,048}                                                \\
                                                         & Beta: $1,024$               &                                                                       \\

            \Xhline{0.1pt}

            \multicolumn{3}{|c|}{\ac{relu}                                                                            \rule{0pt}{3ex} \rule[-1.5ex]{0pt}{0pt}} \\

            \Xhline{0.1pt}

            \multirow{2}{*}{Fully-Connected Layer}       & Weights: $512 \times 1,024$ & \multirow{2}{*}{524,800}                                              \\
                                                         & Bias: $512$                 &                                                                       \\

            \Xhline{0.1pt}

            \multirow{2}{*}{\ac{batchnorm} Layer}        & Gamma: $512$                & \multirow{2}{*}{1,024}                                                \\
                                                         & Beta: $512$                 &                                                                       \\

            \multicolumn{3}{|c|}{\textsb{or}                                                                          \rule{0pt}{3ex} \rule[-1.5ex]{0pt}{0pt}} \\

            \multirow{2}{*}{GroupNorm Layer (32 Groups)} & Gamma: $512$                & \multirow{2}{*}{1,024}                                                \\
                                                         & Beta: $512$                 &                                                                       \\

            \Xhline{0.1pt}

            \multicolumn{3}{|c|}{\ac{relu}                                                                            \rule{0pt}{3ex} \rule[-1.5ex]{0pt}{0pt}} \\

            \Xhline{0.1pt}

            \multirow{2}{*}{Fully-Connected Layer}       & Weights: $256 \times 512$   & \multirow{2}{*}{131,328}                                              \\
                                                         & Bias: $256$                 &                                                                       \\

            \Xhline{0.1pt}

            \multirow{2}{*}{\ac{batchnorm} Layer}        & Gamma: $256$                & \multirow{2}{*}{512}                                                  \\
                                                         & Beta: $256$                 &                                                                       \\

            \multicolumn{3}{|c|}{\textsb{or}                                                                          \rule{0pt}{3ex} \rule[-1.5ex]{0pt}{0pt}} \\

            \multirow{2}{*}{GroupNorm Layer (32 Groups)} & Gamma: $256$                & \multirow{2}{*}{512}                                                  \\
                                                         & Beta: $256$                 &                                                                       \\

            \Xhline{0.1pt}

            \multicolumn{3}{|c|}{\ac{relu}                                                                            \rule{0pt}{3ex} \rule[-1.5ex]{0pt}{0pt}} \\

            \Xhline{0.1pt}

            \multirow{2}{*}{Fully-Connected Layer}       & Weights: $53,796 \times 256$ & \multirow{2}{*}{13,825,572}                                          \\
                                                         & Bias: $53,796$               &                                                                      \\

            \Xhline{0.1pt}

            \multicolumn{3}{|c|}{Softmax                                                                              \rule{0pt}{3ex} \rule[-1.5ex]{0pt}{0pt}} \\

            \hline
            \hline

            \multicolumn{2}{|l}{\textsb{Total}}                                         & \textsb{17,994,852}                                                  \\

            \hline
        \end{tabular}
    }

    \caption{A detailed breakdown of the layers that make up the architecture of the candidate generator \ac{nn} model.}
    \Description{A detailed breakdown of the layers that make up the architecture of the candidate generator \ac{nn} model.}
    \label{table:candidate-generator-model-architecture}
\end{table}

\section{Ranker Experiments}
\label{appendix:ranker}

In this appendix we will perform various experiments to determine the optimal model architecture for the ranker model described in Section~\ref{section:ranking}, as well as the loss function that is used for its training. Furthermore, we will provide a detailed explanation of the chosen \ac{nn} architecture.

\subsection{Embedding Layer Sizes Experiment}
\label{appendix:ranker-embedding-layer-sizes}

\begin{figure}[ht]
    \centering

    \begin{subfigure}{\textwidth}
        \centering
        \includegraphics[height=0.225\textheight]{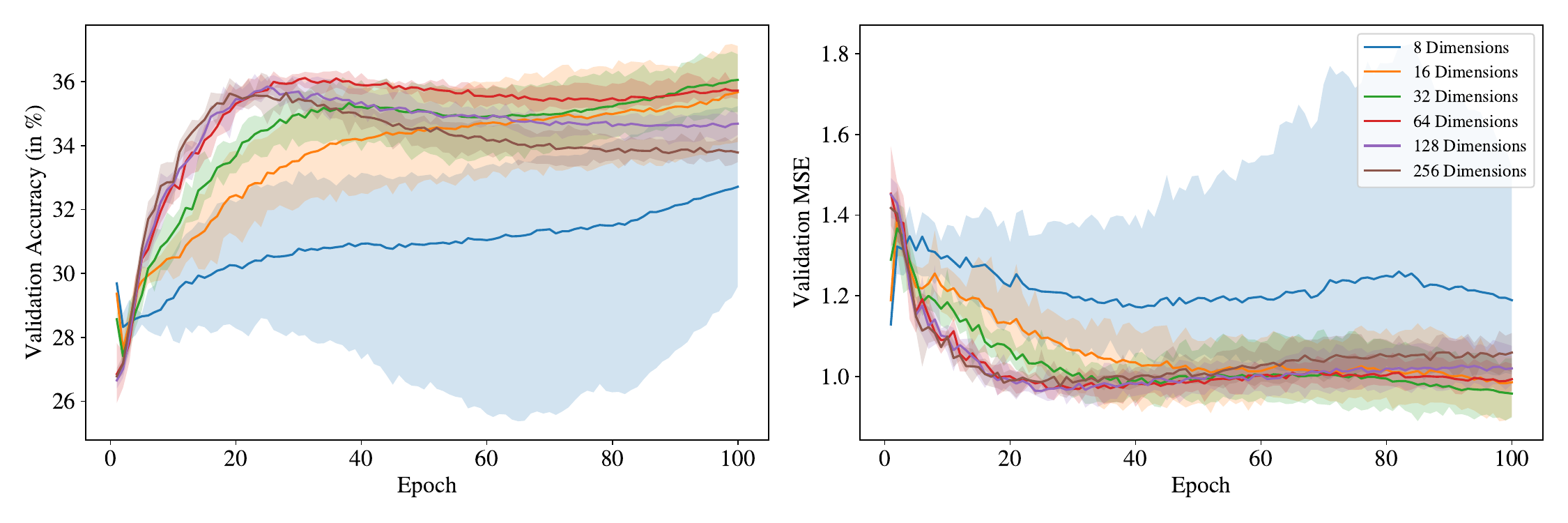}
        \caption{User Embedding Layer}
        \label{subfigure:ranker-number-of-user-embedding-dimensions}
    \end{subfigure}
    \begin{subfigure}{\textwidth}
        \centering
        \includegraphics[height=0.225\textheight]{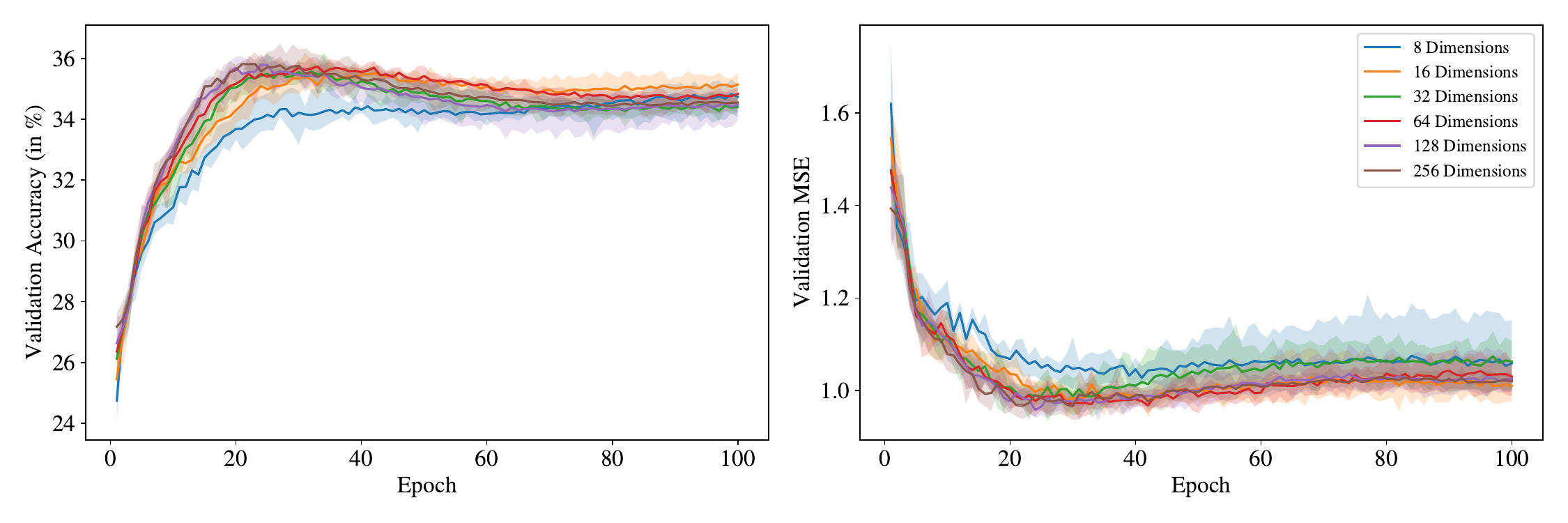}
        \caption{Movie Embedding Layer}
        \label{subfigure:ranker-number-of-movie-embedding-dimensions}
    \end{subfigure}
    \begin{subfigure}{\textwidth}
        \centering
        \includegraphics[height=0.225\textheight]{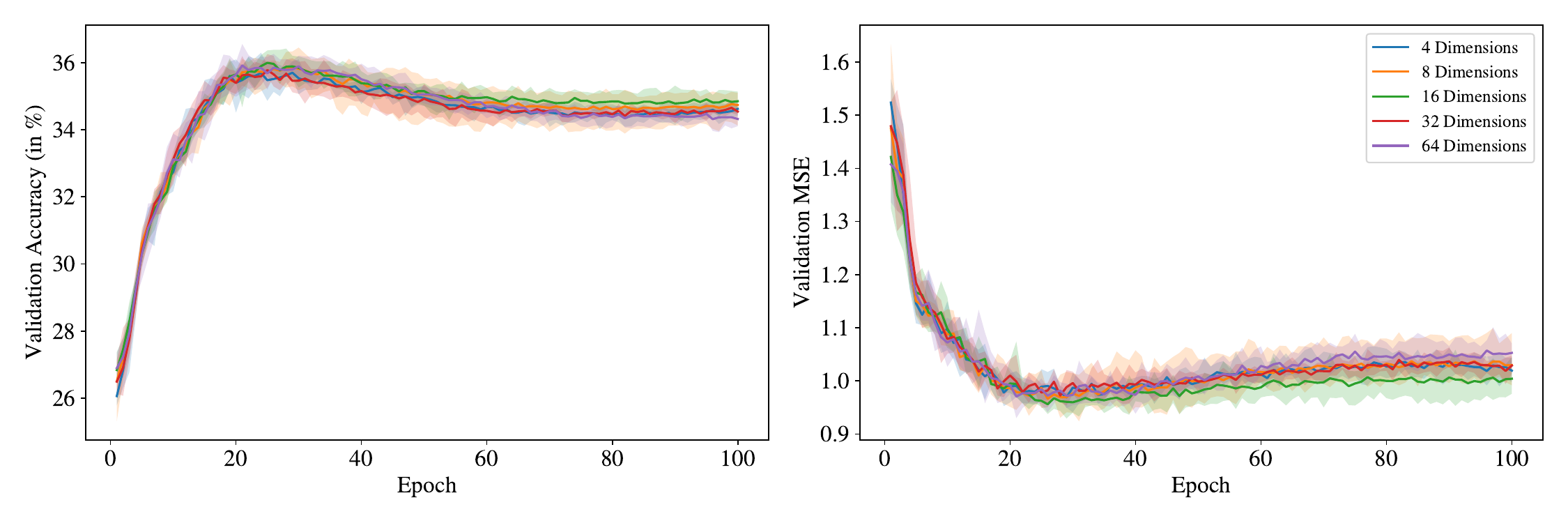}
        \caption{Genre Embedding Layer}
        \label{subfigure:ranker-number-of-genre-embedding-dimensions}
    \end{subfigure}

    \caption{Validation accuracy (\textit{left}) and \ac{mse} results (\textit{right}) vs. number of epochs for different embedding vector sizes for the three embedding layers in the ranker model: \textbf{(\subref{subfigure:ranker-number-of-user-embedding-dimensions})} user embedding layer, \textbf{(\subref{subfigure:ranker-number-of-movie-embedding-dimensions})} movie embedding layer, and \textbf{(\subref{subfigure:ranker-number-of-genre-embedding-dimensions})} genre embedding layer. Each graph shows the minimum and maximum (given by the transparent region), as well as the mean (given by the solid line) of five repetitions of each experiment.}
    \Description{Validation accuracy (\textit{left}) and \ac{mse} results (\textit{right}) vs. number of epochs for different embedding vector sizes for the three embedding layers in the ranker model: \textbf{(\subref{subfigure:ranker-number-of-user-embedding-dimensions})} user embedding layer, \textbf{(\subref{subfigure:ranker-number-of-movie-embedding-dimensions})} movie embedding layer, and \textbf{(\subref{subfigure:ranker-number-of-genre-embedding-dimensions})} genre embedding layer. Each graph shows the minimum and maximum (given by the transparent region), as well as the mean (given by the solid line) of five repetitions of each experiment.}
    \label{figure:ranker-number-of-embedding-dimensions-experiment-results}
\end{figure}

Again, the embedding vector sizes for the three embedding layers for the users, the movies, and the movie genres, must be fine-tuned. Having too large embeddings may result in larger model sizes, overfitting, and longer convergence times. Therefore, we experimentally determined the optimal size of embedding vectors for each embedding layer. As can be seen in Figure~\ref{figure:ranker-number-of-embedding-dimensions-experiment-results}, the optimal embedding sizes are 32 for users, 128 for movies, and 16 for genres. In the case of the user and the genre embedding vector sizes, the experiments are clearly determined, as 32 dimensions and 16 dimensions outperform all other embedding vector sizes both in terms of best final accuracy and \ac{mse}, as well as in best overall accuracy and \ac{mse}, respectively. The results of the movie embedding vector size are a bit ambiguous, as 16 dimensions outperform the other embedding vector sizes in terms of final accuracy and \ac{mse}, however, both 128 dimensions and 256 dimensions yield the highest overall accuracies and \acp{mse}. To balance accuracy and computational complexity, we selected 128 dimensions with a higher overall accuracy and \ac{mse} than the 16-dimension-case and less computational complexity than the 256-dimension case.

\subsection{Number of Hidden Layers Experiment}
\label{appendix:ranker-number-of-hidden-layers}

\begin{figure}[ht]
    \centering
    \includegraphics[height=0.225\textheight]{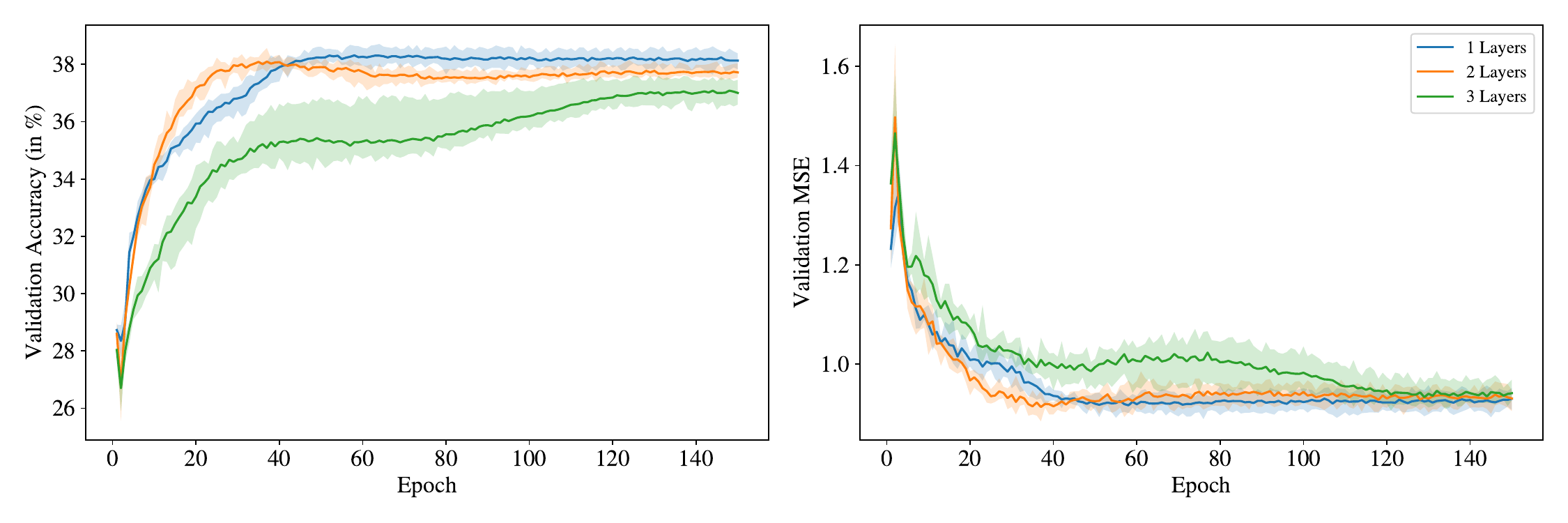}

    \caption{Validation top-100 accuracy (\textit{left}) and \ac{mse} results (\textit{right}) vs. number of epochs for different numbers of hidden layers for the ranker model. Each of the graphs show the minimum and maximum (given by the transparent region), as well as the mean (given by the solid line) of five repetitions of each experiment.}
    \Description{Validation top-100 accuracy (\textit{left}) and \ac{mse} results (\textit{right}) vs. number of epochs for different numbers of hidden layers for the ranker model. Each of the graphs show the minimum and maximum (given by the transparent region), as well as the mean (given by the solid line) of five repetitions of each experiment.}
    \label{figure:ranker-number-of-hidden-layers-experiment-results}
\end{figure}

Similar to the candidate generator, we also determine the optimal number of hidden layers, as shown in Figure~\ref{figure:ranker-number-of-hidden-layers-experiment-results}, resulting in an optimal ranker model with 1 hidden layer. The ranker model with 2 layers converges faster than the ranker model with 1 layer, however, at the expense of a lower final accuracy. Using 3 hidden layers already introduces overfitting and lowers the accuracy further. An optimal model with only 1 hidden layer also requires less computational complexity and is thus beneficial in the \ac{fl} setting.

\subsection{Loss Function Experiment}
\label{appendix:ranker-loss-function}

\begin{figure}[ht]
    \centering
    \includegraphics[height=0.225\textheight]{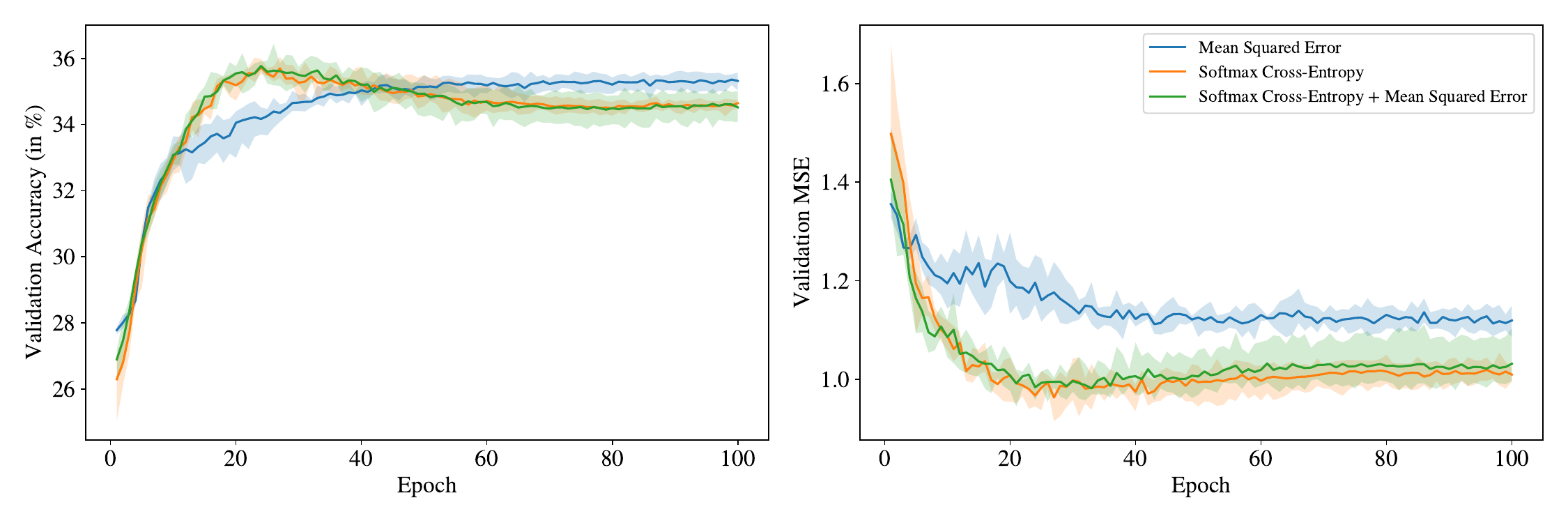}

    \caption{Validation top-100 accuracy (\textit{left}) and \ac{mse} results (\textit{right}) vs. number of epochs for different loss functions for training the ranker model. Each of the graphs shows the minimum and maximum (given by the transparent region), as well as the mean (given by the solid line) of five repetitions of each experiment.}
    \Description{Validation top-100 accuracy (\textit{left}) and \ac{mse} results (\textit{right}) vs. number of epochs for different loss functions for training the ranker model. Each of the graphs shows the minimum and maximum (given by the transparent region), as well as the mean (given by the solid line) of five repetitions of each experiment.}
    \label{figure:ranker-loss-function-experiment-results}
\end{figure}

As the ranker model is trained to perform a classification task, the softmax cross-entropy loss function can be used. However, unlike in a typical classification problem, we want our prediction to be close to the correct value, even if it is wrong (predicting a rating of 3.5, when the actual ground-truth rating is 4.0 is still better than predicting a rating of 0.5, because the deviation from the true rating is smaller). Therefore, other loss functions such as \ac{mse}, which penalize both incorrect predictions and the magnitude of the deviation, may be better suited. To determine this, we conducted experiments using softmax cross-entropy, \ac{mse}, and the sum of the two to combine the best of both approaches. The results are shown in Figure~\ref{figure:ranker-loss-function-experiment-results}. Against our expectations, the \ac{mse} loss function performs worse than the other two in terms of validation \ac{mse}. Here, one would assume that a model optimized on the \ac{mse} loss function should perform best when measuring its performance in terms of \ac{mse}. Although using the \ac{mse} loss function causes the model to not overfit in terms of accuracy like the other loss functions and outperforms them when measuring the final accuracy of the model, the other loss functions converge faster and achieve a better overall accuracy. The softmax cross-entropy loss function and the sum of both yield a similar accuracy, however the softmax cross-entropy loss function is computationally less complex and was therefore selected.

\subsection{Ranker Model Architecture}
\label{appendix:ranker-model-architecture}

The final \ac{nn} architecture that was chosen for the ranker has a 32-dimensional embedding layer for the user, a 128-dimensional embedding layer for the movie, and a 16-dimensional embedding layer for the genres. The genres are then averaged and all inputs, including the embeddings and the movie age, are concatenated. This is followed by a single hidden fully-connected layer. The output of the hidden layer is normalized using a normalization layer, which is followed by a \ac{relu} activation. The hidden layer with its normalization layer and \ac{relu} activation is then followed by an output fully-connected layer, which feeds its logits into a softmax. For the reasons described in Appendix~\ref{appendix:candidate-generator-model-architecture}, a \ac{groupnorm} layer is used for the \ac{fl} experiments and a \ac{batchnorm} layer is used for the non-\ac{fl} experiments. A detailed breakdown of the layers that comprise the \ac{nn} architecture of the ranker model is presented in Table~\ref{table:ranker-model-architecture}.

\begin{table}[ht]
    {\small
        \begin{tabular}{|llr|}
            \hline
            \textbf{Type}                                & \textbf{Shape}            & \textbf{Parameters}                                              \\
            \hline

            User Embedding Layer                         & $162,541 \times 32$       & 5,201,312                                                        \\

            \Xhline{0.1pt}

            Movie Embedding Layer                        & $53,796 \times 128$       & 6,885,888                                                        \\

            \Xhline{0.1pt}

            Genre Embedding Layer                        & $20 \times 16$            & 320                                                              \\

            \Xhline{0.1pt}

            \multicolumn{3}{|c|}{Genre Embedding Average                                                       \rule{0pt}{3ex} \rule[-1.5ex]{0pt}{0pt}} \\

            \Xhline{0.1pt}

            \multicolumn{3}{|c|}{Input Concatenation                                                           \rule{0pt}{3ex} \rule[-1.5ex]{0pt}{0pt}} \\

            \Xhline{0.1pt}

            \multirow{2}{*}{Fully-Connected Layer}       & Weights: $256 \times 177$ & \multirow{2}{*}{45,568}                                          \\
                                                        & Bias: $256$               &                                                                  \\

            \Xhline{0.1pt}

            \multirow{2}{*}{\ac{batchnorm} Layer}        & Gamma: $256$              & \multirow{2}{*}{512}                                             \\
                                                        & Beta: $256$               &                                                                  \\

            \multicolumn{3}{|c|}{\textsb{\quad \ or}                                                           \rule{0pt}{3ex} \rule[-1.5ex]{0pt}{0pt}} \\

            \multirow{2}{*}{GroupNorm Layer (32 Groups)} & Gamma: $256$              & \multirow{2}{*}{512}                                             \\
                                                        & Beta: $256$               &                                                                  \\

            \Xhline{0.1pt}

            \multicolumn{3}{|c|}{\ac{relu}                                                                     \rule{0pt}{3ex} \rule[-1.5ex]{0pt}{0pt}} \\

            \Xhline{0.1pt}

            \multirow{2}{*}{Fully-Connected Layer}       & Weights: $10 \times 256$  & \multirow{2}{*}{2,570}                                           \\
                                                        & Bias: $10$                &                                                                  \\

            \Xhline{0.1pt}

            \multicolumn{3}{|c|}{Softmax                                                                       \rule{0pt}{3ex} \rule[-1.5ex]{0pt}{0pt}} \\

            \hline
            \hline

            \multicolumn{2}{|l}{\textsb{Total}}                                      & \textsb{12,136,170}                                              \\

            \hline
        \end{tabular}
    }

    \caption{A detailed breakdown of the layers that make up the architecture of the ranker \ac{nn} model.}
    \Description{A detailed breakdown of the layers that make up the architecture of the ranker \ac{nn} model.}
    \label{table:ranker-model-architecture}
\end{table}

\section{Extended Federated Learning and FedQ Experiment Results}
\label{appendix:extended-federated-learning-and-fedq-experiment-results}

\begin{figure}[ht]
    \centering

    \begin{subfigure}{\textwidth}
        \centering
        \includegraphics[width=\textwidth]{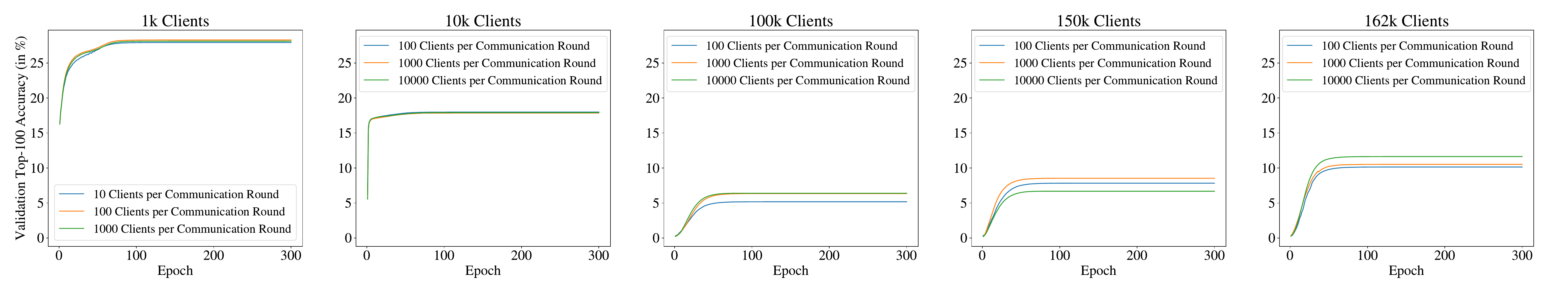}
        \caption{Candidate generator}
        \label{subfigure:federated-learning-experiment-results-candidate-generator}
    \end{subfigure}%
    \hfill
    \begin{subfigure}{\textwidth}
        \centering
        \includegraphics[width=\textwidth]{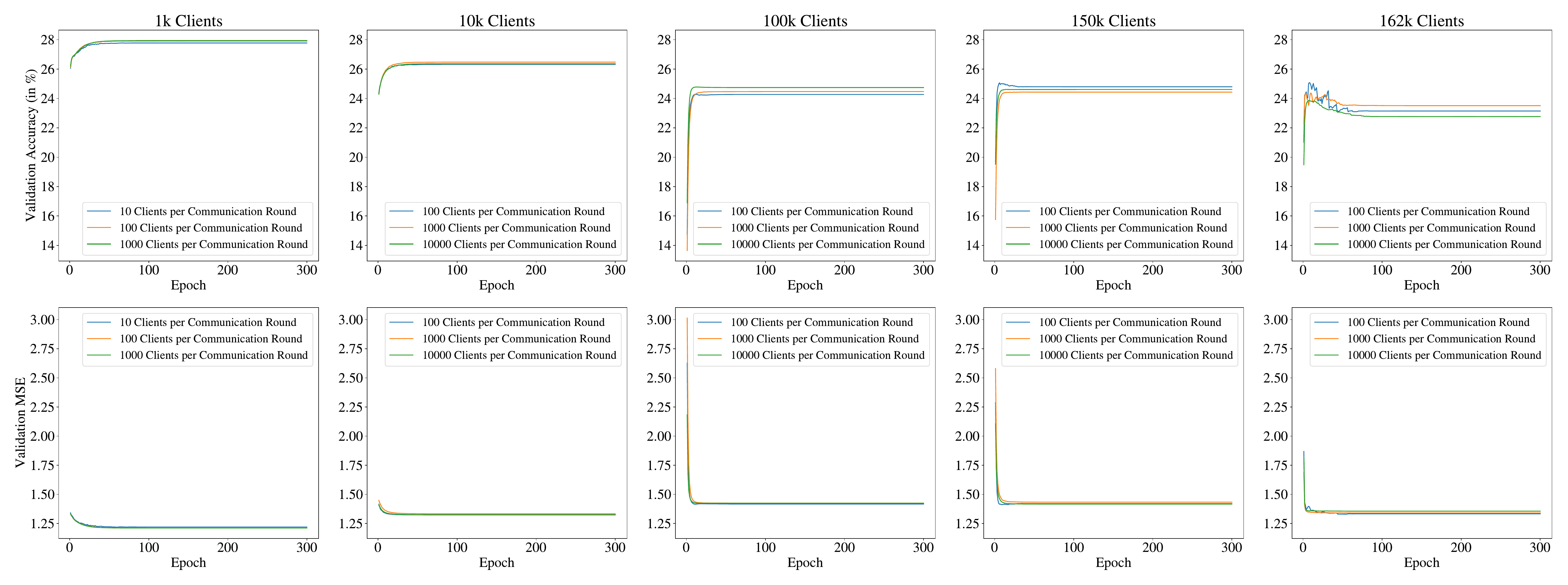}
        \caption{Ranker}
        \label{subfigure:federated-learning-experiment-results-ranker}
    \end{subfigure}

    \caption{\Ac{fl} experiment results for \textbf{(\subref{subfigure:federated-learning-experiment-results-candidate-generator})} the candidate generator and \textbf{(\subref{subfigure:federated-learning-experiment-results-ranker})} the ranker.}
    \Description{\Ac{fl} experiment results for \textbf{(\subref{subfigure:federated-learning-experiment-results-candidate-generator})} the candidate generator and \textbf{(\subref{subfigure:federated-learning-experiment-results-ranker})} the ranker.}
    \label{figure:federated-learning-experiment-results}
\end{figure}

\begin{figure}[ht]
    \centering

    \begin{subfigure}{\textwidth}
        \centering
        \includegraphics[width=\textwidth]{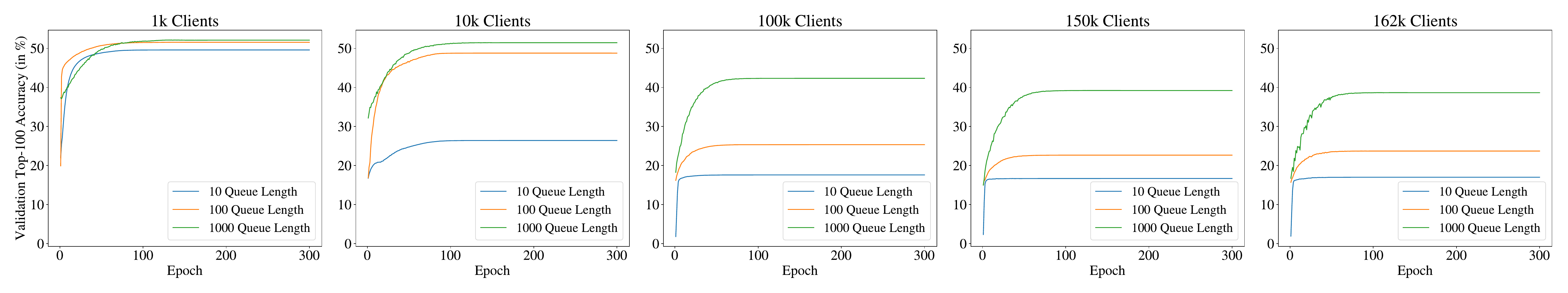}
        \caption{Candidate generator}
        \label{subfigure:fedq-experiment-results-candidate-generator}
    \end{subfigure}%
    \hfill
    \begin{subfigure}{\textwidth}
        \centering
        \includegraphics[width=\textwidth]{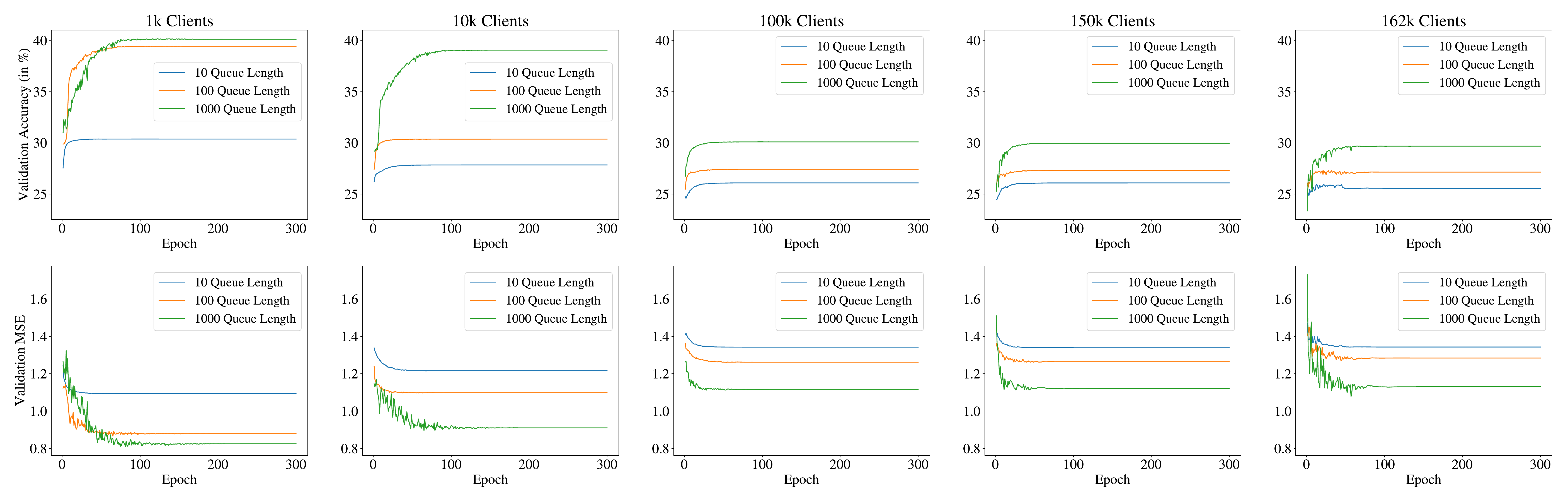}
        \caption{Ranker}
        \label{subfigure:fedq-experiment-results-ranker}
    \end{subfigure}

    \caption{\Ac{fedq} experiment results for \textbf{(\subref{subfigure:fedq-experiment-results-candidate-generator})} the candidate generator and \textbf{(\subref{subfigure:fedq-experiment-results-ranker})} the ranker.}
    \Description{\Ac{fedq} experiment results for \textbf{(\subref{subfigure:fedq-experiment-results-candidate-generator})} the candidate generator and \textbf{(\subref{subfigure:fedq-experiment-results-ranker})} the ranker.}
    \label{figure:fedq-experiment-results}
\end{figure}

Due to the broad range of different numbers of clients for the \ac{fl} and \ac{fedq} experiments, the complete training graphs are poorly legible and were therefore omitted from Section~\ref{section:federated-learning-experiments} and Section~\ref{section:fedq-experiments}. Instead, only the final validation top-100 accuracy for the candidate generator experiments, as well as the final accuracy and \ac{mse} for the ranker experiments were presented. For reference, the complete training graphs are included in this appendix as Figure~\ref{figure:federated-learning-experiment-results} and Figure~\ref{figure:fedq-experiment-results}.

\section{Validation of FedQ on the LEAF Federated Learning Benchmark}
\label{appendix:validation-of-fedq-on-the-leaf-federated-learning-benchmark}

Although \ac{fedq} was developed in the context of a \ac{fedrec}, it is a much more general algorithm that can be employed in other \ac{fl} pipelines that have to deal with small local datasets. To provide further evidence of \ac{fedq}'s efficacy, it was evaluated on LEAF, which is an open-source, modular benchmarking framework for federated settings~\cite{bibliography:leaf-a-benchmark-for-federated-settings}. It consists of (1) multiple open-source datasets, (2) reference implementations for common \ac{fl} methods, and (3) several metrics that measure the statistical properties of the models that are being trained (e.g., accuracy), as well as metrics that measure properties of the \ac{fl} system (e.g., number of communicated bytes and local computation). The reference implementation currently includes scripts for preprocessing the data, the federated optimization algorithms \ac{fedsgd} and \ac{fedavg}, and one or more model architectures for each of the included datasets. We based the evaluation of \ac{fedq} on the following datasets contained in LEAF:

\begin{itemize}
    \item \textbf{\Ac{femnist}}~\cite{bibliography:leaf-a-benchmark-for-federated-settings} is a dataset that was created by the authors of the LEAF benchmark by partitioning the digit and character images of the \ac{emnist}~\cite{bibliography:emnist-extending-mnist-to-handwritten-letters} dataset by the person that wrote it. This partitioning makes the dataset more amenable to \ac{fl}, since writers can be understood as clients. \Ac{emnist} is a dataset that was created from the \ac{nist} Special Database 19~\cite{bibliography:nist-special-database-19-handprinted-forms-and-characters-database}, which is the same database that the popular \ac{mnist}~\cite{bibliography:gradient-based-learning-applied-to-document-recognition} is based on. The \ac{nist} Special Database 19 contains handwritten digits, uppercase letters, and lowercase letters, which is much more data than what is exposed by \ac{mnist}. \Ac{emnist} was created in an effort to create a more challenging benchmark dataset by covering all data contained in the \ac{nist} Special Database 19, while employing the same conversion paradigm used for \ac{mnist} to stay compatible.
    \item \textbf{\Ac{celeba}}~\cite{bibliography:deeplearning-face-attributes-in-the-wild} is a dataset, which contains images of celebrities that were annotated with 40 attributes, including wearing eyeglasses, wearing a hat, wavy hair, and smiling. For the LEAF benchmark, \ac{celeba} was adapted to the federated setting by partitioning it into client datasets based on the celebrity in the image. Furthermore, the classification task was simplified from a multi-label classification task to a binary classification task, which only distinguishes between smiling and not smiling celebrities.
    \item \textbf{Sentiment140}~\cite{bibliography:twitter-sentiment-classification-using-distant-supervision} is an automatically generated dataset that contains Twitter messages that are classified as either positive or negative based on the emoticons contained in them. The dataset therefore presents a binary classification sentiment analysis task, where the input is a sequence of words. For the use in the LEAF benchmark, the messages are partitioned, such that each \ac{fl} client is represented by a different Twitter user.
    \item \textbf{Reddit}~\cite{bibliography:leaf-a-benchmark-for-federated-settings} is a dataset that was created by the authors of the LEAF benchmark. They took comments posted on the social network Reddit in December 2017 and preprocessed them by (1) converting all named and numeric HTML character references to their corresponding unicode characters, (2) removing extraneous white spaces, (3) removing non-ASCII characters, (4) replacing URLs, Reddit user names and Subreddit names with special tokens, (5) converting the text to lowercase, and (6) tokenizing it using NLTK's~\cite{bibliography:natural-language-processing-with-python} tweet tokenizer. Furthermore, users that were determined to be bots, or that had less than 5 or more than 1000 comments were removed, along with their comments. \citeauthor{bibliography:leaf-a-benchmark-for-federated-settings} sub-sampled the dataset for their own experiments, as their reference implementation is not yet capable of training on the complete Reddit dataset. The training task of the dataset is next word prediction with a sequence of previous words as input. Each Reddit user is considered to be an \ac{fl} client.
\end{itemize}

The LEAF benchmark provides two more datasets: \emph{Shakespeare}~\cite{bibliography:communication-efficient-learning-of-deep-networks-from-decentralized-data}, which is a dataset that is based on \emph{``The Complete Works of William Shakespeare''}~\cite{bibliography:the-complete-works-of-william-shakespeare}, where each speaking role represents an \ac{fl} client, and a \emph{synthetic} dataset that is based on the synthetic dataset proposed by~\citet{bibliography:fair-ressource-allocation-in-federated-learning}. The Shakespeare dataset comprises 4,226,158 samples across 1,129 \ac{fl} clients (i.e., speaker roles). On average, each client has 3,743.28 samples with a standard deviation of 6,212.26. \Ac{fedq} is specifically tailored towards federated scenarios with small local datasets, therefore, the Shakespeare dataset is inadequate for evaluating \ac{fedq}'s potency, as almost all clients have plenty of local data (only 8 clients have less than 10 and 114 clients have less than 100 samples in their local datasets). The synthetic dataset was specifically designed by~\citeauthor{bibliography:leaf-a-benchmark-for-federated-settings} to create a more challenging task for meta-learning methods, which does not apply to \ac{fedq}. For these reasons, we decided to only evaluate \ac{fedq} on the four above-mentioned datasets.

To conduct the \ac{fedq} benchmark experiments, we used the LEAF reference implementation and integrated \ac{fedq} as a new federated optimization algorithm. For the \ac{femnist} dataset, we use a simple two-layer \ac{cnn}, which consists of two convolution layers each followed by a maximum pooling layer, followed by two fully-connected layers. For the \ac{celeba} dataset, we utilize a \ac{cnn} with four convolution layers, each followed by a \ac{batchnorm} and a maximum pooling layer, followed by a single fully-connected layer. For the Sentiment140 dataset, we use a stacked \ac{lstm} model with an embedding layer that is initialized with 300-dimensional, pre-trained \ac{glove} embeddings, followed by two \ac{lstm} cells and two fully-connected layers. For the Reddit dataset, we rely on a stacked \ac{lstm} model with an embedding layer that embeds the input words into an 8-dimensional vector space, followed by two \ac{lstm} cells with dropout and a single fully-connected layer. All of these models are part of the reference implementation of the LEAF benchmark. The use of pre-trained \ac{glove} embeddings in the stacked \ac{lstm} model used for the Sentiment140 dataset is an adaptation that we incorporated. The embedding layer of the original reference implementation was randomly initialized and trained on the Sentiment140 dataset using the \ac{glove} vocabulary to embed its input words into a 100-dimensional vector space. Without this adaptation, the model fails to learn anything using the hyperparameters proposed by~\citeauthor{bibliography:leaf-a-benchmark-for-federated-settings}. In fact, the model usually settles in on an accuracy of around 50\% after the first round of federated training and more or less keeps that accuracy for the entire duration of the training. For a binary classification task, an accuracy of 50\% is not better than random chance. As a matter of fact,~\citeauthor{bibliography:federated-meta-learning-with-fast-convergence-and-efficient-communication} also made this adaptation in their LEAF benchmark experiments.

The preprocessed and sub-sampled version of the Reddit dataset used by~\citeauthor{bibliography:leaf-a-benchmark-for-federated-settings} was graciously made available for download. All other datasets were preprocessed using the tools provided in the reference implementation of the LEAF benchmark. We used the settings presented in Table~\ref{table:leaf-dataset-preprocessing-settings}. The statistics of the resulting datasets can be seen in Table~\ref{table:leaf-dataset-statistics}.

\begin{table}[ht]
    {\small
        \begin{tabular}{|l|l|l|l|}
            \cline{2-4}
            \multicolumn{1}{r|}{}                                          & \textbf{\Ac{femnist}} & \textbf{\Ac{celeba}} & \textbf{Sentiment140} \\
            \hline

            \textsb{Client Sample Distribution} \hspace{2.5pt} \lstinline$-s$           & non-\ac{iid}          & non-\ac{iid}         & non-\ac{iid}          \\
            \textsb{Fraction of Data to Sample} \hspace{2.5pt} \lstinline$--sf$         & 100\%                 & 100\%                & 15\%                  \\
            \textsb{Minimum Number of Samples per Client} \hspace{2.5pt} \lstinline$-k$ & 0                     & 0                    & 0                     \\
            \textsb{Training/Test Data Split Mode} \hspace{2.5pt} \lstinline$-t$        & Sample                & Sample               & Sample                \\
            \textsb{Training Data Fraction} \hspace{2.5pt} \lstinline$--tf$             & 90\%                  & 90\%                 & 90\%                  \\
            \textsb{Sampling Seed} \hspace{2.5pt} \lstinline$--smplseed$                & 1691607340            & 1691605746           & 1692132357            \\
            \textsb{Split Seed} \hspace{2.5pt} \lstinline$--spltseed$                   & 1691608842            & 1691605747           & 1692132372            \\
            \hline
        \end{tabular}
    }

    \caption{The settings used to preprocess the \ac{femnist}, \ac{celeba}, and Sentiment140 datasets.}
    \Description{The settings used to preprocess the \ac{femnist}, \ac{celeba}, and Sentiment140 datasets.}
    \label{table:leaf-dataset-preprocessing-settings}
\end{table}

\begin{table}[ht]
    {\small
        \begin{tabular}{|l|l|r|r|r|r|}
            \cline{3-6}
            \multicolumn{2}{l|}{}                                             & \textbf{\Ac{femnist}}    & \textbf{\Ac{celeba}}     & \textbf{Sentiment140}                               & \textbf{Reddit}                                      \\
            \hline

            \multicolumn{2}{|l|}{\multirow{2}{*}{\textsb{Number of Clients}}} & \multirow{2}{*}{3,597}   & \multirow{2}{*}{9,343}   & 99,149                                              & 817                                                  \\ [-3pt]
            \multicolumn{2}{|l|}{}                                            &                          &                          & \footnotesize{\textcolor{darkgray}{(of 660,120)}}   & \footnotesize{\textcolor{darkgray}{(of 1,660,820)}}  \\  [2pt]

            \multicolumn{2}{|l|}{\multirow{2}{*}{\textsb{Number of Samples}}} & \multirow{2}{*}{817,851} & \multirow{2}{*}{200,288} & 240,074                                             & 55,556                                               \\ [-3pt]
            \multicolumn{2}{|l|}{}                                            &                          &                          & \footnotesize{\textcolor{darkgray}{(of 1,600,498)}} & \footnotesize{\textcolor{darkgray}{(of 56,587,343)}} \\  [2pt]

            \Xhline{0.1pt}
            \multirow{4}{*}{\textsb{Samples per Client}} & Minimum            & 19                       & 5                        & 1                                                   & 10                                                   \\
                                                         & Maximum            & 584                      & 35                       & 236                                                 & 1,394                                                \\
                                                         & Mean               & 227.37                   & 21.44                    & 2.42                                                & 68.0                                                 \\
                                                         & Standard Deviation & 88.84                    & 7.63                     & 4.63                                                & 120.27                                               \\
            \hline
        \end{tabular}
    }

    \caption{The dataset statistics of the preprocessed \ac{femnist}, \ac{celeba}, Sentiment140, and Reddit datasets.}
    \Description{The dataset statistics of the preprocessed \ac{femnist}, \ac{celeba}, Sentiment140, and Reddit datasets.}
    \label{table:leaf-dataset-statistics}
\end{table}

\begin{figure}[ht]
    \centering
    \includegraphics[height=0.46\textheight]{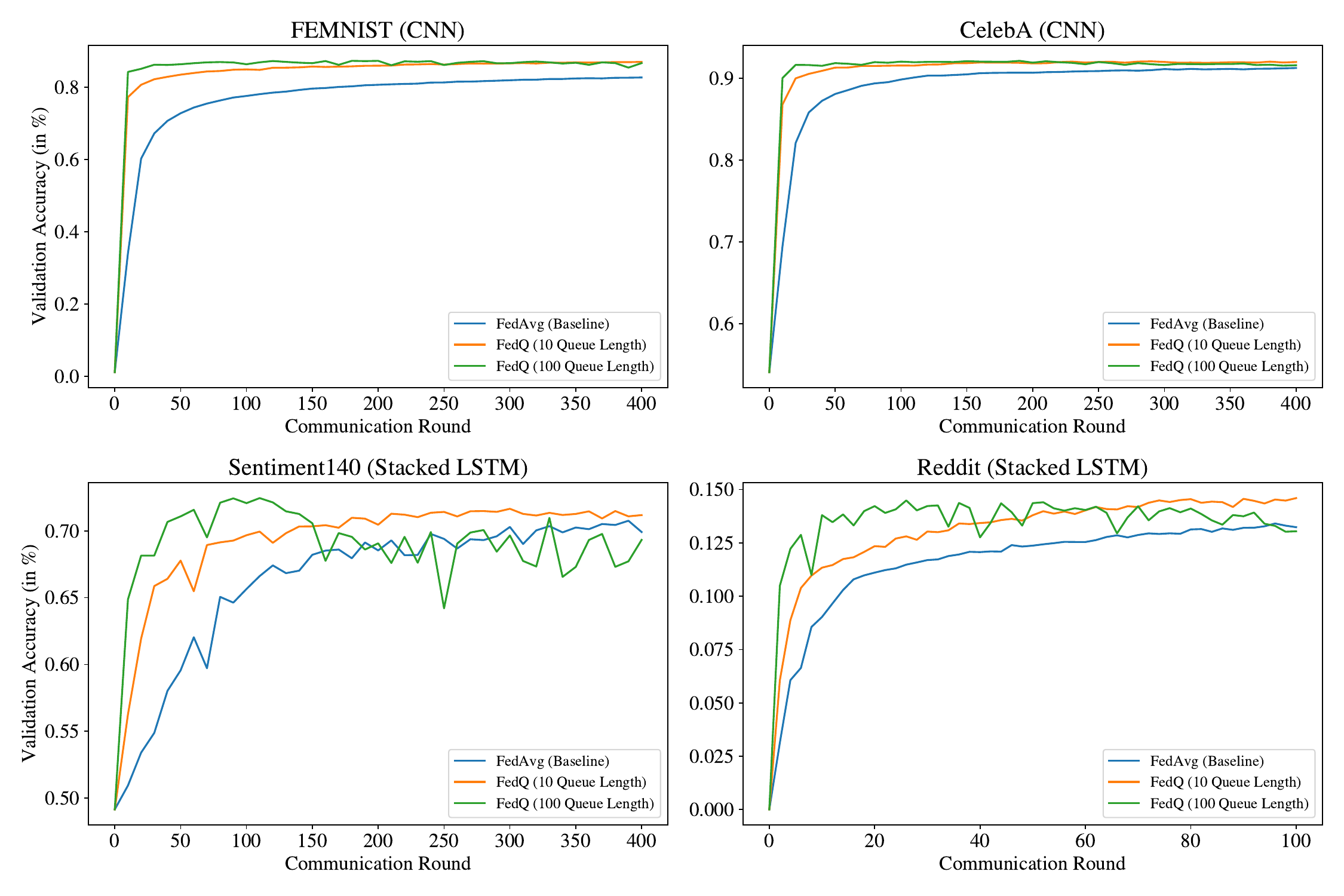}

    \caption{\Ac{fedq} LEAF benchmark experiment results.}
    \Description{\Ac{fedq} LEAF benchmark experiment results.}
    \label{figure:fedq-leaf-benchmark-experiment-results}
\end{figure}

All models were trained using the default random seeds. Most of the remaining hyperparameters, however, deviate from the hyperparameters suggested by~\citeauthor{bibliography:leaf-a-benchmark-for-federated-settings}. Especially the number of clients per communication round was increased to facilitate different queue lengths for \ac{fedq}. The hyperparameters used for each dataset are specified in Table~\ref{table:leaf-fedq-experiment-hyperparameters}. The experiments for each dataset were repeated three times, once with \ac{fedavg} as a baseline against which \ac{fedq} can be compared, once with \ac{fedq} and a queue length of 10, and once with \ac{fedq} and a queue length of 100. The results of the experiments can be seen in Figure~\ref{figure:fedq-leaf-benchmark-experiment-results}.

\begin{table}[ht]
    {\small
        \begin{tabular}{|l|r|r|r|r|}
            \cline{2-5}
            \multicolumn{1}{l|}{}                          & \textbf{\Ac{femnist}} & \textbf{\Ac{celeba}} & \textbf{Sentiment140} & \textbf{Reddit}          \\
            \hline

            \textsb{Communication Rounds}                  & \multirow{2}{*}{400}   & \multirow{2}{*}{400}   & \multirow{2}{*}{400}   & \multirow{2}{*}{100} \\ [-3pt]
            {\footnotesize\lstinline$--num-rounds$}        &                        &                        &                        & {}                   \\  [3pt]

            \textsb{Clients per Communication Round}       & \multirow{2}{*}{1,000} & \multirow{2}{*}{1,000} & \multirow{2}{*}{1,000} & \multirow{2}{*}{500} \\ [-3pt]
            {\footnotesize\lstinline$--clients-per-round$} &                        &                        &                        & {}                   \\  [3pt]

            \textsb{Learning Rate}                         & \multirow{2}{*}{0.01}  & \multirow{2}{*}{0.01}  & \multirow{2}{*}{0.01}  & \multirow{2}{*}{8.0} \\ [-3pt]
            {\footnotesize\lstinline$-lr$}                 &                        &                        &                        & {}                   \\  [3pt]

            \textsb{Batch Size}                            & \multirow{2}{*}{10}    & \multirow{2}{*}{10}    & \multirow{2}{*}{10}    & \multirow{2}{*}{5}   \\ [-3pt]
            {\footnotesize\lstinline$--batch-size$}        &                        &                        &                        & {}                   \\  [3pt]

            \textsb{Local Epochs}                          & \multirow{2}{*}{5}     & \multirow{2}{*}{5}     & \multirow{2}{*}{5}     & \multirow{2}{*}{1}   \\ [-3pt]
            {\footnotesize\lstinline$--num-epochs$}        &                        &                        &                        & {}                   \\  [3pt]

            \hline
        \end{tabular}
    }

    \caption{The hyperparameters that were used for benchmarking \ac{fedq} on LEAF.}
    \Description{The hyperparameters that were used for benchmarking \ac{fedq} on LEAF.}
    \label{table:leaf-fedq-experiment-hyperparameters}
\end{table}

In all experiments, \ac{fedq} with a queue length of 10 had a higher final accuracy than the other two experiments. In the cases of the Sentiment140 and the Reddit datasets, it even manages to clearly outperform \ac{fedq} with a queue length of 100. This is interesting in two ways: First of all, in the \ac{fedq} experiments on our \ac{fedrec}, there was always a benefit when using a larger queue length, albeit with diminishing returns. The LEAF benchmark experiments not only show that using a larger queue length does not always result in a significant increase in performance, but it may even make the training unstable and hinders convergence, as is the case for the Sentiment140 and the Reddit dataset. The second remarkable thing is, that in both cases where a larger queue length causes the training to become unstable, the model is an \ac{lstm}. Of course, no trend can be derived from just these experiments, but this interesting behavior could be explored in future work.

The margins with which \ac{fedq} outperforms the \ac{fedavg} baseline are much smaller as compared to the results achieved with the models of our \ac{fedrec}. Nonetheless, it can be clearly seen that \ac{fedq} has a much faster convergence rate. The \ac{fedavg} baseline reaches its highest accuracy in all cases at the very end of the training window (communication round 400/400 for \ac{femnist}, 390/400 for Sentiment140, 400/400 for \ac{celeba}, and 96/100 for Reddit). Both \ac{fedq} experiments are able to reach or exceed the baseline's highest accuracy in a much shorter time frame: For \ac{femnist}, \ac{fedq} with a queue length of 10 exceeded \ac{fedavg} at communication round 40, while \ac{fedq} with a queue length of 100 already outperformed \ac{fedavg} at communication round 10. For Sentiment140, \ac{fedq} with a queue length of 10 surpasses \ac{fedavg} at communication round 180 and \ac{fedq} with a queue length of 100 at communication round 50. For \ac{celeba}, the communication rounds were 50 and 20, while those for Reddit were 180 and 50 respectively. It should also be noted that, although its training was less stable, \ac{fedq} with a queue length of 100 outperformed the other two experiments in terms of highest accuracy for all datasets except for Reddit. It was also always significantly faster to exceed the highest accuracy of \ac{fedavg} than \ac{fedq} with a queue length of 10. Table~\ref{table:leaf-fedq-experiment-comparison-to-results-from-the-literature} presents the results of our experiments in comparison to the results published by~\citet{bibliography:leaf-a-benchmark-for-federated-settings}. Please be aware that our experiments used different hyperparameters for both the preprocessing of the datasets as well as the training of the models, which renders the results incomparable. Particularly notable is the difference in the Sentiment140 dataset, where \citeauthor{bibliography:leaf-a-benchmark-for-federated-settings} report on four experiments with varying minimum numbers of samples per client ranging from 3 to 100. In our experiments, we have set the minimum number of samples per client to 0 in the preprocessing of Sentiment140, which means that our experiments had a considerably lower number of samples per client on average. We have still included the results for your reference and as another baseline.

\begin{table}[ht]
    {\small
        \begin{tabular}{|l|llr|}
            \hline
            \textbf{Dataset}              & \multicolumn{2}{l}{\textbf{Method}}                                   & \textbf{Result}      \\
            \hline

            \multirow{4}{*}{\Ac{femnist}} & \multicolumn{2}{l}{\Ac{fedavg} (LEAF)}                                & 74.72\%              \\
            \cline{2-4}
                                        & \multicolumn{2}{l}{\Ac{fedavg} (ours)}                                & 82.66\%              \\
            \cline{2-4}
                                        & \multirow{2}{*}{\Ac{fedq} (ours)}      & Queue Length 10              & \textbf{86.98\%}     \\
                                        &                                        & Queue Length 100             & 86.65\%              \\

            \hline

            \multirow{4}{*}{\Ac{celeba}}  & \multicolumn{2}{l}{\Ac{fedavg} (LEAF)}                                & 89.46\%              \\
            \cline{2-4}
                                        & \multicolumn{2}{l}{\Ac{fedavg} (ours)}                                & 91.24\%              \\
            \cline{2-4}
                                        & \multirow{2}{*}{\Ac{fedq} (ours)}      & Queue Length 10              & \textbf{91.98\%}     \\
                                        &                                        & Queue Length 100             & 91.58\%              \\

            \hline

            \multirow{7}{*}{Sentiment140} & \multirow{4}{*}{\Ac{fedavg} (LEAF)}    & $\ge 3$ Samples per Client   & \textasciitilde50\%* \\
                                        &                                        & $\ge 10$ Samples per Client  & \textasciitilde50\%* \\
                                        &                                        & $\ge 30$ Samples per Client  & \textasciitilde60\%* \\
                                        &                                        & $\ge 100$ Samples per Client & \textasciitilde69\%* \\
            \cline{2-4}
                                        & \multicolumn{2}{l}{\Ac{fedavg} (ours)}                                & 69.91\%              \\
            \cline{2-4}
                                        & \multirow{2}{*}{\Ac{fedq} (ours)}      & Queue Length 10              & \textbf{71.18\%}     \\
                                        &                                        & Queue Length 100             & 69.32\%              \\

            \hline

            \multirow{4}{*}{Reddit}       & \multicolumn{2}{l}{\Ac{fedavg} (LEAF)}                                & 13.35\%              \\
            \cline{2-4}
                                        & \multicolumn{2}{l}{\Ac{fedavg} (ours)}                                & 13.23\%              \\
            \cline{2-4}
                                        & \multirow{2}{*}{\Ac{fedq} (ours)}      & Queue Length 10              & \textbf{14.60\%}     \\
                                        &                                        & Queue Length 100             & 13.04\%              \\

            \hline
        \end{tabular}
    }

    \caption{Comparison of the \ac{fedq} LEAF benchmark results against the results published by~\citet{bibliography:leaf-a-benchmark-for-federated-settings}. *Please note that~\citeauthor{bibliography:leaf-a-benchmark-for-federated-settings} do not publish final accuracies for the Sentiment140 dataset. The accuracies shown in the table were read from the graph in Figure 3~\cite{bibliography:leaf-a-benchmark-for-federated-settings} and are only approximations.}
    \Description{Comparison of the \ac{fedq} LEAF benchmark results against the results published by~\citet{bibliography:leaf-a-benchmark-for-federated-settings}. *Please note that~\citeauthor{bibliography:leaf-a-benchmark-for-federated-settings} do not publish final accuracies for the Sentiment140 dataset. The accuracies shown in the table were read from the graph in Figure 3~\cite{bibliography:leaf-a-benchmark-for-federated-settings} and are only approximations.}
    \label{table:leaf-fedq-experiment-comparison-to-results-from-the-literature}
\end{table}

In conclusion, we think these experiments demonstrate that \ac{fedq} is capable of outperforming \ac{fedavg} on a wide variety of data modalities and training tasks. Although the margin with which \ac{fedq} outperforms the baseline varies with dataset and model architecture, its ability to drastically improve convergence speed makes it particularly efficacious.

\section{FedQ and Other Client Chaining Techniques}
\label{appendix:fedq-and-other-client-chaining-techniques}

This appendix section describes further techniques for \ac{fl} client chaining in comparison to \ac{fedq}, that have been developed in parallel to our method, and provides similarities and differences between them. \citet{bibliography:federated-learning-from-small-datasets}, for example, aim to improve \ac{fl} in scenarios where each client only has a small local dataset. They propose a technique called \ac{feddc}, where the central server, instead of aggregating the updated local models of the clients into a new global model, sends each updated local model $M_i$ to a randomly selected client $c_j$, where $i \neq j$. After a few rounds of this daisy-chaining, the resulting models are aggregated analogously to \ac{fedavg}. \citet{bibliography:fedcat} tackle the problem of non-\ac{iid} data in \ac{fl} and propose a technique called \ac{fedcat} that is essentially equivalent to \ac{feddc}. The main difference between \ac{feddc} and \ac{fedcat} is, that in \ac{fedcat} each model is trained by each client before they are aggregated to form a new global model and only the order of the client updates differs, while in \ac{feddc}, depending on the daisy-chaining period, each model is only trained on a random subset of all clients. \citet{bibliography:speeding-up-heterogeneous-federated-learning-with-sequentially-trained-superclients} also try to alleviate the problem of heterogeneous client datasets by proposing a technique called \ac{fedseq}. They perform a pre-training phase, after which they use the resulting model to estimate the data generating distribution of each client. Using the estimated distributions, they generate groups of clients with different local distributions, which they denote as superclients. During \ac{fl}, the clients within each superclient are trained sequentially, where the first client receives the global model and all consecutive clients receive the model of the previous client. The resulting local models of the superclients are then aggregated as in \ac{fedavg}.

All of the proposed techniques have similar goals and try to solve these problems by chaining the local training of multiple clients, but each of the techniques has variations in the training protocol that they follow. Both \ac{feddc} and \ac{fedcat} train as many different models as there are clients in each communication round. \Ac{fedseq} and our method \ac{fedq}, however, only train $\frac{\#clients}{\#clients \, per \, superclient/queue}$ models per communication round. In \ac{feddc}, \ac{fedcat}, and \ac{fedq} the clients for the sequential training are selected randomly, while in \ac{fedseq} they are purposely selected in order to group clients together that have different data generating distributions.

\end{document}